\documentstyle[fleqn,twoside,epsfig]{article}

\def\hitoshi#1{}
\def\michael#1{}
\def\intro#1{}
\def\beq{\begin{equation}}
\def\eeq#1{\label{#1}\end{equation}}
\def\eeqn{\end{equation}}
\def\beqa{\begin{eqnarray}}
\def\eeqa#1{\label{#1}\end{eqnarray}}
\def\eeqan{\end{eqnarray}}
\def\CR{\nonumber \\ }
\def\leqn#1{\ref{#1}}
\def\lsim{\mathrel{\mathpalette\vereq<}}
\def\gsim{\mathrel{\mathpalette\vereq>}}
\makeatletter
\def\vereq#1#2{\lower3pt\vbox{\baselineskip1.5pt \lineskip1.5pt
\ialign{$\m@th#1\hfill##\hfil$\crcr#2\crcr\sim\crcr}}}
\makeatother

\def\half{{1\over 2}}

\def\bar#1{\overline{#1}}
\def\bold#1{\setbox0=\hbox{$#1$}%
     \kern-.025em\copy0\kern-\wd0
     \kern.05em\copy0\kern-\wd0
     \kern-.025em\raise.0433em\box0 }
\def\VEV#1{\left\langle{ #1} \right\rangle}
\def\del{\partial}
\def\dslash{\not{\hbox{\kern-2pt $\partial$}}}
\def\Dslash{\not{\hbox{\kern-4pt $D$}}}
\def\Qslash{\not{\hbox{\kern-4pt $Q$}}}
\def\pslash{\not{\hbox{\kern-2.3pt $p$}}}
\def\kslash{\not{\hbox{\kern-2.3pt $k$}}}
\def\qslash{\not{\hbox{\kern-2.3pt $q$}}}
\def\pairof#1{#1^+ #1^-}

\def\ee{\pairof{e}}
\def\CM{{\hbox{\scriptsize CM}}}
\def\ECM{E_\CM}

\def\L{{\cal L}}

\def\sstw{\sin^2\theta_w}
\def\cstw{\cos^2\theta_w}
\def\mz{m_Z}

\def\mw{m_W}

\def\mt{m_t}

\def\msb{{\bar{\ssstyle M \kern -1pt S}}}

\def\ELER{e^-_Le^+_R}

\def\chone{\tilde{\chi}_1^\pm}
\def\chtwo{\tilde{\chi}_2^\pm}
\def\chonep{\tilde{\chi}_1^+}

\def\chonem{\tilde{\chi}_1^-}

\def\nuone{\tilde{\chi}_1^0}

\def\etal{et~al.}

\def\degrees{$^\circ$}

\begin{document}


\def\Title#1{\begin{center} {\Large #1 } \end{center}}
\def\Author#1{\begin{center}{ \sc #1} \end{center}}
\def\Address#1{\begin{center}{ \it #1} \end{center}}
\def\andauth{\begin{center}{and} \end{center}}
\def\submit#1{\begin{center}Submitted to {\sl #1} \end{center}}
\def\doeack{\footnote{Work supported by the Department of Energy,
                     contract DE--AC03--76SF00515.}}
\def\lblack{\footnote{Work supported in part by the Department of Energy,
                     contract DE--AC03--76SF00098, and by the National
           Science Foundation, grant PHY--90--21139.}}
\def\SLAC{Stanford Linear Accelerator Center\\
    Stanford University, Stanford, California 94309 USA}

\begin{flushright}\begin{tabular}{l} SLAC--PUB--7149\\
           LBNL--38808\\
           UCB--PTH--96/18\\
         June, 1996 \end{tabular}\end{flushright}
\vfill
\Title{Physics Opportunities of $\ee$ Linear Colliders}
\vfill
\Author{Hitoshi Murayama\lblack}
\Address{Department of Physics, University of California\\
Berkeley, California 94720}
\andauth
\Author{Michael E. Peskin\doeack}
\Address{\SLAC}

\vfill
\begin{quotation} \begin{center}
                       ABSTRACT
     \end{center}\bigskip 
We describe the anticipated experimental program of an $\ee$ linear collider
in the energy region 500 GeV--1.5 TeV, emphasizing topics relevant to the 
mystery of electroweak symmetry breaking.
\end{quotation}
\vfill

\vfill

\begin{center}
 to appear in {\it Annual Review of Nuclear and Particle Science}
\end{center}
\vfill

\newpage

\pagestyle{myheadings}
\markboth{\rm MURAYAMA \& PESKIN}{$\ee$ LINEAR COLLIDERS}
 
\noindent
{\huge PHYSICS OPPORTUNITIES OF  \\
$\ee$ LINEAR COLLIDERS}
\bigskip
 
\noindent
{\large {\it Hitoshi Murayama}}
 
\medskip
\noindent
Department of Physics, University of California\\
 Berkeley, California 94720\\
and\\
Theoretical Physics Group, Lawrence Berkeley National Laboratory\\
Berkeley, California 94720
 
\medskip
\noindent
{\large {\it Michael E. Peskin}}
 
\medskip
\noindent
Stanford Linear Accelerator Center, Stanford University\\
Stanford,
California 94309
 
\bigskip
\noindent
{\sc key words}:\quad electron-positron annihilation,
 $W$ boson, Higgs particles, top quark, supersymmetry
 
\bigskip
\hrule
\bigskip
\begin{abstract}
We describe the anticipated experimental program of an $\ee$ linear collider
in the energy region 500 GeV--1.5 TeV, emphasizing topics relevant to the
mystery of electroweak symmetry breaking.
\end{abstract}
 
\bigskip
\hrule
\bigskip
\tableofcontents
 
\bigskip
\section{INTRODUCTION}
 
  Elementary particle physics has always progressed by attacking its
mysteries simultaneously from many different directions. The parton model
of hadronic structure, for example, was developed both in response to
the discovery of limited transverse momentum in high-energy hadron
collisions and to the discovery of scaling in deep-inelastic electron
scattering.  As particle physics has moved to increasingly high energies,
however, the accelerators needed to reach these energies have become
progressively more expensive.  Thus our community has needed to
consolidate its efforts into the most promising channels.  It is inevitable
that this consolidation will continue into the future.
 
  But, it spite of this, it will continue to be important that
experiments confront new phenomena from distinct and complementary
perspectives. The exploration of the 100-GeV mass scale has been carried
out by proton-antiproton experiments at CERN and Fermilab, electron-positron
annihilation experiments at SLAC and CERN, and electron-proton scattering
experiments at DESY.  All of these experiments have contributed pieces
to the major result, the precise
confirmation of the standard model of electroweak
interactions.  In the future, as we explore the 1 TeV mass scale, we hope
that proton-proton and electron-positron collider experiments will both
be available.
 
  A proton-proton collider appropriate to this task, the LHC at CERN
     \cite{LHC}, has
already been approved. 
  The major physics goals of $pp$ experiments at TeV energies have been
summarized in many places, including earlier contributions to 
this series \cite{Chanowitz88,NEllis}, the physics
chapters of the LHC detector technical proposals \cite{cms, atlas}, and 
the classic review paper \cite{EHLQ}.
 In this article, we will present the corresponding review of the 
 major goals of $\ee$ experimentation at the next step of   high energy.
Our discussion will emphasize the unique capabilities of $\ee$
reactions, and the aspects in which $\ee$ experiments complement the
capabilities of $pp$ colliders.

  In the past few years, there have been a number of international 
conferences on physics of $\ee$ linear colliders whose proceedings are
valuable sourcebooks \cite{Finland,Hawaii,Morioka,DESY,DESYC}.  In addition,
a set of  useful review articles on future colliders have been
prepared for a recent study commissioned by the Division of Particles and
Fields of the American Physical Society \cite{Howiesbook}.  Part of our 
task will be to survey the information contained in these volumes.

\subsection{\it `Beyond the Standard Model'}
 
\intro{
In which we explain that the mystery of electroweak symmetry breaking is the
most important open problem of elementary particle physics, and that any
plan for future facilities has to be discussed in terms of its potential to
solve this problem.  (3 pages)}
 
In principle, one could discuss the physics goals of a proposed accelerator
simply by listing the various reactions it can produce and enumerating
the possible results to be obtained from each.  In this review, we will
take a more focused viewpoint.  The physics of the 100 GeV--1 TeV mass
scale is still largely unexplored territory, but it is not the complete
mystery that, for example, the asymptotic behavior of the strong
interactions was in 1960.  We are guided in our approach to this region
by the dramatic success of the standard model of strong, weak, and
electromagnetic interactions, and by the questions that this model
reserves to higher energies.  Indeed, we take the position that
there is a single most crucial problem to be solved by the next generation
of accelerators---to find the mechanism for the spontaneous breaking of
the electroweak gauge symmetry.  In our opinion, any proposal for a new
accelerator must ultimately justify itself by its ability to
uncover crucial clues to this problem.
 
What gives this particular problem such importance?  The first reason is the
contrast between our detailed knowledge of the gauge couplings of the standard
model and our ignorance of the physics of mass generation.  The
 Glashow-Weinberg-Salam $SU(2)\times U(1)$ theory of electroweak interactions
is now tested at the tenth-percent level, most dramatically in the experimental
determination of the $Z^0$ partial widths and asymmetries at LEP and SLC.
\cite{LangPD, Hagiwara95}
 These experiments directly test the central assumptions of the
$SU(2)\times U(1)$ model.  They show that the left- and
right-handed components of the quarks and leptons have
completely different couplings to the fundamental electroweak gauge bosons.
Thus, these components must be viewed as distinct
species at high energy.  At the same time, they show
that the weak interaction coupling constants are universal among
species. This strongly suggests
that the electroweak bosons are the vector
bosons of a gauge theory.  These two facts imply that neither the
elementary fermions nor the elementary vector bosons can obtain mass without
the spontaneous breaking of the gauge symmetry.  However, the $SU(2)\times 
U(1)$ model does not
contain a physical mechanism for breaking its own symmetry, since the
electroweak interactions are weakly coupled.  Some external agent, 
a new particle
or sector of particles, is required.
 
Second, the physics of this new sector should be very close at hand, at an
energy scale within the reach of the next generation of accelerators.  The
gauge relations of the $SU(2)\times U(1)$ model give for the
$W$ boson mass a formula  $\mw = \half g v$, where $g$ is the $SU(2)$ gauge
coupling and $v$ is a mass scale characteristic of the spontaneous
symmetry breaking.  In the simplest model, in which the gauge symmetry is
broken by the expectation value of a single scalar field, $v$ is the
size of this vacuum expectation value.  From the known values of
the $W$ mass and the $SU(2)$ gauge coupling, we have
\beq
           v = 250 \ \hbox{\rm GeV}\ .
\eeq{vvev}
This scale should set at least the order of magnitude for the masses of the
new particles which cause electroweak symmetry breaking.	 To find the
detailed relation between $v$ and these masses, one must study explicit
models of electroweak symmetry breaking, and the answer is somewhat
model-dependent. 
Nevertheless, it is true in all but
the most extreme
models that these particles are accessible to a $pp$ collider at 14 TeV
in the center of mass and to an $\ee$ collider at
1.5 TeV in the center of mass.
 
Finally, the physics of electroweak symmetry breaking is important because
it enters into the discussion of all of the other fundamental problems of the
theory of elementary particles.  We have already explained that this
symmetry breaking is crucial for the generation of quark and lepton masses,
so any explanation of the fermion mass spectrum, and the related problems
of the origin of the quark mixing angles and $CP$ violation, must begin
by assuming a specific mechanism of electroweak symmetry breaking. The
same conclusion holds for problems less obviously connected to mass
generation.  Consider, for example, the possibility of lepton number
violation observed in the process $\mu\to e \gamma$.  If this process were
observed, it would be a spectacular discovery, but its implications for the
broader theory of Nature would be left obscure.  Models of  $\mu\to e \gamma$
include ones based on heavy neutral leptons \cite{BjandW}, on extended
technicolor \cite{EandLmu}, and  on supersymmetric grand unification
\cite{BarbandH}. The broad classes of models for this process, or any
similar exotic process, are distinguished precisely by their assumptions
about the physics of electroweak symmetry breaking.  So it is not enough
to search for anomalies; even to understand the consequences of these
searches, we must go to the electroweak scale and see what is there.
 
This review will be organized around the ability of a proposed $\ee$ collider
to study the implications of various models of electroweak symmetry breaking.
We begin in Sections 2 by providing background material on the
accelerator and detector designs for these colliders. In Sections 3 and 4,
we discuss two exotic standard-model reactions that will be
explored in detail at this collider, $\ee\to W^+W^-$ and $\ee \to t\bar t$.
Both of these reactions have unusual features that should 
already provide
an interesting experimental program, but they are only a prelude
to the real interest of this machine in studying the electroweak scale.
 
In Sections 5--7, we discuss specific models of electroweak symmetry
breaking and their experimental consequences at $\ee$ colliders.  Models
of electroweak symmetry breaking divide generally into two classes---those
models in which the physics is essentially weak-coupling, and those in
which this physics is strong-coupling.  In models of the first class, the
electroweak
symmetry is broken by the vacuum expectation value of an elementary scalar
field, called the Higgs field.  The simplest model contains only one
Higgs field, and one new particle, the Higgs boson.
 This theory is sometimes dignified with the title ``the minimal
standard model,'' but it is not really a model at all; it does not
explain electroweak symmetry breaking and it cannot naturally be incorporated
into a unified model of the fundamental interactions.  More general models
can be built with several Higgs fields and many more free parameters.
However, the only models of this type that are conceptually coherent and
also have the power to explain electroweak symmetry breaking are those that
incorporate an additional symmetry, called
supersymmetry.
In this case, the experimental signatures can be fully worked out and
capabilities of various collider options discussed quantitatively.
In Section 5, we will
discuss experiments at an $\ee$ collider on the Higgs boson and its possible
scalar counterparts.  In Section 6, we will discuss experiments on the
additional new particles predicted by supersymmetry. 

 In Section 7, we will
turn to the second class of 
models in which electroweak symmetry breaking is caused by new
strong-coupling dynamics at the TeV scale.
These models do not contain elementary Higgs fields at all
but instead postulate new forces that lead to electroweak symmetry
breaking.  Because of their strong interactions,
 it is difficult in this case to completely predict the 
properties of the model; thus, many aspects of phenomenology must be 
discussed in a qualitative way.  However, we can still provide an 
overview of the  variety of experimental signatures available.

Finally,  in Section 8, we give a lightning review of other models of 
new physics that can be tested at $\ee$ colliders.

\subsection{\it Special Features of $\ee$ Experimentation}
 
\intro{
In which we explain the general advantages of $\ee$ experimentation:
low and calculable backgrounds, democracy between standard and exotic
processes, and full-event analysis.  (2 pages)}
 
As an introduction to this review, we will discuss in this section three
general features of the experimental environment provided by $\ee$
annihilation.  Electron-positron colliders played a major
role in the discoveries of the 1970's and the confirmation of the standard
model in the 1980's because they offer to experimenters a number of
aspects that simplify the investigation of exotic phenomena.  We will argue
in this review that these features, which are familiar from $\ee$
experiments at present energies, should also be present in the $\ee$
experiments of the future.
 
The first of these features is what is often called the ``cleanliness'' of
$\ee$ reactions, the fact that standard-model event  rates are relatively
low.  At high energy, two somewhat different aspects of the standard model
processes are important, that these processes have relatively
simple topology, and that their rates are precisely calculable.  The
standard-model background processes with the largest cross sections are
photon-photon collisions and radiative annihilation processes ($\ee\to
q \bar q \gamma$); however, these processes are eliminated by simple
cuts on total visible energy and energy balance.  The annihilation process
$\ee \to q\bar q$ is eliminated equally simply by removing two-jet-like
events.  This leaves as the dominant backgrounds for exotic processes
reactions that themselves involve heavy species, in particular,
$\ee\to W^+W^-$ and $\ee\to t \bar t$. We will see that this is the
normal situation in the specific analyses to be discussed below.
  General studies of background
levels at linear collider energies are reviewed, for example,  in 
\cite{LaThuile,Ahn,JLCWS2}, and, briefly, in section 2.2.
 
The second general feature of $\ee$ annihilation is one that we might
call ``democracy.''  The typical values of cross sections in $\ee$
annihilation are set by the point cross section
\beq
      1\ \hbox{\rm R} =  {4\pi \alpha^2\over 3 s}  = 
{86.8\ \hbox{\rm fb} \over (\ECM\
                      \hbox{\rm (TeV)})^2} \ .
\eeq{Rdef}
As long as a given process is kinematically allowed, its cross section will
be of order 1 R times the squares of gauge charges.  Thus, exotic processes
typically occur at the rates of standard-model process.  On the other hand,
the point cross section given in Eq.~\leqn{Rdef} is rather small, and 
this poses a
challenge to accelerator designers.
 
The third general feature of $\ee$ annihilation is one that we (being
Californians) might call ``holism,'' the fact that typically the complete event
is captured, so that its full kinematic information can be used.  In any
study of new physics processes at TeV energies, it is typical that
both the signal and the dominant background processes will contain $W$ bosons.
If these $W$ bosons can be reconstructed, their decay distributions indicate
their polarizations, and this polarization information can become an
important ingredient in the analysis.  We will discuss several examples
in which the decay distributions of heavier particles also come into play.
In addition, $\ee$ colliders offer the freedom to adjust the electron
polarization and the availability of $b$-quark tagging with high efficiency.
We will see how all of these handles can work together to detect and
characterize an exotic reaction.
 
These three themes---cleanliness, democracy, holism---will run through all
of the specific examples of future $\ee$ experiments that we will discuss
below.

\subsection{\it Complementarity of $\ee$ and $pp$ Experiments}
 
\intro{
In which we explain, with a few examples to be amplified later, that
$\ee$ and $pp$ experiments, by accessing different types of processes and
 observables, offer complementary windows into the physics of electroweak
symmetry breaking.   (1 page)}
 
As we noted in the first paragraphs of this article, the argument for a major
new collider must rest not only on the absolute merit of that
accelerator but also on the contribution it will make to the overall program
of high-energy physics.  We must argue, in particular, that the  goals of
experimentation at an $\ee$ collider will not already be met by experiments
at hadron colliders operating in the same time period, including the LHC.
  In fact, as we will see, experiments at $\ee$ and
$pp$ colliders are wonderfully complementary.  As we survey models of
electroweak symmetry breaking in the discussion below, we will see that these
models are typically accessible both to $\ee$ and $pp$ experiments, through
different channels.  In the most important models, the
complete phenomenological portrait is obtained only by combining the
information that these two distinct types of experimentation will make
available.
 
We can illustrate this point, and give examples of the three themes of $\ee$
experimentation, by highlighting some
examples to be discussed in detail later:
 
\begin{enumerate}
 
\item The production of a light Higgs boson is a rare process at $pp$
colliders; this particle can be found at LHC, for example, only by
concentrating on specific decays that give characteristic signatures in
the hadronic environment.  On the other hand,
Higgs boson production has a rate at $\ee$ colliders that is typical of
annihilation processes.  This allows the
observation of the Higgs boson in many distinct decay modes and the
measurement of its branching ratios.  We will discuss these experiments in
Section 5.3.
 
\item  The production cross section for top quarks at the LHC is enormous,
 allowing searches for rare top quark decays to the level of  $10^{-4}$
in the branching ratio \cite{atlas}.  On the other hand, exotic physics
associated with the top quark is more often reflected in modification of the
top-quark couplings to gauge bosons.  The possibility of whole-event analysis
in the $\ee$ environment
allows these couplings to be measured accurately.  We will discuss this
experiment in Sections 4.3 and 7.5.
 
\item Supersymmetry partners of the quarks, gluons, and gauge bosons can be
discovered in $pp$ collisions through a wide variety of signatures.
  However, while it is easy in this
environment to identify anomalies, it is difficult to interpret these
anomalies in terms of a specific underlying supersymmetry spectrum.  On the
other hand, $\ee$ colliders offer specific reactions
 and tools involving whole-event properties by which one can 
measure the underlying supersymmetry parameters.
  We will discuss these experiments in Section 6.2 and 6.3.
 
\item If the Higgs sector is strongly interacting, we will argue below that
one should expect enhanced cross sections for $WW$ scattering that should be
visible both in $pp$ and in $\ee$ experiments. However, $\ee$ experiments
offer another window into the strongly interacting Higgs sector which is often
more sensitive.  This quantity is found in the
detailed analysis of $\ee$ annihilation into $W$ pairs, a process that is,
because of the democracy of reaction rates, a major component of the total
annihilation cross section.  We will discuss this experiment in Section 7.3.
 
\end{enumerate}
 
Through the broad survey of models that we will make in this article, we will
argue that $\ee$ experiments should bring new and crucial information on
the mechanism of electroweak symmetry breaking, over the whole range of ideas
for what that mechanism might be.

\section{THE LINEAR COLLIDER ENVIRONMENT}
 
If we are to discuss the realistic capabilities of $\ee$ colliders to
discover aspects of the new physics of electroweak symmetry breaking, we must
refer to specific machine and detector parameters and discuss the dominant
backgrounds that experiments will need to deal with.  In this section, we will
briefly review these issues.
 
\subsection{\it Design Parameters of Linear Colliders}
 
\intro{
In which we review the main features of linear collider designs, and the
constraints on experimentation which follow from these.  (3 pages)}
 
First of all, what are realistic values of the energy and luminosity to use
in evaluating the capabilities of $\ee$ colliders?   In the energy region
that we are discussing, with $\ECM$  several hundred GeV or greater, the
preferred accelerator configuration is an $\ee$ linear collider.
The physics issues of
the design of linear colliders have been reviewed in an earlier article
of this series \cite{Palmer}, but there  has been tremendous progress
since that
time.  The technology of linear colliders has more  recently been surveyed
in a series of international conferences \cite{LC92,LC93,LC95}, and
in a major international technical review \cite{Loew}.

\begin{table}
\caption[parameters]{Parameters of proposed $\ee$ linear colliders.
In this table, we list
the center of mass energy; the microwave frequency;
$\L_0$,  the nominal luminosity (before accounting for the
 beam-beam interaction);
$\L$, the final predicted luminosity;
 $f$, the pulse frequency;
$N$, the number of particles per bunch;
$n_b$, the number of bunches per pulse;  $\Delta t$, the spacing of 
bunches;
$P$, the beam power; grad., the accelerating 
gradient for the  unloaded
accelerating structure; linac l., the total length of the two linear
accelerators;
 $\sigma^*_i$, 
the nominal bunch size at the collision point; $\delta_B$,  the energy
spread due to beamstrahlung; $n_\gamma$,  the number of photons per $e$
produced in the collision;  $N_{\hbox{\scriptsize pairs}}$,
  the number of $\ee$ pairs
appearing above 150 mrad; $N_{\hbox{\scriptsize had}}$,   the 
number of hadronic events, and $N_{\hbox{\scriptsize jets}}$,
  the number of hadronic 
events with jets of $p_T > 3.2$ GeV.  The last three quantities are 
calculated per bunch collision.}
\centerline{\small
\begin{tabular}{lllllll}
\hline\hline
& \multicolumn{4}{c}{500 GeV} & 1 TeV & 1.5 TeV\\ \cline{2-7}
& \bf TESLA & \bf JLC(X) & \bf NLC & \bf CLIC & \bf NLC & \bf NLC\\
\hline
$E_\CM$ (GeV)
& 500 & 500  & 500 & 500 & 1000 & 1500 \\
RF freq. (GHz)
& 1.3 & 11.4 & 11.4 & 30 & 11.4 & 11.4\\
$\L_0$ ($10^{33}$)
& 2.6 & 5.1 & 5.3 & 3.4 & 10.4 & 10.5 \\
$\L$ ($10^{33})$
& 6.1 & 5.2 & 7.1 & 4.8 & 14.5 & 11.7 \\
$f$ (Hz)
& 10 &  150 & 180 & 1210 & 120 & 120 \\
$N$ ($10^{10}$)
& 5.15  & 0.63 & 0.65 & 0.8 & 1.1 & 1.1 \\
$n_b$
& 800 & 85 & 90 & 10 & 75& 75 \\
$\Delta t$  (nsec)
& 1000 & 1.4 & 1.4 & 0.67 & 1.4 & 1.4\\
$P$ (MW)
& 16.5 & 3.2 & 4.2 & 3.9 & 7.9 &  11.9\\
grad. (MV/m)
& 25 & 73 & 50 & 80 & 85 & 85\\
linac l. (km)
& 29 & 10.4 & 15.6 & 8.8 & 18.7 & 28.0\\
$\sigma^*_x$ (nm) 
& 1000 & 260 & 320 & 247 &  360 & 360\\
$\sigma^*_y$ (nm) 
& 64   &  3.0  & 3.2  & 7.4 & 2.3 & 2.3 \\
$\sigma^*_z$ ($\mu$m)
& 1000  & 90 & 100 & 200  & 100 & 200\\
$\delta_B (\%)$
& 3.3 & 3.5 & 2.4 & 3.6 & 7.4 & 9.0\\
$n_\gamma$
& 2.7 & 0.94 & 0.8 & 1.35 & 1.1 &  1.1\\
$N_{\hbox{\scriptsize pairs}}$ 
& 19.0 & 2.9 & 2.0 & 3.0 & 7.0 &7.0\\
$N_{\hbox{\scriptsize had}}$
& 0.17  & 0.05 & 0.03 & 0.05 & 0.18& 0.23\\
$N_{\hbox{\scriptsize jets}} (10^{-2}$) 
& 0.16 & 0.14 & 0.08 & 0.10 & 1.4 & 3.1
\\ \hline
\end{tabular}
 }

\end{table}

Table 1 summarizes the current design parameters of planned linear
 colliders, as envisioned by the accelerator physics groups at 
DESY, KEK, SLAC, and CERN, as reported in \cite{Loew, Pisin}.
  To facilitate the comparison of options,
we have presented four designs at the common center-of-mass energy of 500 GeV
and then shown the extension of one of these designs to 1 TeV and 1.5 TeV.
Many additional designs, both at 500 GeV and at higher energy, are 
discussed in \cite{Loew}.
Though each of these designs represents a detailed and complex optimization,
it is not difficult to understand the concepts involved in these
designs if we review the general constraints coming from basic physics
considerations.
 
From the experimenter's point of view, a collider is 
pa\-ra\-me\-trized by
the energy and  luminosity that it can deliver.  For an $\ee$ linear collider,
it is easy to imagine strategies for increasing the energy; one can make the
linear accelerator longer, or one can increase the strength of the accelerating
fields.  However, the small size of the point cross section, Eq.~\leqn{Rdef},
indicates that increasing  the luminosity will also be a crucial issue.
To go to an energy two times higher, we require a luminosity four times
higher to study  physics processes with comparable statistics.  The
luminosity of a linear collider is determined by the formula
\beq
      \L =  {1\over 4\pi} {N^2 f\over \sigma_x \sigma_y} \ ,
\eeq{Lform}
where $N$ is the number of particles per bunch, $f$ is the bunch collision
rate, and $\sigma_x$ and $\sigma_y$ are the bunch height and width, assuming
a Gaussian profile.  Though it might seem that the number of particles per
bunch would be fixed by beam-loading limits and other accelerator-related
constraints, a very significant limit comes from the physics of the
electron-positron bunch collisions. 
The tightly bunched beams required for high-luminosity operation create
intense electromagnetic fields as seen by the particles in the opposite
bunch.  These fields can produce coherent, bunch-induced radiation
(``beamstrahlung'') \cite{beamstr,HimandS,Noble}
  and $\ee$ pair creation  \cite{ChenandTel}
at the interaction
point.  Assuming the most favorable case of flat  beams,
 $\sigma_x \gg \sigma_y$,
the average number of beamstrahlung photons per beam
particle  is given by
\beq
     n_\gamma = {2\alpha r_e N\over \sigma_x}\ ,
\eeq{ngform}
where $r_e$ is the classical electron radius.  To minimize collision-related
backgrounds, $n_\gamma$ must be kept to about 1.  Thus, we should rewrite
Eq.~\leqn{Lform} so that $n_\gamma$ appears as a parameter.  We find
\beq
    \L = {1\over 8\pi\alpha r_e m_e c^2} {P n_\gamma\over \gamma \sigma_y} \ ,
\eeq{Lforpower}
where $P = N f \gamma m_e c^2$ is the power in each beam.
 
  This formula for the luminosity
makes clear that the crucial considerations for
the the design of linear colliders are (1) to maximize the efficiency of
the transfer of external electric power to power in the beam, and (2)  to
create and maintain extremely small beam spots.  To the extent that one can
limit the power cost in providing the beam energy, it is possible to allow
less stringent tolerances in beam size.  This has led to two distinct
strategies for the design of linear colliders.  The first, reflected in
the JLC(X) and   NLC
designs in Table 1, has emphasized improving the
efficiency and
beam handling in a linear accelerator design with standard copper
accelerating cavities.    The second, reflected in the TESLA design, has
envisioned the use of superconducting accelerating cavities.  In the design
for a 500 GeV machine, the choice of a superconducting accelerator leads to
significantly milder tolerances in beam size.  However, this advantage
goes away at higher energies due to lower accelerating fields allowed
by the superconducting medium (40 MeV/m, as opposed to about 90 MeV/m in
the copper cavity designs).   Both strategies limit the number of particles
per bunch collision by accelerating trains of bunches.  The
copper-cavity NLC design, for example, contains trains of 90 bunches
accelerated in 1.4 nsec intervals  spaced 120/sec.  The superconducting
TESLA design envisions trains of 800 bunches per second in  1 $\mu$sec
intervals.  The CLIC design in Table 1 uses a more exotic but possibly
more efficient RF source, in which  the electromagnetic
 fields of a comoving relativistic beam transfer power to the high-energy
beam.

 To obtain some idea of the evolution of the machine parameters and physics
backgrounds as the energy of the machine is increased, we have presented  
in Table 1 the
parameters of the NLC design for 500 GeV, 1 TeV, and 1.5 TeV in the 
center of mass.  The first two stages of the NLC design
have been worked out in much more detail in a recent
 report \cite{ZDR,ZDRlite}.	
The 500-GeV and 1-TeV designs
involve essentially the same length of accelerating structure.  The main
difference between the two designs comes in the RF power requirements, that is,
in the assumptions about the efficiency and yield of the klystrons that
produce the microwave power. As of this writing, the klystrons that have been
produced at SLAC and KEK meet the specifications for the 500-GeV design,
and the report \cite{ZDR} envisions a smooth evolution to a 1-TeV machine.
 The 1.5-TeV design is shown in Table 1  as an
increase in the length of the machine, although some of the 
energy increase could also be achieved by improved klystron performance.
 
  The idea that a linear $\ee$ collider is capable of a smooth program of
energy upgrades may be unfamiliar to high-energy physicists used to thinking
about circular $\ee$ colliders.  For circular machines, the RF power
demands for increasing the energy at fixed radius grow as $E^4$ and
 provide an insuperable cutoff.  For linear machines, these
demands grow only as $E$.  It is perhaps worth remembering
that the Stanford Linear Accelerator turned on as a 17-GeV machine and now runs
at 50 GeV, without any increase in length \cite{SLCrev}.
 
  We will see in our discussion below that this idea of the machine upgrade
path corresponds nicely to the physics that the linear collider will explore.
The physics program of the linear collider should begin with programmatic
standard model physics at center of mass energy of 400 GeV---the study of the
top quark at its threshold and the study of the $W$-boson couplings.  In the
weak-coupling models of electroweak symmetry breaking,
 the Higgs boson should also be found already at this energy.  In a
weak-coupling scenario of any complexity, there should be other new particles
at the mass scale of 400-500 GeV; we will argue this specifically in our
discussion of supersymmetric models in Section 6.  In these models, the
possibility of extension to 1 TeV provides a factor of two safety margin in
the estimates for new particle masses.
 
On the other hand, if electroweak
gauge symmetry is broken by an essentially strong-coupling mechanism, there is
no guarantee of new physics easily accessible either to hadron or electron
colliders. By this, we do not mean to imply that accessible signatures are
not expected.  In fact, as we will discuss in Sections 7.4 and 7.5, explicit
realistic models of strong-coupling electroweak symmetry breaking contain
a variety of interesting signatures below 1 TeV.  But there is no 
model-independent argument that this must be so.  If indeed Nature chooses
to hide the electroweak symmetry breaking sector as well as possible, 
 experimenters both at hadron and at
electron colliders must prepare for a long campaign emphasizing high
integrated luminosity.  In this context, a substantial upgrade of the
linear collider would be appropriate.
 
In our physics discussion, we will emphasize the capabilities of the
first-stage linear collider.
We will assume a luminosity of roughly 15,000 R$^{-1}$
 per design year, corresponding
to 50 fb$^{-1}$ per year ($5\times 10^{33}$ cm$^{-2}$ sec$^{-1}$)
 at 500 GeV in the
center of mass and to 200 fb$^{-1}$ per year at 1 TeV.  For the most part, we
will discuss physics studies at 500 GeV.  The reader should understand
that the  results of these studies generally scale smoothly to 1 TeV and
provide the requisite margin of safety for new-particle searches.
Specifically in Sections 7.1 and 7.2, we will discuss advanced experiments
 requiring a center-of-mass energy of 1.5 TeV and luminosity samples of
200 fb$^{-1}$.

\subsection{\it Standard Model and  Background Processes
at High Energy $\ee$ Colliders}
 
\intro{
     In which we review the major standard-model processes and the dominant
sources of background for new particle search experiments. (2 pages)}
 
As we discuss specific particle search experiments and analyses, it will be
useful to understand the most important background processes due to
standard model physics.  In addition, we will discuss backgrounds
associated with the intense bunch collisions required by
the accelerator.
 
There are three types of important standard-model processes in high-energy
$\ee$ collisions.  First of all, there are $\ee$ annihilation processes,
to quark, lepton, and also $W$ and $Z$ boson pairs.  The characteristics
of light quark and lepton pair production are familiar from lower-energy $\ee$
reactions: the hadronic events are two-jet-like and both types of event are
strongly coplanar.  These events are eliminated as background processes
by methods similar to those used in particle search experiments at LEP
(see, for example,  \cite{ALEPHsearch}).
The new processes of $W$ and $t$-quark production, which
could themselves be viewed as exotic processes of the high-energy regime,
make major contributions to the annihilation cross section.  The total
cross sections for these two processes at 500 GeV are 20 R and 1.7 R,
respectively, as compared to 7.6 R for light-quark pair production.  The
pair production of $W$ and $t$ are the major backgrounds to
most of the processes
from beyond the standard model that we will discuss below.

The second type of process is the two-photon reaction.  These reactions are
also familiar from lower-energy $\ee$ experiments, in which
the colliding photons are virtual photons from the Weizs\"acker--Williams
photon distribution associated with each electron.  At linear colliders,
there may be an additional component of the two-photon process arising from
beamstrahlung photons.  In addition, it is important to realize that
the cross section for $W$ pair production in two-photon collisions can be 
very large; it increases from 0.6 R to 92 R as $\ECM$ increases from 500 GeV
to 1.5 TeV. In experiments that focus on annihilation
processes, two-photon
 processes are removed straightforwardly by total  energy and
acoplanarity cuts.  In the $WW$ scattering experiments described in Section
7.2, however, they are a major background and require special discussion.
 
Finally, there are processes in which the electron or positron radiates a
heavier gauge boson.  Of these, the process $\ee\to \gamma Z^0$ is important
even at LEP 2 energies, but even there leads to a highly boosted $Z^0$ that
is lost in the forward region of the detector.  At 500 GeV, the decay
products of the $Z^0$ in this process typically lie within an angle of
150 mrad.  Other peripheral boson-production processes have very small
cross sections and are rarely relevant.
 
\begin{figure}
\psfig{file=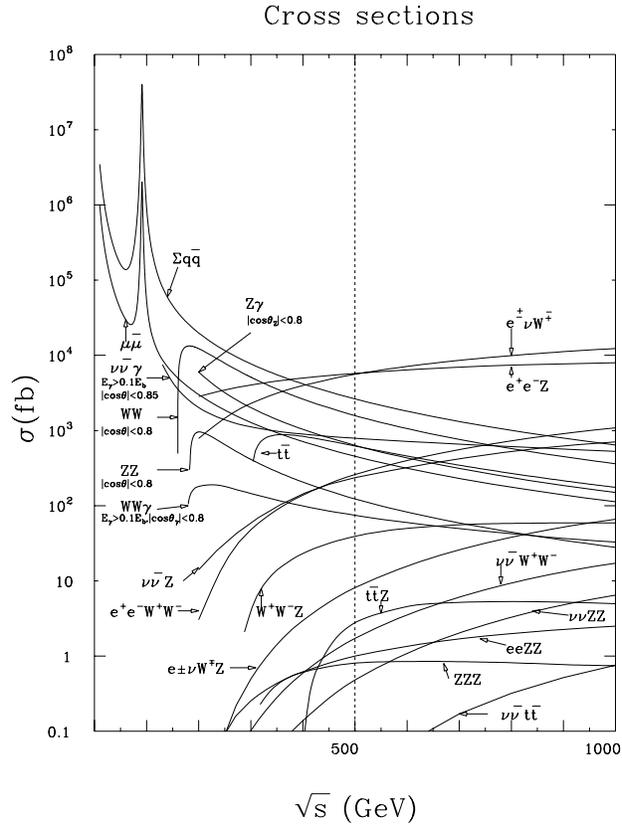,width=\textwidth,angle=90}
\caption[SMProc]{Total cross sections for the 
major standard-model physics processes
 at $\ee$ linear colliders, as a function of center-of-mass energy,
   from \cite{Miyamoto-anom}.}
\label{SMProc}
\end{figure}
 
A summary of all three classes of reactions is given in Figure~\ref{SMProc}
\cite{Miyamoto-anom}, which
plots the total cross sections for a wide variety of standard model
processes versus energy.

We have already noted that the specifications of an $\ee$ linear collider
require substantial photon radiation in the $\ee$ bunch collision process.
At first sight, this situation seems to contrast with that at lower-energy
$\ee$ colliders, where the distribution of collision energies is given by
folding a machine energy spread of about 0.1\% with
the results of initial-state photon radiation.  However, it turns out that the
main difficulty comes in controlling the rate of $\ee$ pair production
due to photon annihilation in the collision region.  The linear collider
designs presented in the previous section typically produce of order
$10^5$ $\ee$ pairs per bunch crosssing. A mask in the detector at
an angle of 150 mrad removes all but a few per bunch collision.  There are
two additional
complications, to be discussed in a moment, but 
once this effect is kept under control, they 
may be seen to be  quite tolerable.

The first of these is the broadening of the spectrum of center of mass
energies due to beamstrahlung. Though  at first sight this is a serious
concern, the effect  is relatively small in realistic designs.  The energy
spread due to beamstrahlung is tabulated as $\delta_B$ in Table 1.
Except at the highest energies, it is comparable to the energy spread due
to initial-state radiation, which is of order $(\alpha/\pi)\log(\ECM/m_e)
\sim 3$\%.
 
The second possible problem is that of hadron
production in relatively low energy
two-photon reactions.  Drees and Godbole \cite{DG91}
suggested that the two-photon reaction might potentially provide an
underlying hadronic event for each high-energy annhilation.  This question
was reexamined in \cite{FandSgam,BCP}, giving the much lower rates tabulated
in the last two rows of Tables 1.  More important, when the 
extra hadrons do occur, they carry very low energy.  At 500 GeV, these 
background processes
typically deposit less than 5 GeV in the detector.

\subsection{\it Characteristics of Linear-Collider Detectors}
 
\intro{
In which we review the general properties of the detectors used in
simulations of linear collider physics, and the main open issues in
detector design.  (2 pages)}
 
Studies of physics processes at linear colliders must assume a particular
detector configuration.  For the most part, though, it has been anticipated
that $\ee$ detectors of the future will resemble those of the past and
present in being conventional $4\pi$ devices that compromise between
tracking and calorimetry.  Many of the studies that we will describe below
use detector models based on the capabilities of existing detectors at
$Z^0$ energies, in particular, ALEPH \cite{ALEPH} and SLD \cite{SLD}.

The main exception to this rule comes in the work of the Japan Linear
Collider (JLC) group.  The JLC studies have incorporated a model detector
about  50\% larger than ALEPH, which includes both enhanced
tracking and calorimetric capabilities  \cite{JLC-1}.
  The resolution of the detector
is projected to be, for the hadron calorimeter,
 $\Delta E/E = 40$\%/$\sqrt{E} + 2$\%,
for the electromagnetic calorimenter,  
$\Delta E/E = 15$\%/$\sqrt{E} + 1$\%, and for the
tracking, $\Delta p_T/p_T = 1.1\times 10^{-4} p_T$/GeV.
 Both types of improvements are directed to 
an important physics capability for the linear collider experiments.
In many processes at the linear collider, 
$W$ and $Z$ bosons are identified in their hadronic decay modes. The JLC 
design achieves a resolution of 3.5 GeV in reconstructed two-jet invariant 
mass at the $W$ mass scale using calorimetry only.  Then, by combining
calorimetry and tracking information, once can achieve a
mass resolution comparable to the natural width of the $W$.  This
makes possible
the separation of $W$ and $Z$ bosons on the
basis of the reconstructed
mass \cite{Miyamoto}.  This separation is useful even
in the light Higgs boson
analyses at low energies, and it becomes a very important tool in the
$WW$ scattering analysis  
described in Section 7.2. 

\begin{figure}[t]
\centerline{\psfig{file=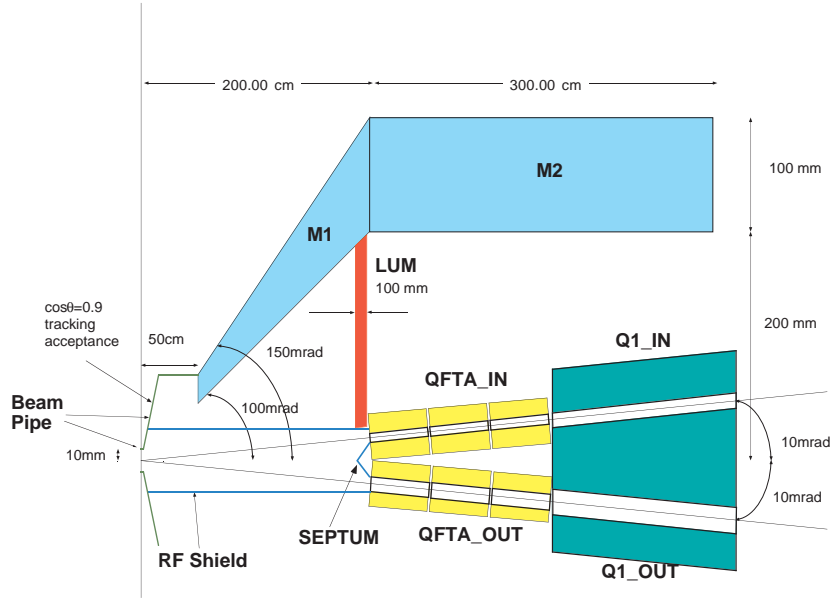,width=\textwidth}}
\caption[intersect]{An illustrative diagram of the NLC intersection region,
showing the positions of the last final-focus quadrupole, the exit hole for 
the oppposite beam, the beamstrahlung mask, and the luminosity monitor.
Note that the figure is stretched by a factor of 10 in the vertical direction.}
\label{intersect}
\end{figure}

Beyond the general layout of the detector, there are four features of
experimentation that deserve special comment.
First, as we have noted in the previous section, linear collider
detectors require a mask protecting them from the substantial $\ee$
pair production at forward angles.  A typical intersection region design
is shown in Figure~\ref{intersect}.  The presence of
 this mask makes the detector
essentially blind to particles produced in the forward and backward directions.
In the simulations we will describe below, particles with $\theta$ within
150 mrad of the beam direction are simply ignored.  Though one might
anticipate that this would cause difficulties in calorimetric energy
reconstruction and missing energy identification, in practice the
interesting  $\ee$ production processes
 are so central that
this cut has very little effect.
 
 The second necessary  feature is a device to  calibrate
luminosity and its spectrum.  We have explained already that the
spectrum of photon radiated from the collision region, and, concomitantly,
the detailed spectrum of $\ee$ center-of-mass energies, depends on the
parameters of the colliding electron and positron bunches.  Most physics
processes
at an $\ee$ linear collider are not sensitive to the initial-state radiation
at this level of detail, but there are a few measurements for which the
knowledge of this distribution is crucial.  The most important of these is the
measurement of the top quark production cross section near threshold,
described
below in Section 4.2.  Frary and 
Miller \cite{FMill} have
 shown that it is possible to 
monitor the spectrum of $\ee$ collision energies experimentally
by measuring the small-angle
Bhabha scattering cross section at angles near the mask.  The position and
size of an appropriate electromagnetic shower detector is indicated in
Figure~\ref{intersect}.

The third aspect of experimentation that deserves a special comment
is the vertex detector.  Because the linear collider experiments focus on
the properties of the Higgs sector, which couples most strongly to
heavy flavors, $b$-tagging is an important tool in many different analyses.
The quality of $b$-tagging assumed in the physics studies
reviewed below is that of current $\ee$ detectors. 
However, because of the extremely small
beam spot sizes expected for linear colliders, one might imagine that
vertex detectors could be
placed much closer to the interaction point. A recent design envisions a
compact tracking system with a 4-Tesla magnetic field to sweep  away  soft
$\ee$ pairs from the bunch collision; this allows a 
CCD vertex detector to be placed   within 2 cm of the interaction 
point \cite{ZDRlite}.
 
The final noteworthy aspect of linear-collider experimentation is the
availability of polarized electron beams.  At low energies, where physics
is dominated by the parity-conserving
strong and electromagnetic interactions, the use of beam polarization has
 limited importance.  However, for energies at the
weak scale and above, the dependence on beam polarization becomes an
essential part of the phenomenology.  We have already noted, in Section 1.1,
that at high energies the left- and right-handed electrons are distinct
species with different $SU(2)\times U(1)$ quantum numbers.  These species
have completely different couplings both to new particles and to the gauge
bosons of the standard model.  Then the 
 differences between the reactions induced
by left- and right-handed electrons can be a key diagnostic tool.  At the
very least, one has the profound effect that the cross section for $\ee \to
W^+ W^-$ is smaller by a factor 30 for right-handed electrons, so that the
control of polarization gives one direct control of this important background
process.
 
There are many 
obstacles to achieving polarized beams in circular colliders \cite{polC}.
But in a linear collider, a beam that is initially polarized longitudinally
naturally retains its longitudinal polarization during acceleration  and
transport.
The degree of polarization to be expected, then, is essentially given by the
properties of the electron source.  For many years, the best cathode
materials allowed an electron polarization of 50\% in the ideal case and
roughly 20 \% in practice.  In 1991, however, groups at SLAC and
Nagoya \cite{Maru,Omori}
 learned to grow gallium arsenide cathodes as a surface layer on 
a substrate ({\it e.g.}, GaAsP) of a slightly different lattice spacing.
The resulting strain  breaks the symmetry between
electron levels with opposite spin and produces a material that could,
in principle, give 100\% electron polarization.  Cathodes using this technique
which are now operating in
the Stanford Linear Accelerator produce a beam polarizations of about 80\%
at the source.  Many of the studies we will review have anticipated further
improvements and have  assumed a beam polarization of 90--95\%.  It is much 
more difficult to produce an  intense polarized positron 
beam \cite{Mikhail}. Fortunately, though, this is not necessary for most
experiments, since in high-energy gauge interactions, the
polarized electron annihilates only on its oppositely polarized
antiparticle.

\subsection{\it $e^-e^-$, $\gamma\gamma$, and $e\gamma$ Colliders}
 
\intro{
In which we review the  possibility  of using linear electron colliders
in alternative modes, including the production of $\gamma$ beams by
Compton backscattering. (2 pages)}
 
With only small modifications, an accelerator and detector designed for
high-energy $\ee$ collisions can also study collisions of several other
types.   Since electrons and positrons can be accelerated by the same
linear accelerator, it requires only a modification of the final-focus
magnets to create $e^-e^-$ collisions.  With some more exotic hardware
in the collision region, an
$e^-e^-$ collider can be converted to an $e\gamma$ or $\gamma\gamma$
collider. 
 
An $e^-e^-$ collider would seem to lose the fundamental advantage of $\ee$
colliders that the initial particles can annihilate with their full energy
into a channel with vacuum quantum numbers.  Nevertheless, there are a few
interesting models in which exotic particles are exchanged in the $t$-channel.
We will discuss such processes in supersymmetric models in Section 6.3 and
in models of the strongly interacting Higgs sector in Section 7.2. From
the technological point of view, the conversion of an $\ee$ collider to
$e^-e^-$ operation is expected to be straightforward \cite{Clemstudy}. 
 With flat
beams, the particle-antiparticle attraction does not make a large contribution
to the luminosity of an $\ee$ collider;  an $e^-e^-$ collider with
the same focusing should have a luminosity not more than a factor 3 lower.
 
An $\ee$ collider always has some luminosity for photon-photon collisions
using the Weizs\"acker--Williams virtual photon field of the electron.
However, it is possible to achieve a much more effective photon beam in a
conceptually simple way \cite{GKST}.  Consider the result of shining an
eV-energy  laser on the high-energy electron beam,
just after the last focusing magnet.  Some fraction of the
photons will be backscattered and achieve energies of the order of the original
electron energy.  These photons, now at high energy, will follow the
electron trajectories ballistically and thus produce a beam spot of the same
size as would have been produced by the electrons.  Thus, if it is possible
to achieve a one-to-one conversion of high-energy electrons to high-energy
photons, the resulting collider should have the same luminosity and almost the
same
energy as the original $\ee$ or $e^-e^-$ collider.
 
To make these observations quantitative, we must consider the kinematics of the
electron-photon collision in more detail.
We introduce a parameter $x$ that is related to
the center of mass energy of the
electron-photon collision by
\beq
       x = {s\over m_e^2} =  {4E \omega\over m_e^2}\ ,
\eeq{xingg}
where $E$ is the beam energy and $\omega$ is the photon energy.
It is  advantageous to make the collisions as relativistic as
possible.  However, it is easy to check that, when $x$ exceeds the criterion
\cite{Telnovx}
\beq
      x_c =   2 + 2\sqrt{2}  \approx 4.8 \ ,
\eeq{Telnovs}
the backscattered photons can annihilate on incoming laser photons to produce
$\ee$ pairs.  Thus, the value given in Eq.~\leqn{Telnovs} is 
the preferred operating
point.  It corresponds to a laser wavelength of 1$\mu$ at 500 GeV 
 $\ee$ center of mass energy.  For fixed $x$, the maximum
backscattered photon energy is $x/(1+x) \cdot E = 0.83  E$ when 
$x = x_c$.
The photon
spectrum is quite hard, and it can be made to peak at the cutoff energy
by a correct choice of polarizations.  For longitudinally polarized laser
photons and beam electrons, the distribution in $y = E_\gamma/E$ 
has the shape
\beqa
   {dn\over dy} \sim
     {1\over 1-y} + 1-y - 4r(1-r) + \Lambda rx(1-2r)(2-y) \ ,
\eeqa{photonspec}
where $r = y/x(1-y)$ and $\Lambda = +1$ when the electrons and the 
photons have both positive or both negative helicity while $\Lambda= -1$ 
in the opposite case.
For the NLC design, the electron beam is totally converted to a high-energy
photon beam with this spectrum for laser pulses of about 1 joule/pulse,
compressed to a picosecond.  A laser meeting this specification with a
repetition rate of 1/sec is now operating in the SLAC experiment 
E-144 \cite{E144};
 a repetition rate of 180/sec (from one or several lasers) would be
required to match the NLC design.
 
For some physics studies, the
scattered electrons, which are are at lower energy but still comoving with the
high-energy photons, lead to important backgrounds and must be
swept away from the 
photon-photon collision point by a magnetic field.  We will see an 
example of this in Section 5.4.
The constraint that this imposes on the collision region geometry
is discussed in \cite{BBC93}.
 
The $\gamma\gamma$ channel has the same property as $\ee$ that the two
colliding particles can annihilate into a state with vacuum quantum numbers.
In the $\gamma\gamma$ reaction, however, processes with $t$-channel
exchanges of light particles can be important, and so there is typically
more background from familiar light particle pair production.  Nevertheless,
we will see several examples in which the $\gamma\gamma$ option contributes
new information beyond that available from $\ee$ annihilation.

\section{$W$ BOSON PHYSICS}
 
The process $\ee\to W^+W^-$ is the largest single component of $\ee$
annihilation into particle pairs
at energies well above 200 GeV. The picture of the $W$ boson
as a gauge boson predicts the $W$ couplings precisely
 from known parameters of the
weak-interaction theory.  Since very little of this
picture is tested experimentally, one might hope to find surprises if
it is probed in detail.  We will show that the linear collider experiments
make this possible.  At the same time, the study of the $W$-boson properties
provides an illustrative example of the analysis techniques 
that the $\ee$
environment makes available.
 
\subsection{\it $W$ Pair Production and Helicity Analysis}
 
\intro{
In which we review the general properties of the reaction $\ee\to W^+W^-$,
emphasizing the possibilities of analyses which use the full event
reconstruction.  (3 pages) }

\begin{figure}
\centerline{\psfig{file=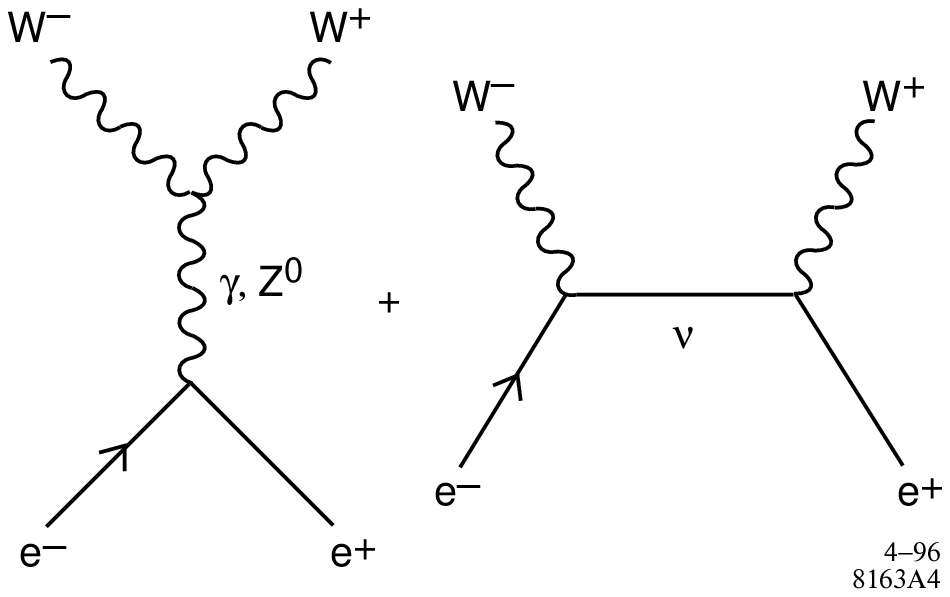,width=0.5\textwidth}}
\caption[wdiags]{Feynman diagrams for $\ee\to W^+W^-$.}
\label{wdiags}
\end{figure}

To begin,
let us review the general properties of the reaction $\ee\to W^+W^-$
within the standard model \cite{HPZH}. 
 The reaction proceeds via the Feynman
diagrams shown in Figure~\ref{wdiags}.
From the second of these diagrams, it is clear that the process
has a strong forward peak associated with $t$-channel neutrino exchange.
The presence of this peak is correlated  with polarization;
it occurs, quite specifically, in the reaction $e^-_Le^+_R \to W^-_L
W^+_R$.\footnote{Throughout our discussion, we will use the 
subscripts  $L$,
$R$, $\ell$, to denote the helicity
 $-1$, $+1$, 0 (or longitudinal) polarization
states of a massive vector boson.}

For the pair production of longitudinally polarized $W$ bosons,
there is a different and more interesting story.  The 
 diagrams of Figure~\ref{wdiags}
 individually violate
unitarity.  It is a wonderful property of the standard model that the sum
of the  diagrams,
adding the $\gamma$ and $Z^0$ exchanges coherently,
 contains the correct cancellations to preserve
unitarity.  In fact, at high energy, the cross section for pair-production
of longitudinal $W$ bosons takes the simple form
\beq
      {  d\sigma\over d\cos\Theta} = {\pi\alpha^2\over 128s}
\biggl({1+4\sin^4\theta_w\over \cos^4\theta_w \sin^4\theta_w} \biggr)
 \sin^2\Theta   \ ,
\eeq{longW}
where $\Theta$ is the production angle in the center-of-mass system.
These facts are explained in the standard model
by the statement that the $W$ obtains a longitudinal
component only by virtue of the Higgs mechanism. The gauge symmetry
associated with the $W$ is spontaneously broken, a Goldstone boson is
created, and this boson becomes the extra, longitudinal polarization state
of the massive $W$.  The longitudinal part of the $W$ then inherits the
properties of the eaten scalar boson, such as the $\sin^2 \Theta$
production cross section shown in Eq.~\leqn{longW}. This 
phenomenon, that a longitudinal gauge boson acquires at high energy 
the properties of a Goldstone boson, is in fact a
general result, called the Goldstone Boson Equivalence Theorem
\cite{equiv,equiv2,LQT,CG}.

\begin{figure}[t]
\centerline{\psfig{file=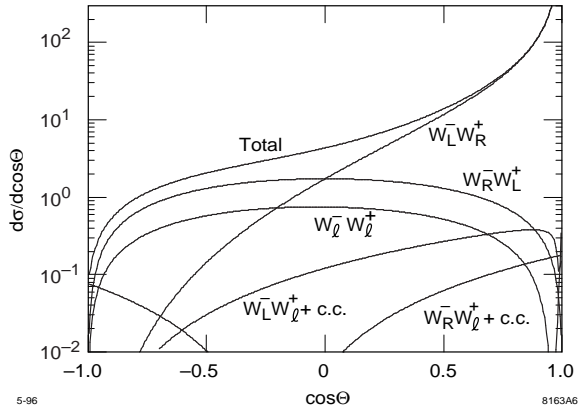,width=0.7\textwidth}}
\caption[wdist]{Angular distributions for $W$ bosons of various 
 helicity in $e_L^- e_R^+ \to W^+W^-$.  The differential cross sections
are given in units of R at $\ECM = 1$ TeV.}
\label{wdist}
\end{figure}
 
Combining these two pieces of physics, we are led to expect a complex
pattern for the cross sections for $\ee$ annihilation to $W$ pairs of
various helicity.  For an initial $e^-_L$, the
predictions of the standard model at a center-of-mass
energy of 1 TeV are shown in Figure~\ref{wdist}.  For an initial $e^-_R$, the
cross section is dominantly longitudinal $W$ pair production, with a rate
1/5 of 
the longitudinal pair production cross section shown in the figure.

Can we test the composition of this
complex mixture of $W$ boson states experimentally? 
This is quite straightforward in the experimental environment of linear
colliders.  By reconstructing events with $W$ pair production and decay,
we will obtain not only information on the distribution in the production
angle $\Theta$ but also information on the individual $W$-boson decay angles.
Since the decay angular distribution encodes the $W$
polarization, the distributions for $W$ pair production can be determined
for each final-state $W$ polarization.

\begin{figure}[t]
\centerline{\psfig{file=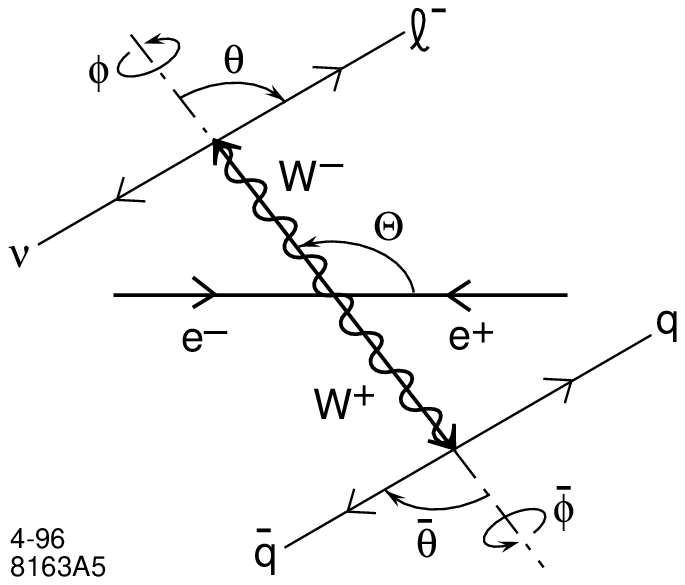,width=0.4\textwidth}}
\caption[wevent]{Production and decay angles in $\ee\to W^+W^-$.}
\label{wevent}
\end{figure}

To understand how the analysis is done, consider, for simplicity,
the case in which one of the $W$'s decays hadronically and the
other $W$ decays leptonically to $e$ or $\mu$. (This sample includes
30\% of all $W$ pair events.)  The
missing neutrino can be reconstructed, even allowing for initial-state
radiation,
and so the whole event is determined.  The event is characterized by the
production angle $\Theta$ and decay angles $\theta$, $\phi$ on each side,
as shown in Figure~\ref{wevent}.   The angle
$\theta$ is related to the $W^-$ helicity $h_W$ through the decay distributions
\beq
 {d \Gamma /d\cos\theta} \sim \cases{ (1+\cos\theta)^2  & $h_W = -1$  \cr
                                       2 \sin^2\theta   & $h_W = 0$ \cr
                                     (1-\cos\theta)^2  & $h_W = +1$ \cr}\ ,
\eeq{Wpoldists}
and just oppositely for $W^+$.  A nontrivial dependence on $\phi$ appears
due to interference between the possible $W$ polarizations.
 There is an observational ambiguity on the hadronic side, since
it is not clear which of the two observed jets originates from the quark
and which from the antiquark.  Nevertheless, each event can be plotted
in a 5-dimensional space of observables  $(\Theta,
\theta_{W^+}, \phi_{W^+}, \theta_{W^-}, \phi_{W^-})$,
and it is possible to compare to
theoretical distributions over this set of five variables.

\begin{figure}
\centerline{\psfig{file=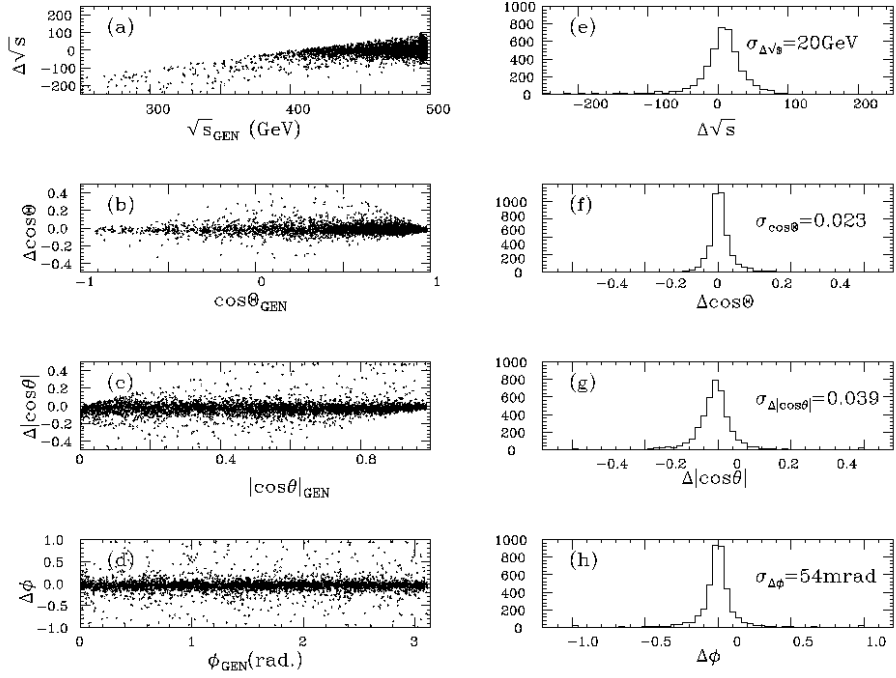,width=\textwidth,angle=90}}
\caption[WWMiyamoto]{Reconstruction of production and decay angles in 
 $\ee \to W^+W^-$, from \cite{Miyamoto-anom}.}
\label{WWMiyamoto}
\end{figure}

Several simulation studies of this kinematic fitting have been 
performed \cite{Miyamoto-anom,Barklow-anomalous,Sett-anom}.
  As an example, consider the  analysis
of  \cite{Miyamoto-anom}.
  Events with the topology of a lepton and two jets are
selected such that the calorimetrically determined hadronic invariant
mass is consistent with the mass of the $W$, the missing energy is 
consistent with being a single massless particle, and the sum of this
momentum vector with that of the lepton gives the $W$ mass to within 
20 GeV.
This yields an event sample of 98\% purity, into which $W$ events of the
required topology are selected with 36\% efficiency.  Kinematic fitting
produces the distributions shown in Figure~\ref{WWMiyamoto} for the
$WW$ center of mass energy $\sqrt{\hat s}$,
$\cos\Theta$, and the leptonic side $W$ decay angles $\theta$ and $\phi$.
These results give a firm foundation for the detailed study of $W$
pair production, and for more exotic reactions which have $W$ pair
production as a standard-model background.
 
\subsection{\it Anomalous Couplings of the $W$}
 
\intro{
In which we present some theory of the possible anomalous
couplings of the $W$, and explain how well these couplings can
be constrained at $\ee$ colliders. (3 pages)}
 
Before going on to more complex reactions, it is worth asking what can be
learned from the detailed study of $\ee\to W^+W^-$.  To make a precise
statement, we will introduce a conventional parametrization of the $WW\gamma$
and $WWZ$ couplings, and discuss the expected size of 
the parameters indicating a deviation from the standard model. 
Electroweak radiative corrections, which typically contribute at the 
level of a few percent, must also be taken into account \cite{Wrad}.
 
For historical reasons, most studies of the $W$ boson couplings assume 
a general vertex
functions of the following form \cite{HPZH}:
\beqa
\L_{WWV} &=& ig_V \bigl( g_{1V} (W^\dagger_{\mu\nu}
 W^\mu V^\nu - W^\dagger_\mu 
      V_\nu W^{\mu\nu}) \CR & &
 + \kappa_V W^\dagger_\mu W_\nu V^{\mu\nu}
  + \lambda_V {1\over \mw^2} W^\dagger_{\lambda\mu} W^\mu{}_\nu V^{\nu\lambda}
    \bigr)\ ,
\eeqa{Wvertex}
where $V$ is $\gamma$ or $Z$, $g_\gamma = e$, 
$g_Z = e\cos\theta_w/\sin\theta_w$,
$W_\mu$ is the $W^-$ field, $W_{\mu\nu} = \partial_\mu W_\nu -\partial_\nu 
W_\mu$, and  $V_{\mu\nu} = \partial_\mu V_\nu -\partial_\nu V_\mu$.
In the standard model at tree level, $g_{1V} = 1$, $\kappa_V = 1$, 
$\lambda_V = 0$ for both $\gamma$ and $Z$.
It is convenient to define $\Delta\kappa_V
= (\kappa_V-1)$.  The expression given in Eq.~\leqn{Wvertex}
 omits possible $CP$-violating couplings and
also, perhaps with not so strong motivation, couplings that violate $C$
and $P$ separately in the gauge boson sector.  (For consideration of these 
latter couplings, see \cite{Valencia}.)
If we ignore possible
$q^2$-dependence of the $W$ form factors, as is done in
Eq.~\leqn{Wvertex},
$g_\gamma = e$ expresses the electric charge of the $W$ boson.  The parameters
$\Delta \kappa_\gamma$ and $\lambda_\gamma$ are then related to the magnetic
dipole moment and the electric quadrupole moment of the $W$:
\beq
      \mu_W =  {e\over 2\mw}(2 + \Delta\kappa_\gamma + \lambda_\gamma ) 
 \ , \qquad 
 Q_W = - {e\over \mw^2}(1 + \Delta\kappa_\gamma - \lambda_\gamma) \ .
\eeq{gandQofW}
Often, $g_{1Z}$ is also taken to have its standard-model value, leaving
still a problem of four
unknown parameters to be constrained experimentally.
 
Should we expect substantial deviations from the standard model in the values
of $\kappa_V$ and $\lambda_V$?  In the older literature, this question is
framed as the question of whether the $W$ bosons are actually the gauge
bosons of a non-Abelian gauge theory.  If there is room to assume that the
$W$ couplings are not necessarily those of Yang-Mills theory,
any constraints on $\kappa_V$ and $\lambda_V$ should be interesting.
However, in this general context, it is difficult to understand
why loop diagrams involving $W$ bosons, which play an important role in the
electroweak radiative corrections tested at LEP and SLC, apparently agree
well with the predictions of the standard model.
 
Over the past few years, a more conservative point of view has evolved in
which the interactions of $W$ bosons are parametrized by gauge-invariant
interactions of the $W$ fields with the electroweak symmetry breaking
sector \cite{Holdom91,EinhornW,Dracula,BLondon}.
  Consider, for example, the effect of  coupling the  $W$ boson
field to a nonlinear-sigma-model field $U$ whose expectation value
$\VEV{U}=1$ signals
$SU(2)\times U(1)$ breaking. 
The  coupling with two derivatives reproduces the conventional $W$ mass
term when $U$ is replaced by  its vacuum expectation value, but allowing 
couplings with four derivatives brings in the more general terms
 \cite{Holdom91,AppelquistWu}:
\beqa
   \Delta \L &=&
      -i L_9 \mbox{Tr} (g' B_{\mu\nu} D^\mu U D^\nu U^\dagger
           + g W_{\mu\nu} D^\mu U^\dagger D^\nu U) \CR
      & & + L_{10} g g' \mbox{Tr} (U^\dagger B_{\mu\nu} U W^{\mu\nu}),
\eeqa{NLSMgen}
where $g$, $g'$ are the $SU(2)\times U(1)$ couplings, and
$D_\mu U = (\del_\mu - i g U W_\mu + i g' B_\mu U$). From Eq. \leqn{NLSMgen},
 one obtains a special case of the
vertex written in Eq.~\leqn{Wvertex}, with $\lambda_V = 0$ and 
\beqa
\kappa_\gamma &=& 1 - g^2 (L_9 + L_{10}) \CR
\kappa_Z &=& 1 - \half (g^2- g'{}^2) L_9 + 2 e^2 L_{10}/(c_w^2 - s_w^2) \CR
 g_{1Z} &=&  1 - \half g^2 L_9/c_w^2  + 
g'{}^2 L_{10}/(c^2_w - s^2_w) \ ,
\eeqa{kapgset}
where $s_w = \sin\theta_w$, $c_w = \cos\theta_w$.

This  point of view, 
 however, suggests rather different values for the
expected anomalies.  A typical model leading to
anomalous $W$ interactions would be one in which the electroweak symmetry
breaking sector contained new strong interactions at TeV energies.
  In such a  model,
the new strongly interacting particles would give virtual corrections
to the $W$ couplings.  A reasonable way to estimate this effect would be
to set the dimensionless parameters $L_9$, $L_{10}$ in Eq.~\leqn{NLSMgen}
 equal to
the values of the corresponding parameters in the 
nonlinear-sigma-model description of QCD \cite{GassL}.
This gives: $L_9 \sim L_{10} \sim  0.045$,  or $\Delta\kappa \sim 10^{-2}$.
It is worth noting 
that $L_{10}$ is related to the
$S$ parameter \cite{PT} of precision electroweak physics through
$S = -L_{10}/\pi$, and that the current constraint on $S$ limits the 
contributions to   
the anomalous couplings from $L_9$ to be of order $10^{-3}$.

Can linear collider experiments meet this extremely challenging criterion
for the appearance of deviations in the $W$ interactions?  Remarkably,
they can. 
There are two aspects of the physics that improve the
sensitivity.  The first is common to all determinations of the $W$ couplings
at high energy:  because anomalous additions to the $W$ couplings
do not respect the gauge cancellations (or, in the language of 
Eq.~\leqn{NLSMgen},
because they multiply higher-dimension operators), these coefficients
multiply terms in the cross section formulae which grow as $(s/\mw^2)$
relative to the leading-order terms.
The second is peculiar to the $\ee$ environment: the full-event analysis
described in the previous section brings the $W$ polarization information into
the analysis as a powerful constraint.

\begin{figure}
\centerline{\psfig{file=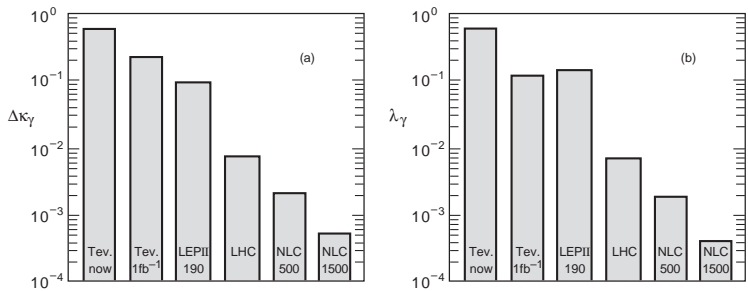,width=0.9\textwidth}}
\caption[WWBarklow]{Comparison of limits on anomalous $W$ couplings
from different colliders, from ref. \cite{DPF-anomalous}.}
\label{WWBarklow}
\end{figure}

The most detailed study of the determination of $W$ couplings at linear
colliders has been done by Barklow and is described in 
\cite{Barklow-anomalous,BarklowA}. His
analysis followed the general strategy described in the previous
section.  Barklow assumes for simplicity the very precise tracking
resolution of the JLC detector; under this assumption, errors in
lepton and jet reconstruction are negligible compared to the
statistical errors.    Reconstructed $W$ events obtained
from the scheme of cuts described above  are fitted to distributions
parametrized by $\Delta\kappa_V$, $\lambda_V$.
  In Figure~\ref{WWBarklow}, taken from ref.
\cite{DPF-anomalous}, the expected
sensitivity of the linear collider experiments is compared to the
estimated sensitivity of other anticipated experiments.  The analyses 
shown in this figure consistently assume a particular two-parameter
formula which relates the $\gamma$ and $Z$ anomalous couplings \cite{HISZ}; 
 thus, it may be 
somewhat optimistic for all colliders shown.

In comparing the sensitivity of experiments at hadron and lepton colliders
to the anomalous W couplings, it is important to note that 
hadron experiments produce $W$ pairs with a wide range of values of the 
$WW$ invariant mass $\hat s$.  Because the anomalous coupling multiply terms
which in the amplitude grow as  $(\hat s/\mw^2)$, the greatest sensitivity to 
anomalous couplings comes at the highest values of $\hat s$.  However, at 
some point these enhancements must be cut off by form factors depending on 
$\hat s$, and the results depend on assumptions about these form 
factors.  At $\ee$ colliders, the center of mass energy is fixed and there is
no corresponding ambiguity.
 
An alternative window into $W$ couplings is provided by
the reactions $e\gamma\to W \nu$ \cite{Yehudai-eg,BRS} and
$\gamma\gamma\to W^+W^-$ \cite{Yehudai-gg,Choi-gg}.
These reactions can be studied at a linear
collider for which Compton backscattering has been used to create a
photon beam, as described in Section 2.4.  For the $\gamma\gamma$ reaction,
complete events can be reconstructed using the same technique that we described
for $\ee\to W^+W^-$.  The sensitivity of this reaction to anomalous couplings
is smaller, because the cross section for producing transversely
polarized $W$ pairs, which is less sensitive to the new interactions, is
more predominant.  Nevertheless, these experiments are expected to give
independent limits on the parameters $\kappa_\gamma$, $\lambda_\gamma$ at the
1\% level  \cite{Miyamoto-anom}.
  The $\gamma\gamma$ reaction can also
be sensitive to a possible $WW\gamma\gamma$ 4-boson anomalous 
vertex \cite{BBd}.

\section{TOP-QUARK PHYSICS}
 
Beyond the $W$  and $Z$,
 there is one more heavy particle of the standard model,
 the top quark.  The reaction $\ee\to
t \bar t$ has a cross section of about 2.0 units of R asymptotically, and
this value is reached rapidly as one crosses the threshold energy of $2\mt$.
 
Using an analysis in the same spirit as that described above for the $W$
boson, it is possible at a linear collider to make a precision study of the
top quark couplings to $\gamma$, $Z$, and $W$.  But, in addition, there are
interesting physics issues associated with the $t\bar t$ threshold region.
For lighter quarks, the energy region just below the threshold contains the
quarkonium states.  For the top quark, this quarkonium region is replaced by a
region of about 10 GeV in width in which the physics is controlled by the
competition between $t \bar t$ binding and decay.  The linear collider will be
the first accelerator with sufficient resolution in $t\bar t$ center of mass
energy to make a detailed study of this region.
 
\subsection{\it Properties of the Heavy Top Quark}
 
\intro{
In which we present a general theoretical perspective on the production
and decay of the heavy top quark. (2 pages)}
 
The top quark is so much heavier than the other quarks that much of the
intuition of ordinary hadronic physics is simply invalid when applied to
$t\bar t$ systems.  To discuss the program of experimental measurements on
the top quark, we must first review the general properties that are
expected for this particle in the standard model.
 
The crucial difference between the top quark and all lighter quarks is that
the top quark is sufficiently massive to decay to an on-shell $W$ boson.
This means that the top quark is not a ``stable particle,'' but rather decays
in a time short compared to typical hadronic scales.  The expression for the
top-quark decay width as a function of its mass, in the limit $m_b =0$, is
\beqa
\Gamma(t\to b W) &=&
 {\alpha_w\over 16}{\mt^3\over \mw^2} \left(1 - {\mw^2\over
 \mt^2}\right)^2 \left(1 + 2 { \mw^2\over \mt^2}\right) \left(1 - 2.9
{\alpha_s\over \pi} \right) \nonumber \\
 &\sim& (1.4 \ \hbox{\rm GeV}) \left({\mt\over 175\
\hbox{\rm GeV}}\right)^3 \ .
\eeqa{topwidth}
The QCD correction \cite{KJGam} is evaluated at $\mt = 175$ GeV; the full
theory of the top quark width is reviewed in \cite{KJGamrev}.  The large 
size of the top-quark 
width is insured by the unexpected $\mt^3$ growth of the
formula given in Eq.~\leqn{topwidth}. 
 This dependence is due to the enhanced coupling
of the top quark to the longitudinal polarization state of the $W$ boson.
Just as in $\ee \to W^+W^-$, the couplings of this state reflect the fact
that it originates as a Higgs boson; the Higgs particle
couples more strongly than a transversely polarized $W$ to the heavy quark.
 
The large width of the top quark has striking implications 
\cite{Bigi81,Kuhn81,BDKKZ}. Because the
top quark
decays before nonperturbative strong-inter\-action processes have time to act,
the top quark is completely a creature of perturbative QCD.  In production
and decay processes, the top quark retains its identity and its spin
orientation.  In the vicinity  of the $t\bar t$ threshold, the spectrum
of top-antitop states is determined by the gluon-exchange potential without
a need to invoke phenomenological confining interactions
\cite{FadinKhoze1} (though the large
width is an essential complication).  Quantitatively, the width of the
top quark takes it off the mass shell by an amount
\beq
               Q \sim   \sqrt{\mt\Gamma_t} \sim 15 \ \hbox{\rm GeV} \ ;
\eeq{topscale}
thus all strong-interaction processes involving top are computable in
perturbation theory using $\alpha_s$(15 GeV) $\sim 0.16$.
 
\subsection{\it The $t\bar t$ Threshold in $\ee$ Annihilation}
 
\intro{
In which we review the theory of the $t \bar t$ threshold cross section,
decay distributions, and forward-backward asymmetry and explain how the
measurements of these quantities will constrain the top quark mass, width,
and couplings. (4 pages).}
 
We begin our more specific discussion of top physics at the $t\bar t$
threshold.
The general properties of the $t\bar t$ threshold are made clear by the
following physical picture:  the $t\bar t$ pair
is produced at
zero separation and then the quarks move outward nonrelativistically.
However, when they reach a separation of $Q^{-1}$
given in Eq.~\leqn{topscale}, they decay via $t\to Wb$. The decay rate
$\Gamma$ is roughly the same as the oscillation frequency in the QCD
potential,
of order
$(\alpha_s^2 m_t)^{-1}$.  Thus, the
QCD potential plays an important role in the physics of the threshold
region, but the top and antitop live for so short a time that no discrete
bound states can form.  Also, on this short time scale, the nonperturbative
confining interaction is irrelevant.

This picture of the $t\bar t$ threshold was made quantitative in a series of
papers by Fadin and Khoze \cite{FadinKhoze1}.
These authors argued that the total
cross section for $t\bar t$ production could be written as a sum over
eigenfunctions of the nonrelativistic Schr\"odinger equation for the QCD
potential,
\beq
     \sigma(\ee \to t\bar t) = {8\pi^2 \alpha^2\over 3 \mt^4} \mbox{Im}\sum_n
                {\psi^*_n(0) \psi_n(0)\over E_n - E_\CM + i\Gamma_t}\ ,
\eeq{tcross}
or more generally, in terms of the Green's function for this potential
problem.  The Green's function is evaluated at an off-shell energy,
shifted by $i\Gamma_t$ because both the $t$ and $\bar t$ are unstable.
The consequences of this formula were worked out for realistic QCD
potentials, and including next-to-leading order QCD corrections 
and the smearing due to initial-state radiation,
in \cite{Kwong,SP}.  For values of $\mt$  below 120 GeV, the 1S quarkonium 
resonance is still clearly  apparent as a peak in the cross section.
However, as the top mass is increased, this state
fades out as a distinct spectral feature.
Naively, it seems that the disappearance of the resonances in the spectrum
of toponium is an unwanted consequence of the large top mass.  But
precisely the reverse is true:  as the top-quark mass increases, the
threshold shape is more precisely determined by perturbative physics
and therefore is a more incisive probe of the fundamental top-quark
properties.

\begin{figure}
\centerline{\psfig{file=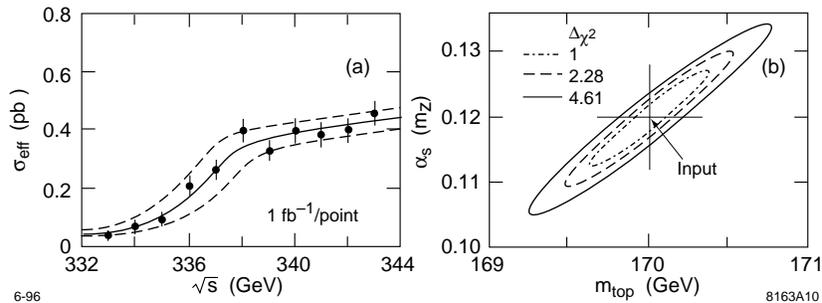,width=\textwidth}}
\caption[topthresh]{Measurement of the top quark mass from the threshold
       shape, from the simulation results of \cite{FMS} with 
                11 fb$^{-1}$ of data, assuming a 
          170-GeV top-quark mass.  The solid curve gives the 
             theoretical expression for the $t\bar t$ threshold, 
         for $\alpha_s(\mz) = 0.12$, 
       including initial-state radiation, beamstrahlung, and a 0.1\%
      energy spread  from the accelerator.
 The dashed curves show the theoretical
  predictions for $\alpha_s(\mz) = 0.11$ and 0.13, from bottom to top.}
\label{topthresh}
\end{figure}

Even including the effect of the top-quark width, the cross section rises
 rapidly at the threshold, and so it is straightforward to obtain a
 very accurate value of the top quark mass.
   Simulation studies of the measurement of the $t\bar t$
production cross section near threshold have been carried out by 
several groups
\cite{Komamiya-ttbar,Igo-Kemenes,FMS,Comas}. 
These analyses include a realistic selection of $t\bar t$ events.  For example,
the analysis of 
\cite{FMS} selects $t\bar t\to 6$  jet events 
through the following set of cuts:  First choose events with visible energy
greater than 200 GeV and  total $p_T$ less than 50 GeV.  Then cluster the
tracks into 6 jets.  Select events with two 2-jet pairs consistent with 
$\mw$ and such that adding another jet gives a mass consistent with $\mt$,
within loose cuts.  Finally, impose a thrust cut, $T < 0.75$.  This 
procedure selects hadronic $t\bar t$ events with 63\% efficiency. The final
cut reduces the dominant background from $W^+W^-$ production to less than
10\% of the 
top quark signal, and of course this background has no threshold.
Under these
conditions, a luminosity
of 10 fb$^{-1}$ scattered over the threshold region, as shown in 
Figure~\ref{topthresh},
still suffices to determine
$\mt$ to an accuracy of 300 MeV.  
 This measurement also determine the
strength of the QCD potential, which can be parametrized by the strong
coupling constant $\alpha_s$ (for example, by the value of $\alpha_s(\mz)$
in the  $\bar{\hbox{\rm MS}}$
 scheme for QCD calculations).  The determinations of
$\mt$ and $\alpha_s$ are correlated; if $\alpha_s$ is known from other
measurements
to $0.002$ (half
the present uncertainty), the error on $\mt$ decreases to 200 MeV. This
should be contrasted with projected determinations of the top quark mass
in hadronic collisions, which are limited to an accuracy of about
2 GeV \cite{amidei}.
 
For such accurate values of $\mt$, it is important to clarify the precise
meaning of the measurement \cite{Aneesh}.
 The value of $\mt$ which enters the top quark
threshold calculations is the `pole mass', the mass appropriate to treating
the top quark as an on-shell state of perturbative QCD.  A more interesting
quantity is the mass of the top quark defined according to the
 $\bar{\hbox{\rm MS}}$
scheme, which can be directly related to the underlying values of the
short distance couplings which are responsible for quark masses. These
quantities are related by
\beqa
      (\mt)_{\hbox{\scriptsize pole}} &=&
 (\mt)_{\bar{\hbox{\scriptsize MS}}}  \bigl[ 1 +
 {4\over 3}{\alpha_s\over \pi} + \cdots \bigr] \CR
     &=& 
  (\mt)_{\bar{\hbox{\scriptsize MS}}}  + (9.7\ \hbox{\rm GeV}
 \pm 2.1\ \hbox{\rm GeV}) \ ,
\eeqa{topmassrel}
where we have included the 2-loop contribution \cite{NGray}, evaluating
the $\bar{\hbox{\rm MS}}$ mass at the pole mass, and we have chosen 
this to be 175 GeV in the numerical estimate.  The error given is the 
magnitude of the 2-loop correction.
The corrections to the $t \bar t$
threshold shape are understood 
at the next-to-leading order in $\alpha_s$
\cite{Sumino-thesis}, but subtle questions remain
about the size of the corrections of order $\alpha_s^2$,
in particular, the effects due to the decrease in the width of a top quark
at off-shell, spacelike momenta \cite{SFHMN,JT93,KM95}. 

The study of the $t\bar t$ threshold allows an accurate measurement of
the top-quark width.  This can be done, first, by measuring the 
threshold shape, which is determined by this width in the way that we have
just described.
From a  fit to the threshold shape  with a 
10 fb$^{-1}$ data sample, one obtains a 20\% measurement of 
the top-quark width.
 But there are two additional techniques available.  The first
involves the momentum distribution of the decaying top and antitop.
The reconstruction of the top-quark kinematics which is implicit in the
cuts defined above allows one to determine this distribution
directly. Thus, one obtains a snapshot of the top-quark wavefunction, in
momentum space, at the given center of mass energy.  This wavefunction
is a linear combination of contributions from nearby $t\bar t$ states; it
contains an increased admixture of distant states, with higher momentum
components, if the top-quark width is large.  The theory of this
momentum distribution is worked out in detail in \cite{SFHMN,KJT}.  The second
of these
probes is the forward-backward asymmetry of the $t\bar t$ system
\cite{MS92}. Though the
nonrelativistic $t\bar t$ system is dominantly produced in an S-wave, the
axial-vector current coupling to the $Z^0$ can also produce P-wave states.
The interference of these components produces the asymmetry.  This
interference effect is sensitive to the overlap of the $t\bar t$
resonances, smeared by the top width, and so it also increases as the
top width is increased.
 
 
\begin{figure}
\centerline{\psfig{file=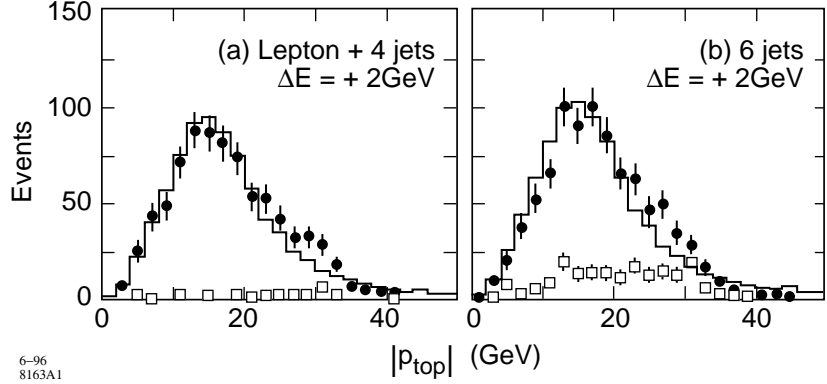,width=\textwidth}}
\caption[topmom]{Reconstructed top-quark momentum distribution in
   the $t\bar t$ threshold region,
    giving the wavefunction of the virtual  $t\bar t$ state, 
 from \cite{FMS}.  The solid squares
 indicate the combinatorial background due to wrong jet assignment; 
     backgrounds from other physics processes are negligible.}
\label{topmom}
\end{figure}

 This strategy was tested in simulation studies of reconstruction of 
the top-quark momentum
distribution \cite{Igo-Kemenes, FMS}.
We will review the
study \cite{FMS} in some detail.  In this work,
top quarks were selected by a set of cuts
 more restrictive than those described above, imposing the criteria that
2-jet combinations sum to $\mw$ within 8 GeV and that 3-jet combinations 
sum to $\mt$ within 15 GeV.  In addition, $b$ jets are identified by vertex
tagging and required to be roughly back-to-back with the associated
$W$ bosons (since the parent top quarks are moving slowly).  These
additional cuts reduce the efficiency to 4.9\% but remove the $W^+W^-$
background and also go far toward resolving the combinatorial ambiguity
in top reconstruction.  A similar analysis can be applied to $t\bar t$
events with one lepton in the final state, and the sign of the lepton
can then be used to measure the forward-backward asymmetry.
The reconstruction of the top-quark momentum distribution,  at
an energy 2 GeV above the nominal 1S peak, is shown in Figure~\ref{topmom}.
In this analysis, which used a top-quark mass of 150 GeV, a 
luminosity sample of 100 fb$^{-1}$ yields the
top-quark width from the momentum distribution and
the forward-backward asymmetry with errors of 4\% and 7\%, respectively,
for the two techniques.  
 
To compare the measurement of the top-quark width that will be available
from a linear collider to that expected from hadron colliders, we should
differentiate two possible sources of a deviation of this quantity from
the standard model.  First, the top width might be larger than the standard
model value due to the presence of new decay modes.  The presence of such
new decay modes will affect the leptonic branching ratio of the top quark,
a quantity which should be measured in  the Fermilab collider experiments
to a few percent \cite{amidei}.  However, such new decay modes can be searched
for directly in the $\ee$ environment by examining the system
recoiling against a reconstructed top quark.  As examples of
analyses with this general strategies, a decay of the top quark
into a  charged Higgs boson plus a $b$ quark with a 5\% branching ratio
can be identified at the
3~$\sigma$ level with 10 fb$^{-1}$ of data, and the decay
 into a top squark and
photino with a 5\% branching ratio can be identified at the 3~$\sigma$ level,
for the mass values ($m_{\tilde t},m_{\tilde \gamma}$) =
 (100,40), with 30~fb$^{-1}$  \cite{DESYtop}.  In general, direct searches for
manifestations of new decay modes are expected to be much more accurate
than probes using the quark total width.
 
On the other hand, even if the top quark decays dominantly to $W^+b$, its
width might be lowered if the Cabbibo-Kobayashi-Maskawa
mixing angle  $V_{tb}$ is not closely equal to 1, or if 
the $t{\bar b}W$ coupling
is enhanced by a nontrivial form factor.
  There are two experiments at hadron colliders that are sensitive to
the strength of the top coupling to $bW$.  The first of these is the
measurement of the subprocess $W^+ g\to t\bar b$ \cite{YuanWg}. However, the
analysis of this experiment has substantial QCD uncertainty as well as the
uncertainty of the gluon distribution.  A more promising method is the
measurement of $\sigma(q\bar q\to t\bar b)/\sigma(q\bar q\to \ell \nu)$
\cite{Willen}.
The measurement suffers from a substantial background due to
$q\bar q\to Wb\bar b$, which accounts for almost 1/3 of the events
under the $t$ mass peak in the $W^+b$ distribution, and an additional
10\% background from $t\bar t$ production in which some jets are not
reconstructed.  If these backgrounds can be subtracted without
introducing a systematic error, this measurement should give a
measurement of the $t{\bar b}W$ coupling corresponding to a 10\% uncertainty
in the top quark width with 12 fb$^{-1}$ of data at the Tevatron collider.
The signal is masked at LHC by the high rate of $gg\to t\bar t$.
 
The comparison of these techniques nicely illustrates the relation of 
$\ee$ and $pp$ experiments.  The $pp$ environment gives a single 
observable which can be determined with great statistical power.  But
the $\ee$ environment allows a variety of measurements which allow 
almost a pictorial view of the interactions of top quarks in their 
binding potential.  To give another example of
the use of this detailed picture, the interaction of the top quark with
the Higgs boson introduces an additional positive Yukawa term into the
$t\bar t$ potential.  For a light Higgs boson with standard-model couplings,
and for $\mt$ = 175 GeV, its strength is 15\% of the strength of the
QCD potential.  For a known value of the Higgs mass (whose measurement
we will explain in Section 5.2) the observation of an enhancement
in the threshold cross section due to this effect measures the $t\bar t h$
coupling \cite{SP,FeigenHtt,JKHiggs}.
 For $m_h$ = 100 GeV and standard-model couplings, the
coupling constant can be determined to 25\% accuracy with 20 fb$^{-1}$
of data \cite{FMS}.  In models in which the top quark has new
interactions associated with electroweak symmetry breaking, this 
coupling can be strong, leading to significant
threshold enhancements.  More generally, the
$t\bar t$ system at threshold is an ideal laboratory for the exploration
of small corrections to the picture of binding provided by QCD.

\subsection{\it Analysis of $t \bar t$ gauge couplings}
 
\intro{
In which we discuss the cross sections and angular distributions
for $t\bar t$ production above threshold, the possibilities for event
reconstruction, and the effects of gluon radiation. (3 pages)}
 
Just as for the $W$ boson, it is interesting to ask whether the top quark
has non-standard couplings to electroweak gauge bosons.  This question can be
addressed directly at $\ee$ colliders by exploiting the naturally large
forward-backward and polarization asymmetries of $t \bar t$ production and
decay.  These asymmetries reflect the very different couplings of the left-
and right-handed components of the top quark to the $SU(2)$ gauge interactions
and the fact already noted that the top quark retains its polarization from
production to decay.

 
\begin{figure}
\centerline{\psfig{file=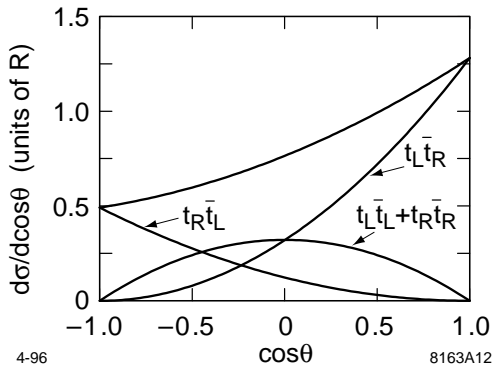,width=0.6\textwidth}}
\caption[topmom]{Angular distribution of top quarks in various 
 polarization states from $e^-_L e^+\to t\bar t$ at $\ECM = 400$ GeV.}
\label{topang}
\end{figure}
 
Though experiments on $t\bar t$ couplings are best done at center-of-mass
energies below 500 GeV, it is easiest to see the essential features of
the phenomenology by thinking first about production at very high energy.
If we consider $\ECM \gg \mt, \mz$,
 the cross section for producing top quark pairs from a left-handed
electron beam is given by
\beq
{d\sigma\over d\cos\theta} = {3\pi\alpha^2\over 4s}\bigl[
 |f_{LL}|^2 (1+\cos\theta)^2
          + |f_{LR}|^2 (1-\cos\theta)^2 \bigr]
\eeq{topcross}
where, with $I^3_H = \half, 0$ for $H = L,R$,
\beqa
     | f_{LH}|^2 &=& \left| -{2\over 3} -
 {(\half - \sstw)(I^3_H - {2\over 3}\sstw)\over \sstw\cstw}\right|^2
	\nonumber \\
 &=& \cases{ 1.4 & $\ELER\to t_L\bar t_R $\cr 0.2 &
 $\ELER\to t_R\bar t_L $\cr}
\eeqa{topfs}
That is, a left-handed electron beam dominantly produces forward-moving,
left-handed top quarks.  The angular distribution for more realistic
conditions, $\ECM = 500$ GeV and $\mt = 175$ GeV, 
is shown in Figure~\ref{topang}. A third component, $(t_L\bar t_L +
t_R\bar t_R)$, is present with a cross section proportional to
$(\mt/\ECM)^2$.  A right-handed electron beam gives
a somewhat larger asymmetry between the two top helicities, and a
total cross section lower by a factor of 2.  

The spin of the top quark can be measured through its decay angular 
distribution.
Returning to Equation \ref{topwidth} and
including the dependence on the angle $\theta$ between the $W$
direction and the top spin, one finds that the factor
$(1+ 2 \mw^2/\mt^2)$ expands to
\beq
{d\Gamma\over d\theta} \sim \left[ (1+\cos\theta)
    + 2{\mw^2\over\mt^2}(1-\cos\theta) \right]
\eeq{topdecayform}
where the first term represents the decay to a longitudinally polarized $W$
and the second term to a left-handed $W$. Alternatively, if the 
$W$ is observed to decay leptonically, the distribution of the
angle $\chi$ between the lepton direction and the top spin is
$(1+\cos\chi)$. 

The QCD radiative corrections to the production \cite{Zerwascrew}
and decay  \cite{KuhnJlep} distributions
have been computed and turn out to be quite small. Formulae
describing the spin correlations in the final decay products from 
$\ee\to t\bar t$ have been presented at the tree level
in \cite{KLY}, and at the one-loop level  in \cite{Stirlingt,Schmidt95}.
The paper \cite{Schmidt95} is especially
explicit and also describes an implementation of these formulae
as a parton-level Monte Carlo program.

To discuss the constraints that can be obtained, we must parametrize
the top quark couplings to gauge bosons. In general, we can write a
gauge boson coupling to the top quark in the form \cite{Yuan92}
\beq
  \L = g_{ttV}\bigl[  F_{1L} \bar t \gamma^\mu t_L V_\mu 
              + F_{2L} {1\over 2m_t}\bar t \sigma^{\mu\nu} t_L V_{\mu\nu}
            + (L \to R)\bigr] \ ,
\eeq{topffs}
 where $V$ is $\gamma$ or $Z$ and $V_{\mu\nu} = \partial_\mu V_\nu -
     \partial_\nu V_\mu$.  For $W$, replace $t_L$ by $b_L$.
This equation defines chiral form factors $F^V_{1L,R}$, $F^V_{2L,R}$.
Conservation of CP requires $F_{2L}=F_{2R}$ for $V = \gamma, Z$.
There is a substantial literature on the experimental
manifestations of CP violation in the top form 
factors \cite{ASoni, topCP, BernrCP}; however,
in realistic models, these effects are typically at the $10^{-3}$ level at
most, and the linear collider would not be expected to provide sufficient
statistics to see the effect (see, however, \cite{topCPlarge}).

The sensitivity of linear collider experiments to deviations from the
standard-model values of the $F^V_i$ has been investigated by several
groups using parton-level simulations and a full-event analysis similar to that
described in Section 3.1 for $W$ pair production \cite{ASoni,Yuan94,SB}.
 The results of these simulations may be summarized by
the statement that 10\% variations of the $F^V_i$ in arbitrary combinations
can be recognized or excluded at the 
95\% confidence level using luminosity samples of 100 fb$^{-1}$, making
this a feasible project for the first-stage of the NLC.  The comparison
of this level of sensitivity to the predictions of models will be
discussed in Section 7.5.
 
The form factors in the top-quark decay amplitude can also be studied 
at threshold, and with higher statistics, by using the fact that, in 
pair production of nonrelativistic fermions, the spin in the final state
follows the spin of the initial electron.
The theory of the top-quark polarization near 
threshold, taking into account the details of the $b\bar t$ binding, is
presented in \cite{Harlan}.  Alternatively, this study can be done by 
noting that, in production above threshold, the top-quark spin is 
still strongly aligned with the electron spin direction as measured in 
the top rest frame \cite{Parke}.

\section{THE HIGGS SECTOR (WEAK COUPLING)}
 
Up to this point, we have discussed tests of the standard model in the
pair-production of $W$ bosons and top quarks.  We have emphasized that
these standard-model processes have interesting qualitative features
and provide many experimental handles in the search for anomalies.
These features add to the general promise of the $\ee$ environment for
new particle searches.
 
However, in presenting the motivation for a new accelerator, one must also
ask how the window that it provides corresponds to general expectations
for where new physics can be found.  This necessarily brings us into the
detailed study of theoretical models.  For the reasons presented in
Section 1.1, we will concentrate here on models of electroweak symmetry
breaking.  In Sections 5-7, we will review the most important models of
this phenomenon, explaining, for each class of models, the relevance of
linear-collider experiments.
 
\subsection{\it Higgs Bosons at $\ee$ Colliders}
 
\intro{
In which we give a general theoretical orientation on the properties of
Higgs boson (or bosons) in weakly-coupled models of electroweak symmetry
breaking. (2 pages)}
 
If the electroweak symmetry breaking occurs in a weakly-coupled theory,
the symmetry breaking must arise from the vacuum expectation values
of elementary scalar fields.  In general, three components of the scalar
fields combine with the $W^\pm$ and $Z^0$ to form the longitudinal 
components of these vector bosons, while the remaining scalar fields
are massive scalar particles.  In models of this type, these particles,
called Higgs bosons,
are the direct manifestations of the symmetry-breaking mechanism and therefore
deserve intensive study. 

 In the minimal standard model, the theory contains one
multiplet of scalar fields with four degrees of freedom.  After symmetry
breaking, one neutral scalar Higgs boson is left over.  In more complex
models, there may be additional multiplets of scalar fields; then 
the spectrum of physical Higgs bosons will also be more interesting.
For example, supersymmetric models require at least two scalar-field
multiplets.  Then one finds five physical Higgs fields---two neutral scalars
$h^0$ and $H^0$, a neutral pseudoscalar $A^0$, and charged scalars
$H^\pm$.\footnote{More precisely, assuming that CP is a good symmetry at the 
weak interaction scale, $h^0$ and $H^0$ are CP-even while $A^0$ is 
CP-odd.}  In general, these particles are linear combinations of 
components of the original two Higgs fields $\phi_1$ and $\phi_2$.
One mixing angle in particular, the angle $\beta$ defined by 
\beq
       \tan \beta = \VEV{\phi_2}/\VEV{\phi_1} \ ,
\eeq{tanbetadef}
appears as a parameter in many phenomenological relations.

The mass of the Higgs boson of the minimal standard model is not 
predicted by the theory.  This mass is constrained
 by direct searches at LEP to be
above 65 GeV \cite{LEPHiggs}, and it is constrained to be below roughly 700 GeV
by the consistency requirements for nonlinear scalar field theories
\cite{trivial}.
The lower end of the spectrum corresponds to a scalar field with 
weak self-interactions---$m_h$ is of order $\mz$ when the Higgs self-coupling
is of the order of the weak-interaction coupling constant---and
the high end corresponds to a field with strong self-interactions.
It requires a model that can explain the electroweak symmetry breaking
with specific weak or strong  coupling dynamics to predict
the Higgs-boson mass.  In such models, one
 typically finds values at the low or high extremes of
this range.

Supersymmetric models, for example, favor Higgs-boson masses at the 
low end of the allowed range.  In the case of two Higgs multiplets and
no additional $SU(2)$-singlet fields---the conditions that define the
`Minimal Supersymmetric Standard Model' (MSSM)---these
models predict that the 
lightest scalar $h^0$ has a mass below 130 GeV \cite{Okada,HHemp,EllisRZ}.
This bound is relaxed in models that contain additional fields. However,
these models also restrict the Higgs boson masses from 
a more general principle.
Supersymmetric models are consistent with the grand unification of 
gauge couplings and seem to fit together naturally with this idea. 
If the Higgs bosons are elementary at the grand unification scale, the 
extrapolation of their properties back to the weak interaction scale yields
an upper limit to the mass of the $h^0$ at about 200 GeV \cite{CMPP}. 
In supersymmetric models, one finds a stronger bound,
150 to 180 GeV, depending on 
whether the gauge group below the grand unification scale is the
 standard-model group or some extension
 of it \cite{Sher89,Drees89,MO92,KKW93}.  

In this section,
we will concentrate on the situation in which the mass of the lightest
Higgs boson is in this lower part of the range.  To discuss concrete
situations, we will typically consider a Higgs boson above 90 GeV, the
reach of LEP II experiments, and below  140 GeV.

Because of the central role of the question of electroweak symmetry 
breaking and the variety of theoretical models available, it will
not suffice for the next generation of colliders simply to identify 
a particle that is plausibly the Higgs boson of the minimal standard
model.  We must establish experimentally
that this boson has the properties required of a 
Higgs boson---that it is a scalar particle, that it arises from a 
field with a vacuum
expectation value, and that this vacuum value contributes to the 
$W$ and $Z$ masses.  These properties are determined by measuring the 
form and strength of the $ZZh$ and $WWh$ vertices.  If $\phi$ is a neutral
component of a scalar field, the gauge-invariant weak interaction Lagrangian
may not contain a $ZZ\phi$ coupling; however, it contains couplings of 
two scalars to one $Z$ boson and a coupling
\beq 
         \Delta \L =   \half( (gI^3)^2 + (g'Y)^2) Z_\mu Z^\mu \phi^2\ , 
\eeq{ZZcouple}
where $I^3$ is the weak isospin of $\phi$.
If $\phi$ obtains a vacuum expectation value $w$, this interaction
yields a $ZZ\phi$ vertex
\beq 
         \Delta \L =    (g^2 + g^{\prime 2})(I^3)^2  w  Z_\mu Z^\mu \phi 
                  =    (2I^3)^2 {w\over v^2} \mz^2 Z_\mu Z^\mu \phi \ , 
\eeq{Zonecouple}
where $v$ is given by Eq.~\leqn{vvev}.
  This  reproduces the Higgs coupling to $Z$ in the minimal standard model
for $Y = -I^3 = \half$ and $w = v$.  In models in which the 
relation  $\mw^2 = \mz^2 \cstw$ is natural, the $WW$ coupling and
$ZZ$ couplings to $\phi$ are simply related by
\beq 
g_{ZZ\phi}/g_{WW\phi} = \cstw.
\eeq{ZZWWrel}

\begin{figure}
\centerline{\psfig{file=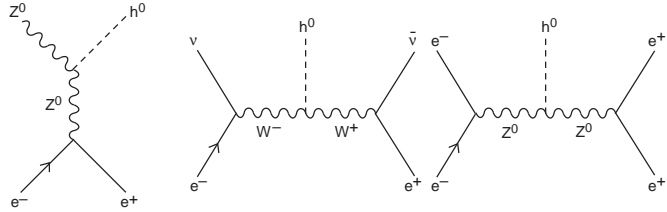,width=0.8\textwidth}}
\caption[Higgsproc]{Processes for the production of Higgs bosons at 
   $\ee$ linear colliders.}
\label{Higgsproc}
\end{figure}

The phenomenology of Higgs bosons at both hadron and $\ee$
colliders has been reviewed 
in the useful book \cite{HHG} and, more recently, 
in the survey \cite{DPF-Higgs}.
At $\ee$ colliders, the most promising processes for the production of 
Higgs bosons are those shown in Figure~\ref{Higgsproc}.  All three of 
these processes 
involve the $ZZh$ and $WWh$ couplings. Above $\ECM = 1$ TeV, the 
$Z$ and $W$ fusion processes have large cross sections, of order
100 fb \cite{BCKP}.  However,
at energies below 500 GeV, and for Higgs-boson masses below 300 GeV,
the process $\ee \to Zh$ has a cross section of 40--80 fb (0.2 R), 
comparable to that of the fusion process.  It also offers distinct
experimental advantages:  The 
$Z$ boson can be reconstructed,
and then the Higgs boson can be identified, independent of its 
decay mode, as the state recoiling against it.   For Higgs bosons 
lighter than 200 GeV, it is
better to run the machine at lower energies (say, $\sqrt{s} = 300$~GeV)
to increase the cross section for this process.  We will see that this
technique allows the identification of the Higgs boson, the measurement
of its crucial coupling to $ZZ$, and the systematic study of its
decay branching ratios.

We will not discuss in detail the case of Higgs bosons of mass 200--700 GeV,
since this situation is not favored in any model of electroweak symmetry
breaking.  Nevertheless, it is quite straightfoward to find a Higgs boson
in this mass range, both at $\ee$ and at hadron colliders.  Such a Higgs
decays dominantly to $WW$ and $ZZ$, in a 2:1 ratio of branching fractions.
At an $\ee$ collider, the weak bosons can be reconstructed in their hadronic
modes.  A data sample of 60 fb$^{-1}$ at $\ECM = 1$ TeV is quite sufficient to 
discover a 500 GeV Higgs boson \cite{HKMlost,Kurihara2}.  The more difficult
case of a very heavy Higgs boson will be discussed in Section 7.2.

Higgs bosons can also be created at hadron colliders, through a variety of
production mechanisms.  A Higgs boson in the mass range 150--700 GeV
can be found straightforwardly in 
the decay $h^0 \to ZZ \to 4$ leptons.
 For Higgs boson masses in the
lower range favored by weak-coupling theories, the hadron experiments are
most sensitive to Higgs bosons produced through gluon fusion and
decaying by $h^0\to \gamma\gamma$.
The ATLAS and CMS detectors at the LHC will have impressive capabilities
to see this $\gamma\gamma$ decay
even in the high luminosity environment  \cite{cms, atlas}.
With 100~fb$^{-1}$ of data, 
a year's running at the design luminosity, they should discover the
standard model Higgs boson.  These experiments also can find at least 
one Higgs scalar over most of the parameter space of the MSSM by 
combining many different signatures, such as $h^0 \to
\gamma\gamma$, $A^0, H^0 \to \tau^+ \tau^-$, $H^0 \to 4 \ell$.
On the other hand, only in a limited region of the parameter space of the 
MSSM can one observe a Higgs decay that involves the coupling 
of Eq. \leqn{Zonecouple}; thus, it is unlikely that we could establish at
the LHC that the 
 new particle discovered in this way is indeed
responsible for generating the $W$ and $Z$ masses.

\subsection{\it Detection of Light Higgs Bosons}
 
\intro{
In which we review search techniques for finding a light Higgs boson
($m_h < 2m_W$) and the multiple Higgs bosons of the supersymmetric
Higgs sector.   (3 pages)}

The Higgs boson of the minimal standard model, in the mass range below
140 GeV, decays mainly into 
$b\bar{b}$. With smaller branching fractions, it also decays 
into $WW^*$, $\tau\tau$, $c\bar{c}$, $gg$.  
(The mode $WW^*$ refers to one $W$ on-shell and one virtual $W$ observed as
$q\bar q$ or $\ell \nu$.) The branching fraction
into $\gamma\gamma$ is of order $10^{-3}$, which is
probably too small to allow observation of this mode in $\ee$ annihilation.
(See, however, Section 5.4.)  This general pattern also holds for light
scalar bosons in more general models \cite{HHG}.
 In this subsection, we will mainly 
focus on the observation of a light boson $h$ in the $b\bar{b}$ final state.
 Other decay modes are
discussed in Section 5.3.

\begin{figure}
\centerline{\psfig{file=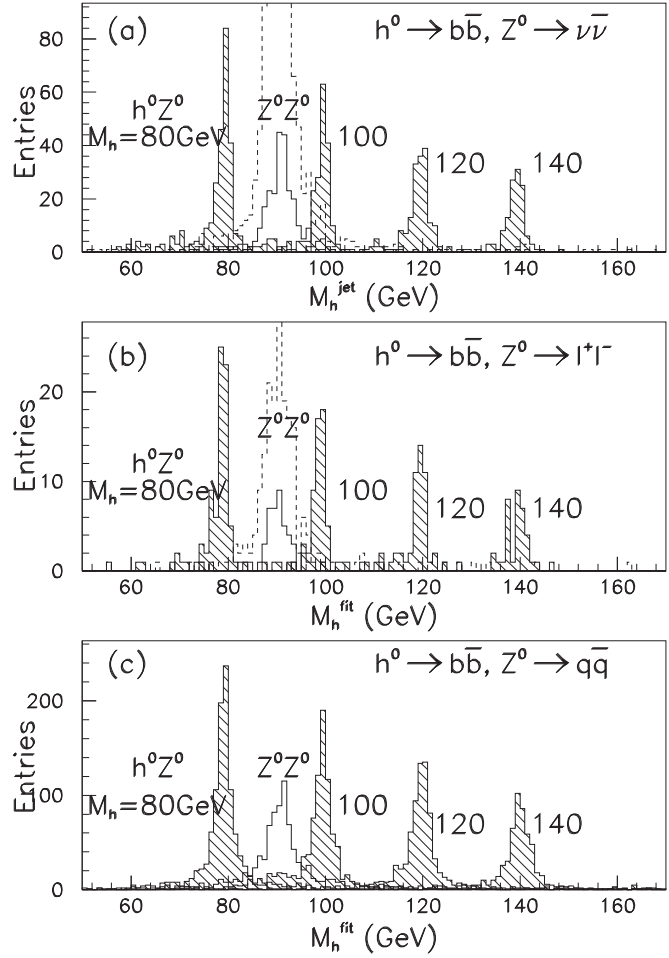,width=0.8\textwidth}}
\caption[Zh]{Simulation of the detection of the Higgs boson in the 
        process $\ee\to Z^0h^0$, from \cite{JLC-1}.  The various hatched
       peaks show the signal expected for a series of values of the 
      Higgs-boson mass from 80 GeV to 140 GeV.
    The $h^0$ is assumed to decay dominantly to $b\bar b$; the 
       three figures  show the cases of $Z^0$ decay to 
 (a) $\nu \bar{\nu}$, (b) $l^+ l^-$, and (c) $q\bar{q}$. The dashed and solid
      unhatched peaks show the standard-model background without  and with 
   a $b$ lifetime cut. The simulation
      assumes 30 fb$^{-1}$ of data at 300 GeV in the center of mass.}
\label{Zh}
\end{figure}
 
It  is a remarkable feature of the $Zh$ production process
that one can use all three
types of $Z$ decay modes---$l^+ l^-$, $\nu \bar{\nu}$ and
$q\bar{q}$. Thus, this process gives three independent signals of the
discovery. Figure~\ref{Zh} shows simulation results, taken from \cite{JLC-1},
for the searches in all three
final states, assuming the Higgs couplings of the minimal standard model.
 The main backgrounds are $WW$, $ZZ$,
$q\bar{q}$, $t\bar{t}$, $e\nu W$, $\nu\bar{\nu} Z$ final states. 
The $l^+ l^- b\bar{b}$ mode has the lowest cross section; here, $ZZ$ 
is the main
background, and can be discriminated from the signal in the recoil mass
distribution (or, equivalently, in the $E_{l^+ l^-}$ distribution) as long
as $|m_h - m_Z| \gsim 10$~GeV. In the $\nu\bar{\nu} b\bar{b}$ final state,
the signal can be seen in the $m_{jj}$ distribution, again if
$m_h$ is not too close to
$m_Z$. If we use four-jet final states, or if $m_h \simeq m_Z$,
$b$-tagging is necessary to reduce the backgrounds. 
Even in the worst case, and with a cross section
 of about 150--200~fb at $\sqrt{s} = 300$~GeV, it was shown that
an integrated luminosity of 1~fb$^{-1}$ is already more than enough for
the discovery \cite{Janot-Hawaii}.
It is noteworthy that a Higgs boson that decays invisibly can be
detected with the same analysis in the $Z\rightarrow l^+ l^-$ channel.
 
The mass of the Higgs boson can be determined from the di-jet invariant
mass of $b\bar{b}$ system,  or, more accurately, by the 
recoil mass in the process with $Z
\rightarrow l^+ l^-$. With the capabilities of existing detectors, it is 
possible to measure $m_h$ to 180 MeV with 50~fb$^{-1}$ of data
\cite{Janot-Hawaii}.
 
Once the  Higgs boson has been found and its mass determined, we would like to
establish that this  particle is indeed associated with a field that
obtains a vacuum expectation value and contributes to the $W$ and $Z$
masses.  The general method that answers this question was discussed
in the previous subsection.  We need to measure the form and strength of
the $ZZh$ coupling, which can be inferred from the cross section and 
angular distribution of the discovery reaction $\ee\to Zh$.  The angular
distribution predicted for the coupling given in Eq.~\leqn{Zonecouple} is
\beq
   {d\sigma\over d\cos\theta} \sim   2 + \beta_Z^2  \gamma_Z^2 \sin^2\theta \ .
\eeq{higgsang}
where $\beta_Z$, $\gamma_Z$ are the velocity and boost of the final-state
$Z$.  In the high-energy limit, this distribution tends to $\sin^2\theta$.
This is the characteristic angular distribution of the production of a 
pair of scalars in $\ee$ annihilation. It indicates that the $h$ is being
produced in association with the Goldstone boson that is eaten to form
the longitudinal $Z$.  Since the $Z$ is reconstructed, its longitudinal 
polarization can  be verified directly from the angular distribution
of its decay into leptons or jets.  The angular distributions are described
in detail in \cite{BargerZerwas}.
The process has a small
polarization asymmetry, proportional to $(1-4\sstw)$,
 which establishes that the the $h$ is produced through a virtual $Z^0$.
  Finally, the 
total cross section can be measured independently of any assumption
about the branching ratios of $h$ by 
using the leptonic decays modes of the
$Z$.  This should give a measurement of the $ZZh$ coupling to 4\%
accuracy with 50~fb$^{-1}$ of integrated
luminosity \cite{Janot-Hawaii}.  By comparing the cross-section normalization
to the prediction from Eq.~\leqn{Zonecouple}, we can see whether the $h$ field
is responsible for the whole $Z$ mass, as in the minimal standard model,
or only for a part of it.

The cross sections for production of the MSSM Higgs bosons are 
presented in \cite{Gcrcrew}.  If the heavier scalar Higgs  $H^0$ is
relatively light, the $Z$ receives only a fraction of its mass from 
the $h^0$, and the cross section is correspondingly suppressed.
 In the region where this suppression is large,
the $H^0$ should also be within the reach of a 
500~GeV collider.  Since the $H$ field
has the vacuum expectation value which contributes the remainder of the
$Z$ mass in this model, the sum of the cross sections into the final
states $Zh$ and $Z H$ is approximately the
same as the $Zh$ production cross section in the minimal standard model.
 The analysis in
\cite{Janot-Hawaii} shows that in this case one can find 
two clear peaks due to $h^0$ and $H^0$ in the
recoil mass distribution.  

\begin{figure}
\centerline{\psfig{file=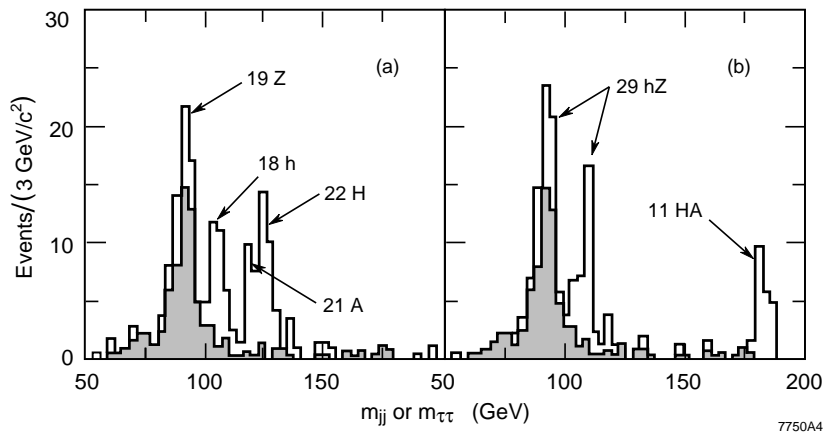,width=\textwidth}}
\caption[bbtautau]{Reconstructed masses of Higgs and $Z^0$ bosons, in an 
        analysis optimized to 
        identify  the process $\ee \to b\bar b \tau^+ \tau^-$, from 
        \cite{Janot-Hawaii}.  The simulation assumes 10 fb$^{-1}$ of data
         at 400 GeV in the center of mass.  The two figures correspond to
          (a) $m_A = 120$ GeV, (b) ; $m_A = 180$ GeV.  The shaded area
         shows the standard-model background, which comes dominantly from 
          $\ee\to Z^0Z^0$.}
\label{bbtautau}
\end{figure}
 
If $H^0$ is heavier, it essentially does not
contribute to the $Z$ mass, and the 
properties of the $h$ revert to those of the Higgs boson of the 
minimal standard model.  In this limit, the four states
$H^0$, $A^0$, $H^+$ and $H^-$ become almost degenerate, and can be
looked for in the final states $H^0 A^0$ and $H^+ H^-$.  Their decay
branching fractions are sensitive functions of many parameters; they
depend on  $\tan
\beta$, on whether decay into top quark pairs  is kinematically allowed, 
and on 
whether they can decay into neutralinos or charginos.  The possible
decay modes include: $H^0 \rightarrow Z Z^{(*)}$, $H^0, A^0 \rightarrow
b\bar{b}$, $\tau^+ \tau^-$, $t\bar{t}$, $\nuone \nuone$,
$\tilde{\chi}_1^+ \tilde{\chi}_1^-$, $\tilde{t} \tilde{t}^*$,
$H^+ \rightarrow c\bar{s}$,
$t\bar{b}$, $\tilde{\chi}_1^+ \nuone$, $\tilde{t}\tilde{b}^*$. 
 Separate searches have to be
performed for each cases, but the studies have shown no problems in
looking for these final states.  Simulation studies of the detection of 
the charged Higgs boson are presented in \cite{Eerola,Sopczak}.
 A particularly interesting case for the neutral bosons is that
in which the $A^0$ mass is below 200 GeV, and there are no exotic modes
which compete with the decays into $b\bar b$ and $\tau^+\tau^-$.  In that
case, a set of cuts that isolate the $b\bar b  \tau^+\tau^-$ final state
should show all four neutral Higgs bosons in the same 
analysis \cite{Janot-Hawaii}.  Simulation
results for this case are shown in Figure~\ref{bbtautau}.
 
\subsection{\it Measurement of the Higgs-Boson Couplings}
 
\intro{
In which we review techniques for measuring the branching ratios of the
Higgs boson into $b\bar b$, $\tau^+ \tau^-$, $c \bar c$, $gg$, $WW^*$.
  (3 pages)}
 
Once a Higgs boson is discovered and has been 
confirmed to play a role in the mass
generation for $Z$ and $W$, it is also interesting to test whether
it is responsible for the masses of the quarks and leptons.  This 
can be done by measuring its couplings to 
$t$, $b$, $\tau$, {\it etc.}  The possible variation of these couplings
within the MSSM is discussed in \cite{GHBR,DjBR}; more general models of the 
Higgs sector allow even a wider variation.
 
First of all, one can measure relative branching ratios of the Higgs
boson. Table~2 shows the expected accuracy of relative branching ratio
measurements with 50~fb$^{-1}$, from the study \cite{Hildreth-Hawaii}.
The $b$ branching fraction is expected to be dominant for a light Higgs.
It is straightforwardly obtained by vertex tagging. 
The  vertex detector is assumed to be of the quality of the current SLD
vertex detector.  
 The $\tau$ branching
ration is expected to be about 6\% of the $b$ branching fraction; it can
be obtained experimentally by selecting events with isolated tracks.

The $WW^*$ mode, in which the Higgs boson decays to one
$W$ on-shell and one virtual $W$ \cite{MKeu84},
 requires a more subtle analysis.
In the minimal standard model, the branching ratio for this mode 
rises from 1\% to 40\% over the mass range from 100 to 140 GeV.
The most powerful technique for measuring this branching ratio involves
dividing the event into six jets, selecting
events in which no pair of jets is too 
close in angle.\footnote{More specfically,
using the JADE jet-finding
algorithm \cite{JADE}, $y_{\hbox {\rm cut}} > 8\times 10^{-4}$.}
Then the jets are combined in pairs to find a combination consistent
with the $Z$ mass and a combination consistent with the $W$ mass.  The 
remaining two jets give a distribution in jet-jet
 invariant mass peaked below its
kinematic limit of $(m_h-m_W)$.  This signal appears on a background of
twice the number of events which is roughly flat in this variable. The 
definite Higgs boson mass or $Z$ recoil energy provides a cross-check to the
analysis.

The $c\bar c$ and $gg$ decay modes of the Higgs boson, which have branching
fractions at the few-percent level, can be recognized as decays to 
jets that do not contain particles with the characteristically
long $b$ lifetimes. It is challenging to separate the charm contribution
on the basis of lifetime given the large background from decays to $b\bar b$,
and this was not attempted in \cite{Hildreth-Hawaii}.
Techniques for improving the accuracy claimed in Table~2, and possibly
resolving the the $c\bar c$ mode, are discussed in  \cite{Nakamura-Iwate}.


\begin{table}
\caption[Hildreth]{The errors
in relative branching fraction measurement,\cite{Hildreth-Hawaii} calculated
assuming Standard Model coupling for the Higgs boson and 50~fb$^{-1}$ of
integrated luminosity at
$\sqrt{s}=400$~GeV. }
\begin{center}
\begin{tabular}{ccc}\hline\hline
&$m_h = 140$~GeV & $m_h = 120$~GeV\\
Relative Branching Fraction & Expected
error & Extrapolated error\\ \hline
$h \rightarrow b\bar{b}$ & $\pm 12$~\% & $\pm 7$~\% \\
$h \rightarrow WW^*$ & $\pm 24$~\% & $\pm 48$~\% \\
$h \rightarrow c\bar{c}+gg$ & $\pm 116$~\% & $\pm 39$~\%\\
$h \rightarrow \tau^+ \tau^-$ & $\pm 22$~\% & $\pm 14$~\%\\
\hline
\end{tabular}
\end{center}
\end{table}

So far, we have discussed only the measurement of Higgs boson branching 
fractions.  However, it is also possible to obtain the total width 
of a Higgs boson $h$ by combining various measurements that we have 
discussed.  We have explained in the previous section that the 
$ZZh$ coupling can be determined from the total production cross section.
At the same time, the branching ratio for $h^0$ to $ZZ^*$ can be found
from the measurement of the 
branching ratio to $WW^*$ and the relation \leqn{ZZWWrel}.
By  comparing these
values, one finds the total width of the $h^0$ to an accuracy comparable to the
accuracy with which the $WW^*$ branching ratio has been determined.

Finally, it is possible to measure the Yukawa coupling of $h^0$ to the
top quark, thereby testing whether the top quark mass originates from the
Higgs vacuum expectation value.  For the light $h^0$ under discussion,
the most promising process is $e^+ e^- \rightarrow t\bar{t} h^0$
\cite{DKZ92} which has a cross section of a few fb. It
can determine the $h^0 t\bar{t}$ coupling to 20\%
accuracy  with 50~fb$^{-1}$.  We have also noted at the end of 
Section 4.2 that the measurement of the total cross section for $t\bar{t}$
production in the threshold region can provide an independent measurement
of this coupling with comparable accuracy.
 
\subsection{\it Measurement of the Higgs-Boson Coupling to $\gamma\gamma$}
 
\intro{
In which we review the experiment $\gamma\gamma \to h^0$. (1 page)}
 
The Higgs decay width into $\gamma\gamma$ and $gg$ is of
special interest since it appears at the one-loop level.  Thus,
{\it any\/} particles which obtain their masses from electroweak symmetry
breaking can contribute in the loop. It happens that the dominant
contributions come from particles too heavy to appear in direct
decays of the $h^0$ \cite{HHG, GHphoton}.  Therefore, the measurement of 
these widths can  signal the existence of new heavy particles.
Since the branching ratio for $h^0 \to \gamma\gamma$ is expected to be
of order $10^{-3}$, this process is unlikely to be measured through $h^0$
production in $\ee$ annihilation.   However, using the $\gamma\gamma$
collider mode discussed in Section 2.4, the Higgs boson
can be produced as an $s$-channel resonance decaying, for instance, into
$b\bar b$.  The cross section is proportional to the combination
$\Gamma(h\to \gamma\gamma) \cdot BR(h\to b \bar b)$.  The branching ratio
will already have been determined in $\ee$ annihilation.  More important,
the mass of the $h^0$ will already be known from $\ee$ experiments, and we
can tune the energy of the $\gamma\gamma$ collider so that the photon-photon
luminosity spectrum peaks at $m_h$.

The main background to the Higgs signal is the 
continuum production of $b\bar{b}$.  However, helicity conservation
implies that, for the photon helicities $(+,+)$ and $(-,-)$ that 
produce a $J=0$ resonant state, the $b\bar{b}$ cross section is 
suppressed by the factor $m_b^2/s$.
 This virtue is somewhat diluted by the resolved-photon process \cite{Eboli} 
in which a gluon from the photon structure
function produces $b\bar{b}$, and by  continuum production with radiation
of an additional gluon \cite{Borden-bbg,Jikia-bbg}.  However the
study of \cite{Borden-bbg} showed that the Higgs signal can still
be observed well above the background.  

Simulation studies of the 
Higgs-boson reconstruction were performed in \cite{BBC93,WatanabeGG,JikiaT}.
In \cite{BBC93}, it was found  that   the reaction
$e\gamma \to e Z$, with an initial electron from the Compton-scattered
beam, is an  important background for Higgs-boson masses below 150 GeV.
 To suppress this 
background, a magnetic field must be introduced 
to displace the scattered electron beam away from the photon-photon
collision point.  When that is done, the $\gamma\gamma\to h^0$ signal
stands out above the 
remaining background processes.
 The total cross section can be
measured at the 6--10\% level with
20~fb$^{-1}$. As a benchmark, this is sufficient
to exclude the contribution of 
a fourth generation of quarks to the 
decay vertex at the $5\sigma$ level.

 For Higgs bosons
heavier than $2\mz$, this cross section can be measured with 10\% accuracy
in a similar sample by reconstructing the $h^0$ from the decay 
$h^0 \to ZZ$ \cite{BBC93}. However, 
as the mass of the Higgs is increased further,
the signal gradually  disappears  below
the background due to the reaction $\gamma\gamma\to Z Z$, which appears
at the one-loop level in the electroweak theory, and is lost altogether
for $m_h>350$ GeV \cite{JikiaLoop}.

If the Higgs boson is in the low-mass range, it is also most likely to 
be observed at the LHC in its $\gamma\gamma$ decay mode.  Thus, it is
worthwhile to say a few words about the comparison of the $\ee$ and 
$pp$ measurements.  In the $pp$ experiments, the Higgs boson will be 
produced dominantly via $gg\to h^0$; thus the measured rate is 
proportional to the quantity
\beq
     \Gamma(h^0\to gg) {1\over \Gamma_{\mbox{\rm tot}}(h^0)}
 \Gamma(h^0\to \gamma\gamma)\ .
\eeq{pprate}
In principle, this measurement could agree with the prediction of the 
minimal standard model, but there is information in any discrepancy.
From this one measurement, however, it is unclear which of the 
three factors  in Eq.~\leqn{pprate} is responsible for the deviation.
The observable cross section for the $\gamma\gamma$ signal at the LHC
varies from a few fb to over 100 fb over the 
parameter space of the MSSM, or even over that part of the parameter space
in which the $h^0$ is inaccessible at LEP II \cite{Wells}. 
 From one number, it is
difficult to learn the correct story.  However, by combining this number
with the values of the second and third factors in Eq.~\leqn{pprate}---measured
respectively in $\ee$ and $\gamma\gamma$ experiments---and with 
information on exotic channels of Higgs decay, one can assemble the 
complete picture of the Higgs boson couplings \cite{DPF-Higgs}.

For heavier Higgs bosons, the $\gamma\gamma$ process has another virtue:
It can make use of the full center-of-mass energy of the collision to 
produce a single Higgs boson.  This is especially
attractive in the search for heavy Higgs states in the extended Higss
models such as the MSSM.  If the heavy Higgs states lie well above $\mz$, then
in the  $\ee$ mode they
are produced only in pairs, $H^0 A^0$ or $H^+ H^-$.  On the other hand,
$H^0$ and $A^0$ can be produced as $s$-channel resonances in the 
$\gamma\gamma$ mode.  The same analysis as the $h^0$ case applies if
they primarily decay into $b\bar{b}$. 
Simulation studies are needed for
the other possible decay modes such as $t\bar{t}$ or $hh$. 
 It is also possible that the heavy Higgs states may decay
mainly invisibly into neutralinos, as emphasized in \cite{GKO95}
 
Finally, the production of Higgs bosons at a $\gamma\gamma$ collider offers
a special experimental handle to determine whether a particular Higgs boson
is CP-even or CP-odd \cite{Gunion-CP,Zerwas-CP}. 
If $\vec E$ and $\vec B$ are the electromagnetic field strengths, 
a CP-even Higgs boson couples to the combination $(E^2 - B^2)$ while a 
CP-odd Higgs boson couples to $(\vec E \cdot \vec B)$.  The first of these
structures couples to linearly polarized photons only if the polarizations
are parallel, the latter only if the polarizations are perpendicular.
If a particular Higgs boson is a mixture of CP-even and CP-odd components,
as can occur in models in which there are new sources of CP-violation in
the Higgs sector \cite{WeinCP}, the interference of these terms gives
 rise to an asymmetry
in the total rate for Higgs production between the helicity states
$(+,+)$ and $(-,-)$ \cite{Gunion-CP}.  Polarization asymmetries of this 
sort could be studied at interesting levels with event samples of 
about 100 fb$^{-1}$.

\section{SUPERSYMMETRY}

Though it is appealing that the electroweak symmetry should be broken by 
expectation values of scalar fields, it is very difficult to build a 
fundamental theory which includes this mechanism.  In an ordinary 
scalar field theory, loop diagrams give additive contributions to the 
scalar mass, and thus the (mass)$^2$ of a scalar field naturally is driven
to a value of order $\alpha M^2$, where $M$ is the largest scale in the 
theory.  In a grand unified theory, $M$ is of order the unification or even
the Planck scale.  Any scalar field with such enormous mass is irrelevant
to electroweak symmetry breaking.

The only known solution to this problem is that of postulating an underlying 
symmetry that links bosons and fermions, {\it supersymmetry}.  In the standard
model, the fermions are forbidden from obtaining mass except
through $SU(2)\times U(1)$ breaking.  In a supersymmetric model, this 
is also true for scalar fields in the model, and so it is possible for
elementary Higgs scalar particles  to naturally remain at the weak scale
rather than being driven to up to the unification scale.  

We have no space for a full review of supersymmetric models here.  Excellent
reviews can be found in \cite{Nillrev, HandKrev, HaberTASI, Murarev,newSSrev}. 
 We should,
however, point out two serendipitous features of supersymmetric models 
of particle interactions.  The first comes in the relation among the 
standard-model coupling constants.  The simplest grand unification models
predict a relation between $\alpha$, $\alpha_s$, and $\sstw$ that is not
obeyed by the values of these quantities measured in the precision experiments
at LEP. However, the assumption that the supersymmetric partners of the 
known particles appear at the weak scale changes the extrapolation to large
scales and results in a successful prediction.  The current status of this 
prediction is reviewed in \cite{lpol}. 

 The second success of supersymmetric
models comes in providing a mechanism for electroweak symmetry breaking.
To explain this phenomenon, one must explain why it is that the Higgs boson
(mass)$^2$ is not only small but also negative.  In supersymmetric models,
one of the Higgs mass parameters receives a negative correction from loop
diagrams proportional to the top-quark Higgs coupling.  Thus, if the 
top quark is the heaviest standard particle, the Higgs field potential
energy naturally has a symmetry breaking form.  This phenomenon is reviewed
in \cite{IRrev}. 

An important consequence of this mechanism is that the $Z$ and $W$ masses
become connected to the scale of superpartner masses. Unless there is a
fine adjustment of parameters to make the $Z$ and $W$ masses especially
small, these mass would be expected to be roughly equal to the masses of
the $W$, $Z$, and Higgs superpartners.  Several groups have tried to 
make this connection quantitative and have used it to bound the
masses of supersymmetric particles \cite{nat,nat2,nat3}.  Among 
their limits are bounds on the 
masses of the $W$ and gluon partners
\beq 
             m(\widetilde w) \lsim 250 \hbox{\rm GeV} \ , \qquad
             m(\widetilde g) \lsim 800  \hbox{\rm GeV} \ , 
\eeq{natlim}
for some reasonable limits on allowable fine adjustment.  This argument
implies that the superpartners should be found at the next generation
of high-energy colliders.	 

In addition to being well-motivated, supersymmetric models have another
importance for understanding the role of future colliders.  Because these 
theories contain only weak-coupling phenomena, we can analyze their 
consequences in detail by direct calculation.  This allows us to appreciate,
in a way that is not possible for theories with strong-coupling dynamics,
the wide variety of phenomena that  these models
 make available to experiment.  By showing the level of detail at which 
we can observe these phenomena, we  illustrate the analytic power of linear
colliders.  If the physics 
of electroweak symmetry breaking is interesting and complex, even if it is
of a different character, we expect that the lessons we learn here will
carry over to exploration of the new sector that this implies.

\subsection{\it The Experimental Investigation of Supersymmetry}
 
\intro{ In which we present a theoretical perspective on supersymmetry and
the challenges which this class of models poses to experiments at
high energy colliders. (3 pages)}

The basic implication  of supersymmetry is that each particle in Nature is
accompanied by a particle with the same standard model quantum numbers,
differing in spin by $\half$ unit.  Thus, quarks and leptons have scalar
partners (squarks and sleptons), gauge bosons have spin-$\half$ partners
(gauginos), and so forth.  If supersymmetry is exact, the partners have the
same mass as the original particles, and this is clearly excluded. However,
it is reasonable that supersymmetry could be spontaneously broken.  In this
case, the renormalizable couplings of particles and superpartners will be 
constrained by the symmetry
to be equal to the corresponding standard-model couplings, 
but the superpartner masses and soft interactions may take a more 
general form.  

Both aspects of supersymmetry theory are important to test in experiments.
First, we must find the supersymmetry partners of quarks, leptons, and gauge
bosons, and we must verify that they have the quantum numbers and couplings
predicted by supersymmetry.  Second, we must investigate the properties of 
the supersymmetry-breaking mass terms and interactions. In most models, these
originate at very high scales, and so their measurement can give
new information on the nature of the grand unified or other underlying 
theory.  In Sections 6.2 and 6.3, we will discuss techniques for discovering 
superpartners and measuring their properties at linear colliders.
In Section 6.4, we will discuss the signficance of these measurements
for tests of unifying theories.

Hadron colliders are also powerful tools for discovering supersymmetric
particles, particularly the squarks and gluinos which are produced with
large cross section in gluon-gluon collisions.  Reviews of
the expectations for supersymmetry experiments at the LHC, for example,
can be found in \cite{atlas,bslep,bjets,BrevSUSY}.  It is likely that the
LHC can observe signatures of a gluino up to gluino masses approaching   
2~TeV,
well beyond the reach of planned $\ee$ colliders for any superpartner.
However, it is a typical property of models that the squarks and gluinos
 are much heavier than the color-singlet particles that are easy to 
study at $\ee$ colliders.  In addition, it is a prediction of the theory
that the supersymmetry signatures at $pp$ colliders are complex
and difficult to interpret, while supersymmetry phenomena at $\ee$ colliders
are much simpler to study in detail.  We will return to this 
point in Section 6.4.
 
In this review, we will discuss only the most 
popular framework for  supersymmetry phenomenology.  We will
assume the minimal particle content (that is, the MSSM) and 
we will assume the presence of an exact $R$-parity symmetry, under which
all of the particles of the standard model have $R= +1$ while their 
superpartners have $R=-1$.   Therefore, the lightest superpartner is 
stable.  Cosmological arguments require this lightest particle
to be neutral and make it unlikely to be the sneutrino. Because of 
our assumption of  exact $R$-parity, the superparticles are always produced
in pairs. Each decays into the lightest superpartner
 directly or in a cascade, giving the experimental signature of 
missing $p_T$ and/or large acoplanarity.  Models with broken 
$R$-parity are discussed in \cite{Rviol,DandGR}; these necessarily involve
either lepton or baryon number violation at the weak interaction scale and 
so give different, but sometimes quite spectacular, signatures.

\subsection{\it Gauge-Boson Superpartners at $\ee$ Colliders}
 
\intro{ In which we review the properties of the chargino and neutralino
sectors of supersymmetric models, explain the difficulties caused by
their mixing problems, and explain how this mixing problem can be
resolved in $\ee$ experiments. (4 pages)}
 
The naturalness argument leading to Eq. \leqn{natlim} implies that the 
superpartner of the $W$ boson is likely to be relatively light. Indeed,
this particle could well be the lightest charged superpartner and thus an
interesting object of study at an $\ee$ collider. 

In supersymmetric models, the $W$ partner ($\tilde W$ or ``wino'')
 is generally not a mass 
eigenstate.
Instead, it mixes with another superpartner with the same electric charge, 
the partner of the charged Higgs boson ($\tilde H^+$ or ``higgsino'').
  The mass matrix of
these fields has the form
\beq
	( \tilde{W}^-\, \tilde{H}_1^- )
	\left( \begin{array}{cc}
		M_2 & \sqrt{2} m_W \sin \beta \\
		\sqrt{2} m_W \cos \beta & \mu
	\end{array} \right)
	\left( \begin{array}{c}
		\tilde{W}^+ \\ \tilde{H}_2^+
	\end{array} \right) ,
\eeq{chmatrix}
where the fermion fields $\tilde{W}^\pm$, $\tilde{H}^\pm$ are left-handed
spinors, $M_2$ is the $SU(2)_L$ gaugino mass,
 $\mu$ is the supersymmetic Higgs mass term, and 
$\tan\beta$ is the vacuum angle defined in the previous section.  The
eigenstates are called ``charginos'' and denoted
$\tilde{\chi}_{1,2}^\pm$ for the lighter and
heavier states, respectively.  If $\mu \gg M_2$, $\chone$ is
to a good approximation a 
wino of mass $M_2$.  If $M_2
\gg \mu$, $\chone$ is approximately a higgsino of mass $|\mu|$.
If $(M_2 - |\mu|)$ is of order $\mw$, the mass eigenstates are 
mixtures of the two.  This occurs in 
 a relatively 
large part of parameter space for  charginos light enough to be found at
LEP II, but only in a more limited region for heavier charginos.
 
In a similar way, the superpartners of the photon and $Z$
mix with 
the partners of the two neutral Higgs boson fields.  This leads to a 
$4\times 4$ mixing problem in the space of the four fields
 $(\tilde B, \tilde W^0, \tilde H_1^0, \tilde H_2^0)$, where $\tilde B$,
$\tilde W^0$ denote the superpartners of the $U(1)$ and neutral $SU(2)$
gauge bosons.  The four mass eigenstates are called ``neutralinos''
 and denoted by
$\tilde{\chi}_{1,2,3,4}^0$ from the lightest to the heaviest.

To reduce the number of parameters on which the chargino and neutralino
masses depend, the assumption is often made that the 
 gaugino masses are in the ratio of the corresponding gauge coupling
constants:
\beq
\frac{M_1}{\alpha_1} =
\frac{M_2}{\alpha_2} =
\frac{M_3}{\alpha_3} ,
\eeq{gutMrel}
where $\alpha_1 = \frac{5}{3} \alpha/\cos^2 \theta_W$. This relation 
follows from the assumption that the three gauginos are unified into a 
single multiplet with a common mass at the grand-unification scale.
We will use this relation as a convenient reference point in our discussion,
but one should not forget that it is important also to test it experimentally.
Using this assumption to eliminate $M_1$,
 we can write  the mass matrices both for charginos and neutralinos
in terms of the three parameters $M_2$, $\mu$, and $\tan\beta$.  Then, in 
the two limiting  cases just described for charginos, the neutralinos have the
following spectrum:  In the case $\mu \gg M_2$, the lightest two neutralinos
are approximately $\tilde B$, with mass $M_1 \approx 
\half M_2$,  and  $\tilde W^0$, with mass $M_2$.  In the case $M_2 \gg \mu$, 
the two light neutralinos are approximately higgsinos, both with mass
close to $|\mu|$.

\begin{figure}
\centerline{\psfig{file=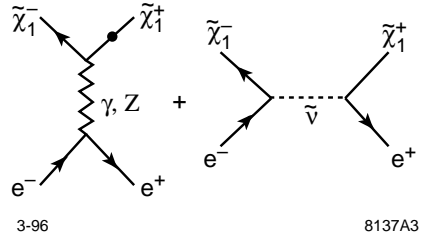,width= 0.5\textwidth}}
\caption[chardi]{Feynman diagrams for 
chargino pair production.}
\label{chardi}
\end{figure}

The charginos are pair-produced in $\ee$ annihilation through the diagrams
shown in Figure~\ref{chardi}. The amplitude receives contributions from 
$s$-channel $\gamma$ and $Z$ exchange, and from 
 $t$-channel $\tilde{\nu}_e$ exchange.
  The first of these processes
is present for both electron polarization states, but the $\tilde{\nu}_e$ 
exchange is present only for left-handed electrons, since the vertex is 
related to the usual weak-interaction vertex by supersymmetry. The production
cross section is of order 100 fb from an $e^-_R$ beam and
of order 1000 fb from an $e^-_L$ beam, with some destructive interference
if the  $\tilde{\nu}_e$ is in the same mass region.  A very useful  
compilation of the formulae for this and other supersymmetry production
processes in $\ee$ annihilation is given in \cite{Baercomp}.

The chargino decay depends both on the makeup of the mass eigenstate and on 
the masses of other superpartners.  If the decay
$\chone \rightarrow \nuone W^\pm$ is kinematically
allowed, this channel usually dominates.
 Then the ratio of leptonic and hadronic
final states is determined by the $W$ branching fractions, and it is possible
to reconstruct the $W$ from its 2-jet decay.
  If this mode is forbidden,
$\chone$ decays into three-body final states $l\nu \nuone$ and $q\bar{q}
\nuone$.  The amplitudes contain both off-shell $W$-exchange as well as
slepton or squark exchanges, and  the decay branching ratios are
sensitive function of their masses.

The detection of chargino pair-production is quite straightforward.
One selects events with large missing energy,  large acoplanarity
to eliminate background from two-photon events, and sufficient visible
energy to be inconsistent with $\ee\to e\nu W$.  These cuts eliminate
most of the background from $\ee\to W^+W^-$, and tighter kinematic cuts
can be placed if necessary.  An explicit simulation is described in 
\cite{Grivaz-Hawaii}; the cuts suggested there have an efficiency of 
25\% in the case where both charginos decay hadronically and 10\% if 
one chargino decays leptonically.
The discovery reach
with 20~fb$^{-1}$ is almost indistinguishable from the kinematic 
limit over most of parameter space, unless 
 $m(\tilde{\nu}_e) \sim \ECM/2$. Once the chargino is  found,
we will discover whether or not it decays to an on-shell $W$; then 
one can optimize the cuts for higher efficiency to study the properties 
of this particle.
 
\begin{figure}
\centerline{\psfig{file=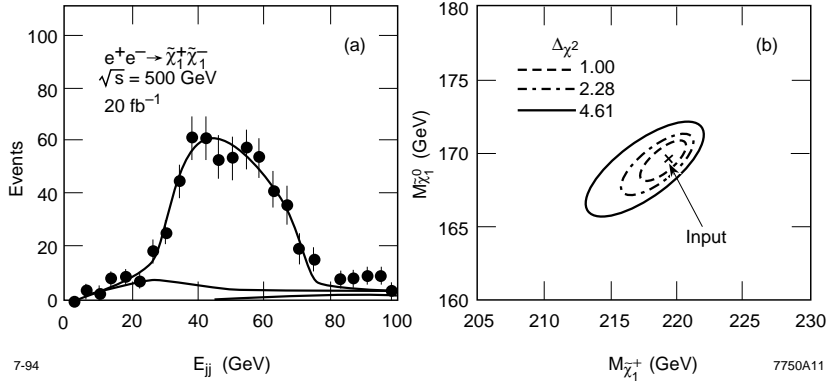,width=\textwidth}}
\caption[chargino-mass]{Chargino mass measurement, 
from \protect\cite{Tsukamoto}:
(a) Di-jet energy distribution from chargino pair
production, (b) $\chi^2$ contours for the fit to the $\chone$ and $\nuone$
masses.  A sample parameter set $M_2 = 400$~GeV,
 $\mu =
250$~GeV, $\tan \beta = 2$ was chosen.}
\label{chargino-mass}
\end{figure}

The mass of the chargino can be measured by selecting events with 
one hadronic and one leptonic decay and 
identifying the endpoints
of the 2-jet energy distribution.  These endpoints directly reflect the
kinematics of the decay  $\chone \to q\bar{q}\nuone$ and thus 
determine  both $m(\chone)$ and $m(\nuone)$.
Figure~\ref{chargino-mass}
shows a simulation study
of this measurement \cite{Tsukamoto}; for that set of parameters, 
both masses are determined at the  
3\% level with 20~fb$^{-1}$ of data.  At the same time, one can compare
the production rates for hadronic and leptonic decays 
and determine the relative
branching ratio.
Since these two are the only available channels, one can derive from this
the absolute branching fractions and the total cross section for
chargino production.
 
However, the measurement of the mass of the $\chone$ is only the beginning of 
what is needed to understand the physics of the gauge-boson partners.
We have seen that the lightest chargino is in general a mixture of 
wino and higgsino components; to completely determine the chargino state,
we must find the mixing angles.  This is not a purely academic problem,
 for two reasons. First, the 
mixing angles are functions of the underlying parameters $(M_2,\, \mu,\,
\tan\beta)$, and  their measurement can play a major role in determining 
these parameters.  More importantly, all heavier superparticles will 
eventually decay into charginos and neutralinos, and thus the observable
signatures of their decays will be determined by the chargino and neutralino
mixing pattern.

Fortunately, we have by no means exhausted the tools available to us in 
$\ee$ annihilation.  Consider, for example, measuring the cross section
for chargino pair production from a right-handed electron beam.  For 
this initial state, the sneutrino diagram in  Figure~\ref{chardi}
vanishes.  But also the $s$-channel diagram in this figure undergoes some
simplification.  If, for a moment, we ignore the $Z$ mass and convert the 
$\gamma$ and $Z$ to $SU(2)\times U(1)$ states, the $e^-_R$ couples only to
the $U(1)$ gauge boson.  This, in turn, does not couple to the $\tilde W^\pm$
but only to the $\tilde H^\pm$.  Thus, the diagrams for chargino 
pair production couple only to the higgsino components of the chargino. 
Actually, since the mass matrix \leqn{chmatrix} is not symmetric, it requires
two mixing angles, one for $\tilde\chi^+_{1L}$, one for $\tilde\chi^-_{1L}$.
The first of these angles gives the cross section for backward $\chonep$
production, the second for forward $\chonep$ production.  By measuring the
cross section and the forward-backward asymmetry, both mixing angles can 
be determined.

 The realization of this strategy does not require asymptotic conditions.
For a 200 GeV chargino
at $\ECM = 500$ GeV, the cross section for $e^-_Re^+ \to \chonep\chonem$
varies from zero to 150 fb as one  moves from the pure wino to pure higgsino
case. The forward-backward asymmetry of $\chonep\chonem$ production cannot
be measured directly because each chargino decays to an unobserved 
neutralino.  It is possible to approximate this observable, however, by
selecting events with one hadronic and one leptonic decay and measuring 
the forward-backward asymmetry of the $q\bar q$ system.  In practice, one 
must impose a cut to remove events in which the total momentum of the 
$q\bar q$ system has $\cos\theta > 0.8$ to suppress background from 
$W$ pair production.  Even with this restriction, it was shown by 
simulation that this quantity is highly correlated with the forward-backward
asymmetry of the chargino pairs \cite{FPMT}.  In that study, a point in 
parameter space was chosen where the chargino $\chtwo$ could also be observed,
so that the masses of the two charginos and the cross section and 
forward-backward asymmetry for $e^-_Re^+ \to \chonep\chonem$ could be 
used to determine the four parameters in Eq. \leqn{chmatrix}. 
 With 30 fb$^{-1}$ of 
data, the two mixing angles could be independently determined to an 
accuracy of 5\degrees. 

A similar analysis applied to the process
$e^-_Le^+ \to \chonep\chonem$ can determine
 $m(\tilde{\nu}_e)$. Once the chargino mixing angles are determined, 
this mass is the only unknown parameter in the formula for the cross section.
The sensitivity to the sneutrino contribution at $\ECM = 500$ GeV
extends almost to a
sneutrino mass of 1 TeV.  Thus, if the lepton partners are not observed 
at the first stage of the linear collider, 
this measurement can give an idea of how much the energy must be raised
to find them.

We should note that the analyses we have described here assumed that the 
chargino $\chone$ is the lightest charged superpartner.  The signatures
may be more complicated if the 
chargino is heavy
enough to  decay into sleptons, {\it e.g.}, by
$\chone \rightarrow l^\pm \tilde{\nu}_l$ followed by $\tilde{\nu}_\ell
\rightarrow \nu_\ell \nuone$.  Fortunately, in this case, the 
charged slepton can always be observed;
the mass splitting between the sneutrino  $\tilde{\nu}_\ell$
and the corresponding 
slepton  $\tilde{\ell}_L$ obeys an  inequality
\beq
m^2(\tilde{\ell}_L)
\leq m^2(\tilde{\nu}_\ell) + 0.77 m_Z^2
\eeq{lepineq}
 which follows directly
from supersymmetry and predicts a rather small splitting.  In a situation
such as this, the best strategy would be to decrease the energy so that only
the sleptons could be produced, study these with care, and then use the 
properties of the final-state sleptons to isolate the charginos.
At an $\ee$ collider, we can always study a novel spectroscopy systematically
in this way, gaining precision information about the lightest particles
and then using this information to disentangle the complex signatures of the
heavier states.

\subsection{\it Quark and Lepton Superpartners at $\ee$ Colliders}
 
\intro{ In which we review methods for measuring the masses and couplings of
lepton and quark superpartners in $\ee$ experiments. (3 pages)}
 
In supersymmetric models, the sleptons often have masses comparable to those
of the charginos.  Thus, these particles may also be light enough to be 
observed even at the first stage of the linear-collider program.  There are
six distinct slepton states, since the left- and right-handed components
of $e$, $\mu$, and $\tau$ each have separate superpartners.  The $\tau$
partners $\tilde\tau_L$, $\tilde\tau_R$ can mix, with an off-diagonal
element in the mass matrix proportional to $m_\tau$.  This effect is 
unimportant for the electron and muon partners, which are thus associated
with definite chirality.  The muon and $\tau$ partners are pair-produced
by $s$-channel $\gamma$ and $Z$ exchange.  For the electron partners,
there is another contribution from $t$-channel neutralino exchange.  The cross
sections are 
of order 100~fb  at $\sqrt{s} = 500$~GeV, and can be larger
for selectrons due to the $t$-channel contribution.  All of the processes
have large polarization asymmetries, with the $e^-_R$ beam favoring 
$\tilde \ell_R$ production and vice versa.
 
If sleptons are lighter than the chargino, they decay directly into leptons
and $\nuone$: $\tilde{\ell} \rightarrow \ell \nuone$.
  Even if the sleptons can 
decay to charginos, the branching ratio into this mode typically 
remains  substantial.  The signature of this decay is particularly simple,
since it gives acoplanar leptons with no other visible
energy.  The main background comes from  $W$ pairs decaying into leptons and
neutrinos.
 
The discovery of sleptons is relatively easy close to the kinematic 
limit.  The analysis in \cite{BV} shows one can discover smuons at the
5~$\sigma$ level up to 225~GeV with a collider of $\sqrt{s}=500$~GeV and
an integrated luminosity of 20~fb$^{-1}$, as long as the mass difference
beween the smuon and the $\nuone$ is greater than 25~GeV, despite the 
$\beta^3$ threshold behavior of the cross section for scalar particles.

\begin{figure}
\centerline{\psfig{file=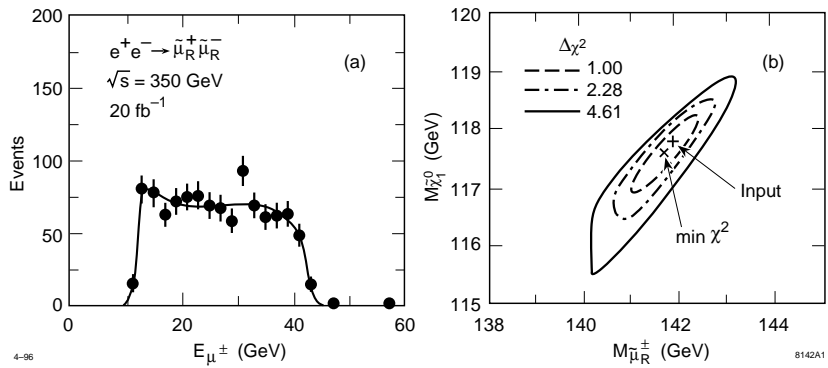,width=\textwidth}}
\caption[smu-mass]{Smuon mass measurement, from 
\protect\cite{Tsukamoto}: (a) Energy distribution of the final muon from
$\tilde{\mu}_R$-pair production, including standard model backgrounds.
 (b) $\chi^2$ contours for the fit to the $\tilde\mu_R$ and $\nuone$
masses.}
\label{smu-mass}
\end{figure}

Once the sleptons are discovered, we would like to measure their 
properties.  The mass measurement is very simple.  Because the sleptons
are scalars and they decay to a two-body final state, the final lepton has a
flat energy distribution over the kinematically allowed range
\begin{equation}
\frac{m_{\tilde{\ell}}}{2}
	\left(1 - \frac{m_{\nuone}^2}{m_{\tilde{\ell}}^2} \right)
	\gamma (1-\beta)
< E_l <
\frac{m_{\tilde{\ell}}}{2}
	\left(1 - \frac{m_{\nuone}^2}{m_{\tilde{\ell}}^2} \right)
	\gamma (1-\beta) \, ,
\end{equation}
where $\beta$ and $\gamma$ are the velocity and boost of the slepton.
This distribution has sharp discontinuities at the endpoints, which 
directly indicate the masses of the slepton and neutralino.
A simulation of this measurement for the case of the $\tilde\mu_R$
is shown in  Figure~\ref{smu-mass} \cite{Tsukamoto}.
We see from the $\chi^2$ distribution that the masses of  
the smuon and of the $\nuone$ are determined to 1\% accuracy.
The exceptionally low level of background was achieved by employing 
a right-handed electron beam (assumed to have $P = 95$\%) to
decrease the cross section for $W$ pair production.
 
The angular distribution of sleptons can be inferred from the lepton
angular distributions up to a two-fold ambiguity.  For the case 
of smuons, this gives another check of the spin from the characteristic
$\sin^2\theta$ distribution.  For selectrons, the measurement the 
forward peak due to $t$-channel neutralino exchange can be used as an
alternative way to constrain the neutralino mixing problem.  At the 
point in parameter space studied in the simulations of \cite{Tsukamoto},
this led to a measurement of the ratio $M_1/M_2$ with 5\% accuracy for a 
data sample of 50 fb$^{-1}$, giving a crucial and necessary test of the 
grand unification relation, Eq. \leqn{gutMrel}.  On the other hand, if the 
chargino is not found at the first stage of the linear collider, the 
assumption of Eq. \leqn{gutMrel} 
implies an upper bound on the chargino
mass,  $m(\chone) <  2 m(\nuone)$, which can be confirmed as the 
energy of the collider is increased.

An alternative way to study the effects of $t$-channel neutralino exchange
is to produce selectron pairs through the reaction $e^-e^- \to 
\tilde e^- \tilde e^-$. This process also offers an environment with very
low background in which to search for the selectron at the extremes of 
parameter space \cite{Cuypers}.  Another interesting feature of the $e^-e^-$
production mode is its ability to search for lepton flavor violation, in 
the process $e^-e^- \to \tilde e \tilde \mu$, at interesting 
levels~\cite{Hallandco}.

As the sleptons become heavier, the 
left-handed sleptons may decay into charginos, by 
 $\tilde{\ell}_L \rightarrow \nu_\ell \tilde{\chi}_1^-$ or $\tilde{\nu}_\ell
\rightarrow \ell \tilde{\chi}_1^+$.  In this case, the sneutrino has 
decays with significant visible energy, making it straightforward to measure
its mass accurately \cite{ZDRlite}.
 
For very heavy sleptons, the $e^-\gamma$ collider option can extend the 
reach for the selectron search beyond that of the $\ee$ mode.
As long as the neutralinos are relatively light, the selectron can be 
produced in the process
 $e^- \gamma \rightarrow \tilde{e}^- \nuone$ up to 80\% of the collider
center-of-mass energy. The $\nu_e W^-$ background 
to this process can be suppressed by the use of 
beam polarization \cite{se-at-egamma}.

\begin{figure}
\centerline{\psfig{file=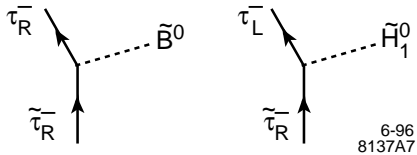,width=0.5\textwidth}}
\caption[Nojiriproc]{Contributions to the decay $\tilde\tau_R \to 
  \tau \nuone$, showing the correlation with $\tau$ polarization.}
\label{Nojiriproc}
\end{figure}

In the case of the $\tilde \tau$, there are two further interesting 
features.  First, as noted above, the mass eigenstates of the $\tilde\tau$
can be mixtures of $\tilde\tau_L$ and $\tilde\tau_R$.  It is possible
to measure the mixing angle using the production cross sections 
from the two polarized beams \cite{Nojiri}.  Second, the $\tilde\tau$
has a additional and wonderful observable, the polarization of the 
final-state $\tau$ from $\tilde\tau$ decay.  The $\tau$  polarization can be 
measured as is done at LEP, from the energy distributions of its  
$\pi$ and  $\rho$ decay products.  For the simplest case in which 
the $\tilde\tau$ is pure $\tilde\tau_R$, the $\tau$ polarization 
indicates the mixture 
of gaugino and higgino in the final-state neutralino, as shown in 
Fig.~\ref{Nojiriproc}.  If the $\tilde\tau$ is known from the cross
section measurements to be a mixture of components, this can be 
taken into account in the analysis.  Since the size of the $\tau$ coupling to
the higgsino depends on $\tan\beta$, this measurement can be used to 
determine $\tan\beta$ if the neutralino mixing is known from other 
observations.

Similarly detailed studies can be made for $\tilde{t}$, $\tilde{b}$ and
other types of $\tilde{q}$ as well.  We refer to 
\cite{squark,Bartlstop}
for further details.

\subsection{\it Tests of Supersymmetric Unification}

\intro{ In which we review the great questions of principle in supersymmetric
models
and explain how these can be addressed by the complementary information
available from hadron and lepton colliders.  (2 pages) }
 
One of the wonderful properties of the supersymmetric models is their 
connection to models of grand unification, and to more ambitious models
of gravity and string theory.  Thus, the measurements of superparticle
masses and properties are important not only in their own right but
also as a window into these deep but speculative ideas.  Because 
supersymmetric models are expected to be weakly coupled from the TeV 
scale to the Planck scale, it is not unreasonable that masses observed
in collider experiments can be extrapolated to such high energies.  This
is a straightforward renormalization-group analysis, which is already
know to work
well for the values of the standard-model gauge couplings \cite{lpol}.
It is, then, worth reviewing how well we can measure those quantities that
are the necessary inputs to this analysis.  A more complete discussion of
these issues can be found in \cite{Peskin-Iwate}.

First of all, it is important to note that supersymmetry makes quantitative
predictions of relations among coupling constants.  The experimental 
verification  of these relations would provide important confirmation that 
the new physics observed at high-energy colliders indeed arises from 
supersymmetry.  At a linear collider, several tests of this type are
possible.  In the chargino study described in Section 6.2, the determination
of the chargino mass matrix leads to a determination of the parameter
$\mw$ in Eq. \leqn{chmatrix}.  This parameter is equal to the $W$ mass
by virtue of the equality of the Higgs-Higgs-$W$ coupling and the 
$h \tilde H \tilde W$ coupling.  It is also possible to test
the equality of the $e\nu W$ coupling and the $e\tilde\nu\tilde W$ coupling.
These tests check the supersymmetric coupling constant relations at the 
20\% level with 100 fb$^{-1}$ of data~\cite{FPMT}.
 We know of no comparable experiments
that are possible at hadron colliders.

To test the gaugino-mass-unification relation, Eq. \leqn{gutMrel}, and 
to examine the question of unification relations for the masses of 
quark and lepton partners, it is necessary to have accurate determinations
of the superpartner masses.  We have already discussed the mass measurements
for color-singlet superpartners.  If squarks lie within the energy reach
of the linear collider, their masses will also be measured accurately.
For gluinos and for heavy squarks, however, we will need to rely on 
measurements made at the LHC. 
  Using the unification relation Eq. \leqn{gutMrel}, the gluino mass
reach of 1.7 TeV quoted in \cite{atlas} would be equivalent to a 
$\chone$ mass of 500 GeV, so that the LHC would cover roughly the same
region of the parameter space of a unified model as the linear collider at 
1 TeV in the center of mass.   We have already noted that the LHC offers
powerful capabilities to recognize the supersymmetric particle
production, particularly for squark and gluino production.  However, 
the signatures of supersymmetry visible at hadron colliders are, for the 
most part, integral quantities such as cross sections for missing energy
and multilepton events.  We know of only one observable in which the 
gluino produces a peak in a mass distribution \cite{BGH93}, and 
even in that case, the location
 of the peak relative to the gluino mass is model-dependent.
  The wealth of data available from 
supersymmetry observations at the LHC can be used to determine the 
squark and gluino
masses, through experiments described in \cite{atlas, bjets}, but the 
interpretation of these experiments requires knowledge of the decay patterns
of the superparticles.  In theoretical models, these decay patterns are
complicated because they involve cascade decays through the spectrum of 
charginos and neutralinos \cite{cascade,cascade2}.
  Thus, the measurement of the chargino and 
neutralino mixing angles at a linear collider will be important
in the interpretation of the LHC data.  They  may be essential for the LHC 
experiments to produce the precision mass measurements needed for the study
of unification.

\begin{figure}
\centerline{\psfig{file=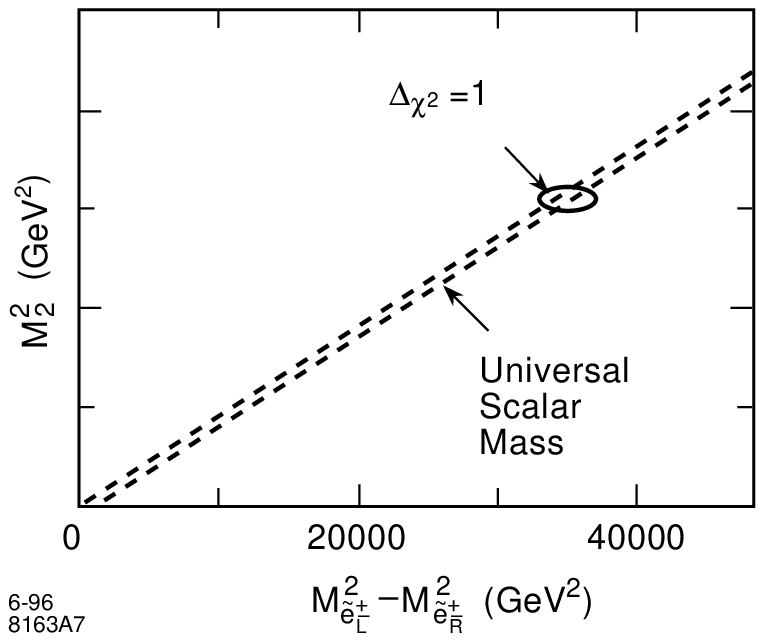,width=0.7\textwidth}}
\caption[GUT-relation]{Test of the mass relation of Eq. \leqn{leptonmassrel},
  assuming 50 fb$^{-1}$ of data, from \cite{Tsukamoto}.}
\label{GUT-relation}
\end{figure}

Once the spectrum of superparticles is known, we will be able to 
extrapolate the
mass pattern to very high energies and look for regularities.  We have
already mentioned the test of the ratio $M_1/M_2$ which can be obtained
in the study of selectron production.  With information from the LHC, we 
can learn the relation of the gluino to the lighter gauginos.  For the 
quark and lepton partners, a very important question is that of whether 
the masses are universal among species, or follow some different pattern,
when extrapolated to the unification scale.  The equality of the $\tilde e_R$
and $\tilde \mu_R$ masses at the unification scale implies equality for the
physical masses.  If this equality is violated even at the level of a few 
percent, a level of accuracy we have shown should be reached at a linear
collider, this strongly constrains sources of lepton flavor violation 
up to the grand unification scale and actually excludes the simplest 
$SO(10)$ unified theories.  The relation of the left- and right-handed
slepton masses is slightly more involved.
The masses of the left-handed sleptons receive a radiative correction
from the loop diagram which includes a lepton and a wino.  This implies
that, if the masses of $\tilde\ell_R$ and $\tilde\ell_L$ are to be equal
 at the 
unification scale,  their physical masses must  obey the 
relation
\beq 
       m^2(\tilde\ell_L) - m^2(\tilde\ell_R) = (0.5 M_2)^2 \ .
\eeq{leptonmassrel}
The simulation study \cite{Tsukamoto} addressed the question of how well
this relation could be tested at a linear collider; the result, for 
50 fb$^{-1}$ of data, is 
shown in Figure~\ref{GUT-relation}. 

Thus, if we are lucky, the discovery of supersymmetry at the next generation
of hadron and lepton colliders could be the beginning of a fascinating 
study of  the fundamental structure of the unified theory at very 
small distances.  The linear-collider experiments would have a central role
in this investigation.

\section{THE HIGGS SECTOR (STRONG COUPLING)}
 
We now turn to models in which electroweak symmetry breaking does not 
involve a fundamental Higgs boson, but rather is the result of strong-coupling
dynamics.  This class of models realizes
the original notion of Higgs, who imagined gauge symmetry breaking as
proceding by a mechanism analogous to that of superconductivity \cite{Higgs}.
In this section, we will survey the components of these models and their
 signatures at linear colliders.

\subsection{\it Strongly Coupled Higgs Sectors}
 
\intro{
In which we present a general theoretical perspective on models in which
electroweak symmetry breaking arises from a strongly interacting sector.
(2 pages)}
 
Ideally, we would discuss models with strong-coupling electroweak symmetry
breaking in the same way that we discussed supersymmetry in Section 6,
by constructing a minimal model with the essential illustrative features
of this class and then analyzing the consequences of that model in detail.
Unfortunately, for strong-coupling models, an approach of this type is not
straightforward.  Strong-coupling models of electroweak symmetry breaking
divide into classes according to the particular dynamics assumed, and
some of these cases can be studied in detail, but in no case is the
story as clean as in weak-coupling models.
 
There is a strong-coupling model
whose theoretical basis is well-understood, and which does naturally
lead to electroweak symmetry breaking at a scale well below an assumed
scale of unification.  This is the minimal technicolor model
\cite{tech2,technicolor},
which postulates a new set of strong interactions similar to those
of QCD, at an energy scale of roughly 1~TeV.  One assumes that this
theory has chiral symmetries as in QCD, and that these are broken by the
same mechanism, fermion-antifermion pair condensation
due to the strong QCD attraction.
If the elementary fermions
of this new gauge theory are assigned
$SU(2)\times U(1)$ quantum numbers
similar to those of quarks, this chiral symmetry
breaking leads to spontaneous $SU(2)\times U(1)$ breaking, in which
the parameters of the symmetry-breaking sector are determined as
properties of the technihadrons.  For example, the
Higgs field vacuum expectation value Eq.~\leqn{vvev} is reinterpreted as the
pion decay constant $f_\pi$ of the new strong interactions.  Insofar
as this theory exactly mimics the dynamics of QCD, its predictions can be
worked out in detail.
 
On the other hand, the minimal technicolor model does not agree with
experiment, for several reasons.  First, it contains no mechanism for
giving mass to the quarks and leptons.  This problem can be solved
by introducing additional gauge particles, called extended technicolor
(ETC) bosons, which can convert technicolor fermions to ordinary quarks and
leptons \cite{EandLmu}.
However, this modification typically results in unacceptably large
predictions for flavor-changing neutral current 
processes \cite{elp,savasandjohn},  and in a
value of the top-quark mass bounded from above at about 100 GeV \cite{Rabytop}.
  In addition, the
corrections of this model to precision electroweak physics are
large enough to be excluded by the most recent measurements (see, for
instance, \cite{Hagiwara95}).
As a cure for these problems, most enthusiasts of techhnicolor models
would consider the dynamics of technicolor to be  rather different from
QCD, either a non-asymptotically free gauge theory near an ultraviolet
fixed point \cite{holdomup}
 or an asymptotically free gauge theory with very slow running
of the coupling constant (``walking technicolor'') \cite{walking}.
 Because little 
is known about the dynamics of the
underlying gauge theories of these types there is considerable room for
assumption or guesswork.
 
A particularly interesting line of speculation is that the fermions that
condense in pairs to break $SU(2)\times U(1)$ are precisely the top and
antitop \cite{Nambutop,MTY,BHL}.
 This idea has given rise to a more general class of models,
called topcolor, in which the top quark or the third generation of fermions
has special gauge interactions not shared by the lighter fermions
\cite{topcolor}.
The spectrum of models that realize this idea blends smoothly into
the class of technicolor models in which strong ETC interactions enhance the
top-quark mass to its observed value \cite{topplus,topassist}.
 This idea of new gauge forces
coupling to the third generation leads to a number of interesting
signatures both at hadron and lepton colliders; we will review some of these
in Section 7.5.
 
One might also react to this confusion of models by asking for experiments that
are sensitive to new strong interactions in the Higgs  sector in a
model-independent way.  To imagine what such experiments would look like,
we can start from the minimal requirement for a theory of electroweak
symmetry breaking by strong interactions. Every such theory begins as a
strong-interaction theory with a global symmetry $SU(2)\times U(1)$ that is
spontaneously broken.  When the global symmetry
$SU(2)\times U(1)$ is promoted to a local symmetry by coupling in the
weak interaction gauge bosons, those particles obtain mass.  The
observed relation $\mw = \mz \cos\theta_w$ is not obvious in this
general context, but it is imposed straightforwardly \cite{SSVZ}
 if we assume also
that the original theory had an $SU(2)$ global symmetry that is unbroken,
under which the weak interaction currents form an isotriplet.  Then the
underlying strong interaction theory has global symmetry $SU(2)\times SU(2)$,
spontaneously broken to the ``custodial'' $SU(2)$.  This is just the
symmetry structure of QCD with two flavors, and that is the reason for the
successes (such as they are) of the minimal technicolor model.
 
In this class of models, it is possible to probe  aspects of the
new strong interactions by studying the reactions of $W$ bosons.  This
follows from a remarkable theorem, true for the most general models of this
kind, called the Goldstone Boson Equivalence Theorem.
In the original
strong-interaction theory with global $SU(2)\times U(1)$ symmetry, the
spontaneous breaking of this symmetry leads to three Goldstone bosons.  In
technicolor models, these are the analogues of the pions in QCD.  When the
weak-interaction gauge bosons are coupled in and the $W$ and $Z$ bosons
obtain mass, the Goldstone bosons disappear from the spectrum while the
 vector bosons obtain a longitudinal polarization state.  The theorem states
that, at high energy, this new polarization state is exactly the eaten
Goldstone boson, or, more precisely, that the scattering amplitudes of
the longitudinal gauge bosons reproduce those of the Goldstone bosons, up
to corrections of order $(\mw/E)^2$  \cite{equiv,equiv2, LQT,CG}.
 
\begin{figure}
\centerline{\psfig{file=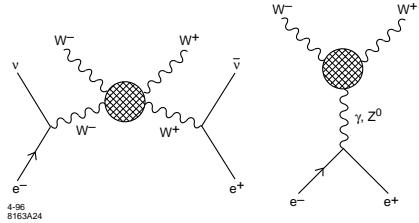,width=0.5\textwidth}}
\caption[Wprocesses]{Processes useful for measuring the Goldstone-boson
      scattering amplitudes.}
\label{Wprocesses}
\end{figure}
 
This theorem suggests two ways to measure amplitudes of the new strong
interactions experimentally. These are illustrated schematically in
Figure \ref{Wprocesses}.  The first is to measure the scattering
of $W$ or $Z$ bosons, a process related by the theorem to the 
Goldstone-boson 
scattering amplitudes,
 the analogue for the new interactions of $\pi\pi$ scattering.
 The second is to measure the pair production
of longitudinal $W$ bosons in $\ee$ annihilation.
This gives the analogue
for the new interactions of the timelike pion form factor.
In Sections 7.2 and 7.3, we will
explore the application of both of these techniques at
$\ee$ linear colliders. We will then turn in Sections 7.4 and 7.5 to
more model-dependent
probes of a possible strong symmetry-breaking sector.

\subsection{\it $WW$ Scattering at $\ee$ Colliders}
 
\intro{
In which we review the capabilities of $\ee$ colliders to measure the
cross sections for $W^+W^-$ and $W^+W^+$ scattering at TeV energies.
(2 pages)}
 
The scattering of $W$ bosons can be observed at high-energy colliders 
through processes such as $\ee \to \nu\bar\nu +  W^+ W^-$ or 
$q\bar q \to q\bar q + WW$,
as illustrated in Figure~\ref{Wprocesses} \cite{JonesandP,CahnDawson,CG}.
This process has been studied in great detail in the hadronic environment;
for recent reviews, see \cite{Baggercrew,Chivrev}.  
  For instance, the ATLAS Technical Design Report \cite{atlas} includes a 
study showing effective rates in like-sign $W^\pm W^\pm$ scattering
processes of order 20 events per LHC year, over a standard-model background
of about 40 events.  In some  particular models of $WW$ scattering,
the $WW$ invariant mass contains a resonance at some value, giving a 
clear signal of an effect above background. The analogous effect is seen
in $\pi\pi$ scattering in QCD at the rho resonance.  However, the more
typical situation in parametrizations of the $WW$ scattering is that this
cross section has a broad, featureless shape such as is seen in 
S-wave $\pi\pi$  scattering in QCD.  In this situation, the effect just
described for the LHC would be rather marginal, and a complementary 
experiment with completely different systematics would be crucial to 
establish the effect.

It is interesting, then, to carry out the analogous experient at an 
$\ee$ collider, using the reactions $\ee\to \nu\bar \nu W^+ W^-$ and
$\ee\to \nu\bar \nu Z^0 Z^0$ \cite{HKMlost,KandNaj,BHee}. 
 In the $e^-e^-$ collision mode, the reaction
$e^-e^- \to \nu\nu W^-W^-$ is an equally interesting probe.  The 
final-state $W$ and $Z$ bosons can be observed in their hadronic 
decay modes to maximize statistics; with the calorimeter of the JLC
detector, $W$ and $Z$ bosons can be distinguished on the basis of their
reconstructed masses, at least at the statistical level needed to measure
the ratio of cross sections for the two processes. The size of the 
longitudinal $W$, $Z$ signal is order 1~fb.

  The main backgrounds to the vector-boson scattering
process come from the production
of transversely
 polarized $W$, $Z$ pairs due to interactions of virtual photons radiated
from the electron and positron.  The most important of these are
the processes
$\gamma\gamma\to W^+W^-$, which has a cross section of 1--2~pb at
$\sqrt{s} = 1$--1.5~TeV, and from $\gamma W \to WZ$, which has a cross
section of about 100~fb. These large cross sections for the background
seem daunting, but the backgrounds can be removed by simple cuts.  In 
$\gamma\gamma$ fusion, even in this case where the photons are virtual,
 the initial particles typically have small transverse momentum,
while in $WW$ scattering the longitudinal $W$'s radiated from the electron 
lines
typically have a transverse momentum of order $m_W$.  Thus, it is useful
to cut
on the transverse momentum of the final $W$ pair, at $p_T(WW) > 50$ GeV.
 The background can be decreased further by vetoing
events with hard forward electrons.  These two cuts remove the $\gamma\gamma$
reactions almost completely and bring the $WZ$ production to within a factor
2 of the signal \cite{HKMlost} 
 At this level, the calorimetric discrimination of $W$ from
$Z$ reduces the $WZ$ reaction to a small background to the $WW$ and $ZZ$ 
signals.

\begin{table}[t]
\caption[eeWW]{Total numbers of $W^+W^-, ZZ \rightarrow  4$-jet
signal $S$ and background $B$ events, after cuts, calculated for  a 1.5~TeV
$e^\pm e^-$ linear collider with  integrated luminosity 200~fb$^{-1}$. The 
statistical significance $S/\sqrt B$ is also given.
The hadronic branching fractions of $WW$ decays and a realistic $W^\pm/Z$
misidentification probability are included. The significance
is improved by using polarized $e^-_L$ beams \cite{BHee}.}
\begin{center}
\small
\begin{tabular}{lcccc}
\hline\hline
 & SM  & Scalar & Vector   & LET  \\ \cline{2-5}
channels & $m_H=$ & $M_S=$ & $M_V=$ &\\
& 1 TeV & 1 TeV & 1 TeV &\\
\hline
$S(e^+ e^- \to \bar \nu \nu W^+ W^-)$
& 330   & 320   & 92  & 62  \\
$B$(backgrounds)
& 280    & 280   & 7.1  & 280  \\
$S/\sqrt B$ & 20 & 20 & 35 & 3.7 \\
\hline
$S(e^+ e^- \to \bar\nu \nu ZZ)$
&  240  & 260  & 72  & 90   \\
$B$(backgrounds)
& 110    & 110   & 110  & 110  \\
$S/\sqrt B$ & 23 & 25& 6.8& 8.5\\
\hline
\hline
$S(e^- e^-_L \to \nu \nu W^- W^-)$  & 54 & 70 & 72 & 84 \\
$B$(background) & 400 & 400 & 400 & 400\\
$S/\sqrt B$ & 2.7 & 3.5 & 3.6 & 4.2 \\
\hline
$S(e^-_L e^-_L \to \nu \nu W^- W^-)$  & 110 & 140 & 140 & 170 \\
$B$(background) & 710 & 710 & 710 & 710\\
$S/\sqrt B$ & 4.0 & 5.2 & 5.4 & 6.3 \\
\hline
\end{tabular}
\end{center}
\label{eeWW}
\end{table}

This strategy for isolating vector-boson scattering at an $\ee$ collider
was studied by simulation in \cite{BHee}.  This study did not 
include a realistic detector simulation but simply used the 
parametrization of the JLC detector. However, it did include the complete
matrix elements for all relevant $2\to 4$ particle processes; for example,
$\gamma\gamma\to W^+W^-$ was included as subprocess of $\ee \to \ee W^+W^-$.
Following the framework of \cite{Baggercrew}, the authors considered
four particular models of the vector-boson scattering amplitude:
the minimal standard model with a Higgs boson of mass 1 TeV, a model with a 
broad scalar resonance at 1 TeV, a model with a vector resonance
at 1 TeV, and the ``LET'' model in which the $WW$ 
interactions are precisely those predicted by the low-energy theorem 
for pion-pion scattering, carried over to $WW$ scattering
using the Equivalence Theorem.  The first two of these models are
rather similar. The third
mimics the most naive technicolor models.  The fourth is a more
 pessimistic scenario.  The results for the signal/background estimates
for these four  cases, assuming a relatively high energy
  $\sqrt{s} = 1.5$~TeV and an
integrated luminosity of 200~fb$^{-1}$, are shown in Table~\ref{eeWW}.  The
statistical significance of the signals is comparable to that achievable
at the LHC.  The background estimates are presumably more solid than those
made for the hadronic environment, since all important backgrounds are 
electroweak processes whose rates are precisely calculable.  The enhancement
of signal over background is improved with the use of polarized beams,
as shown for $e^-e^-$ reactions in the last two lines of the Table.
 
It may also be possible to study $WW$ scattering at a $\gamma\gamma$ 
collider, by using the fact that a high-energy $\gamma$ has a large
probability to branch into $W^+W^-$. One then observes the process
$\gamma\gamma\to W^+W^-W^+ W^-$, with two $W$'s at high transverse
momentum \cite{Brodsky, Cheung}

We conclude this discussion of the $WW$ scattering signal with two 
comments.  First, while the studies we have cited for 
hadron and lepton colliders have considered a wide range of models of 
pion-pion scattering in the new strong interactions, they have all assumed
that the pion-pion scattering is elastic.  If the new strong sector 
contains other relatively light particles (so-called pseudo-Goldstone
bosons), this need not be true.  Then the weak vector bosons might 
primarily scatter into pairs of these exotic particles rather than 
scattering to final-state $W$ and $Z$ pairs \cite{CGR}.  In this case,
it is extremely difficult to isolate the vector-boson scattering signal,
and one must, alternatively, search for the pair production of 
new particles.  In the hadronic environment, this could be a problem; the 
new particles may  be recognized if they decay hadronically, especially if 
 the dominant decays do not include top quarks.  In the 
$\ee$ environment, however, there is no difficulty in recognizing these
exotic states.  We will discuss search techniques for pseudo-Goldstone 
bosons in Section 7.4.

Second, because it is so difficult to observe
the vector boson scattering signal
either at hadron or electron colliders, it is important
to buttress the observation of $WW$ scattering through new 
strong interactions by showing that there is no light 
Higgs particle that contributes significantly to the $W$ and $Z$ masses.
We have argued in Section 5.2 that an $\ee$ linear collider can discover
any such light particle without relying on the assumptions of any model;
 conversely, it can rule out the existence of
such a particle definitively.  As for the LHC, though this collider can 
find a light Higgs boson in a large class of models, it cannot exclude the
existence of such a particle except in specific model contexts.

\subsection{\it $\ee\to W^+W^-$ as a Window to Higgs Strong Interactions}
 
\intro{
In which we discuss the `Higgs Pion Form Factor' as an indicator of
new strong interactions in technicolor models and more general models
of electroweak symmetry breaking.  (3 pages)}

In the $\ee$ environment, there is a second relatively model-independent 
signature of
new strong interactions coupling to the $W$.  This is the analogue
 of the pion form factor in the new strong sector. 

 In QCD, the process
$\ee\to \pi^+\pi^-$ contains the rho resonance and, in fact, receives a
cross section 
enhancement of about a factor of 20 from the resonance pole in the 
pion form factor.  If the known strong interactions were copied at the 
TeV scale, the analogue of the rho in the new strong interactions would
lead, through the Equivalence Theorem, to a similar enhancement in 
$\ee\to W_\ell^+W_\ell^-$, where $W_\ell^+$ is the longitudinally polarized
$W$ boson.   On the other hand, we have emphasized in Section 3 that
the process $\ee\to W^+W^-$ is one of the major components of $\ee$ 
annihilation at linear collider energies, and that tools exist to study
this process in exquisite detail.  We thus expect that effects which 
correspond to a percent enhancement of the rate for $W$ pair production,
or a few percent enhancement of the rate of $W_\ell$ pair production,
should be observable experimentally.  This means that linear-collider
experiments have a very large dynamic range in which they are sensitive
to this particular amplitude arising from new strong interactions.

We will use the term ``Higgs pion form factor'' to refer to the 
form factor for the 
production of pairs of Goldstone bosons of the new strong interactions by 
the vector current of custodial $SU(2)$.  If the new strong sector 
contains a vector meson with the $SU(2)$ quantum numbers of this current,
the corresponding form factor should have a pole at the vector-boson  
mass, leading to a large enhancement of Goldstone-boson pair 
production.  In a technicolor model, the new strong interactions
involve new strongly interacting fermions, and the desired vector bosons
appear as spin-1 $L=0$ quark-model bound states of these fermions.
Vector states with these quantum numbers can also appear in other types of 
models, for example, models in which the constituents in the new strong
interactions are vector particles \cite{VeltV}.  One might expect more
generally that, in a strongly interacting theory, the vector current
should always be the interpolating field for some composite particle.

\begin{figure}[t]
\centerline{\psfig{file=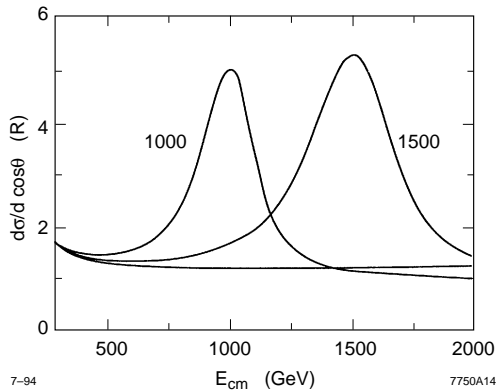,width=0.6\textwidth}}
\caption[mytrho]{Effect on the unpolarized differential cross section for
        $\ee\to W^+ W^-$ at $\cos\theta = 0$, plotted as a function 
           of $E_\CM$, of technicolor rho resonances at 1 TeV and 
            1.5 TeV, compared to the cross section in the minimal standard
           model with a light Higgs boson, from ref. \cite{MyFinn}.}
\label{mytrho}
\end{figure}

In models with such strong enhancements, the effect of the vector 
resonance can be seen in the rate for $\ee\to W^+W^-$ without any special
final-state or polarization analysis \cite{MySC,Iddir}.
  We should only note that
the pion form factor is specifically an enhancement of longitudinal
$W$ pair production.  Looking back at the distributions shown in 
Figure~\ref{wdist}, we see that, away from the forward peak, the longitudinal
$W$ pair production accounts for about 1/4 of the differential
cross section summed over polarizations. Taking into account this
dilution of a factor of 4, we show in Figure~\ref{mytrho} the effect on the 
differential cross section for $\ee\to W^+W^-$ of a rho resonance 
scaled up from the familiar strong interactions to a mass of 1 TeV or 
1.5 TeV \cite{MyFinn}.  A more complete analysis of production and decay
distributions can observe a technirho resonance with a mass of up to 
2 TeV, or 
exclude it  at the 95\% confidence level, already 
at $\ECM = 500$ GeV \cite{HikasaRes}.

There are, however, models with high-energy strong interactions
in which there is no prominent vector resonance
coupling to Goldstone-boson pairs.  The minimal standard model with a 
heavy Higgs boson is a model of this type, and one might imagine that
other models
with only scalar constituents might share this property. Curiously, there
is no model of this type that satisfies the criterion we set out in
Section 1.1, that it explain the origin of electroweak symmetry breaking.

\begin{figure}[t]
\centerline{\psfig{file=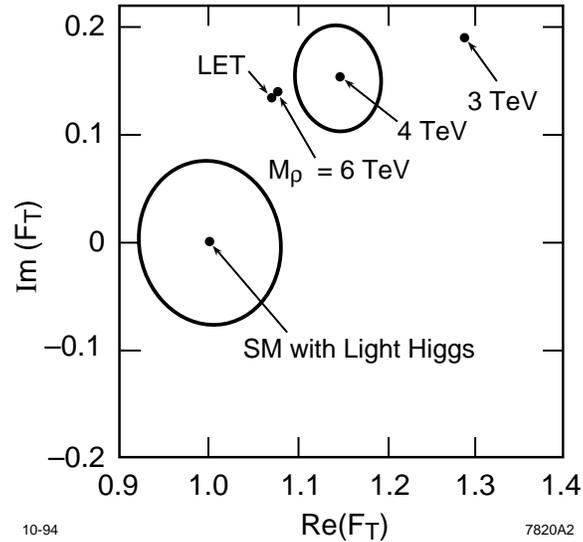,width=0.7\textwidth}}
\caption[timstrho]{Confidence-level contours for the real and imaginary
      parts of the Higgs pion form factor $F_\pi$ at $\sqrt{s} = 1.5$ TeV
     with 200 fb$^{-1}$ of data, from \cite{BarklowA}.  The simulation 
       data was evaluated against a theory of this form factor that included
       both a vector resonance at a mass $M_\rho$ 
       and model-independent $WW$ scattering, as 
       described in the text.  The contour about 
  the light Higgs value is a 95\% confidence contour; the contour about 
the point $M_\rho = 4$ TeV is a 68\% confidence contour.}
\label{timstrho}
\end{figure}

Such a model can still have
observable effects on the Higgs pion form factor.
 From unitarity and the assumption that Goldstone-boson
scattering is dominantly an elastic two-body process at TeV energies, 
it can be shown that the vector form factor takes the form \cite{BJD}
\beq
    F_\pi(q^2) = P(q^2) \exp\left[{s\over \pi} \int ds'{\delta_1(s')\over 
           s'(s'-q^2)}\right]\ ,
\eeq{myFform}
where $\delta_1(s)$ is the pion-pion scattering phase shift 
in the channel $I = J=1$ with the quantum numbers of the vector
current.  The factor $P(q^2)$ is a polynomial in $q^2$ such that $P(0)=1$.
To obtain a concrete prediction, set $P(q^2)=1$ \cite{MySC}.
In QCD, this approximation reproduces the observed pion form
factor to about 20\% accuracy.
The phase shift $\delta_1$ was modelled using the prediction of the 
low-energy theorem for pion-pion scattering, plus a vector resonance 
at a specified mass.  As this mass is taken to infinity, an  
irreducible contribution remains from $W^+W^-$ rescattering through the
interactions predicted by the low-energy theorem.  That contribution
gave a 15\% shift  of the Higgs pion form factor at 1.5 TeV, mainly 
contributing to its imaginary part.
The assumption that $P(q^2) =1$ was questioned in 
\cite{BWill}, and the authors of that paper proposed a phenomenological
model in which the polynomial has a zero in the TeV energy region.
It seems difficult to resolve this controversy  without reference to a 
plausible underlying model of the dynamics.

Nevertheless, it is interesting to see whether an enhancement
of this general size
can be observed experimentally.
This issue was studied in \cite{BarklowA}, using the methods for the 
study of $\ee\to W^+W^-$ that we have described in 
Section~3.1. Assuming a high $\ee$  center-of-mass energy of 1.5 TeV
and a large event sample of 200 fb$^{-1}$, comparable to what is needed
for the study of $WW$ scattering,  the real and imaginary parts of the
Higgs pion form factor  can be constrained within the  limits
illustrated in Figure~\ref{timstrho}. 
 It should be noted that the 
sensitivity to the imaginary part of the form factor depends on the
ability to make  separate
cross section measurements for left- and right-handed polarized beams
(with 90\% polarization assumed). The theoretical values of the form factor
come from the model of \cite{MySC}, using very high values of the 
vector resonance mass.
  (The values of the vector resonance
mass actually predicted in technicolor models lie far off the page to the
right.)   At the endpoint marked LET, the 
only effect is $WW$ rescattering according to the low-energy theorem.
If this contribution  to the the Higgs pion form factor is present in the 
data, the value $F_\pi = 1$ corresponding to the minimal standard model
with a light Higgs boson will be excluded at a very high level of 
confidence.
It is remarkable that, even in this very  pessimistic case, the 
precision study of $\ee\to W^+W^-$ can provide clear evidence for the 
presence of new strong interactions coupling to the weak vector bosons.
This window into the dynamics of the new strong sector is completely 
complementary to the $WW$ scattering experiments discussed in the 
previous section, and it is available only at $\ee$ colliders.

\subsection{\it Pseudo-Goldstone Bosons}
 
\intro{
In which we discuss the phenomenology of exotic scalar particles from
the Higgs sector.  (1 page)}
 
Up to this point, we have 
discussed relatively model-independent signatures of a strongly
coupled Higgs sector.  In this section and the next, we will discuss
signatures of specific models or mechanisms.  Even if there is no
preferred model of the new strong interactions at 1 TeV, model-dependent
phenomena can be interesting to look for if they make it possible to 
confirm or exclude specific approaches to model-building. Signatures associated
with specific mechanisms for generating the quark and lepton masses are
espcially important targets for future colliders.

In Section 7.1, we discussed briefly the status of technicolor models of 
electroweak symmetry breaking.  These models have the appealing feature 
that they give a clear physical explanation for the spontaneous breaking 
of $SU(2)\times U(1)$.  However, in their simplest versions,
 they also have numerous phenomenological 
problems.  It is possible to pursue the idea of technicolor by formulating
more complicated models which include methods to solve these problems.
We find it interesting that those mechanisms typically lead to new and 
distinctive experimental signatures at relatively low energies.

In technicolor models, the pseudoscalar bound states of techni\-fer\-mi\-ons
and their antiparticles must include the Goldstone bosons which are 
eaten by the $W$ and $Z$ as these obtain mass.  However, there may be
many other such bound states.  These states are massless at the level of
the pure technicolor theory but receive mass from the standard model 
gauge couplings and other effect that break the symmetry among technifermions.
Hence, they are called pseudo-Goldstone bosons.  These particles
typically have masses in the range of a few hundred GeV \cite{Chivukrev}.
  The colored bosons have 
larger masses than color-siglet bosons \cite{MeandJohn},
 giving rise to the same sort of 
complementarity between 
searches in $\ee$ and $pp$ collisions that we have seen in the case of 
supersymmetry.

At $\ee$ colliders, the search for colorless pseudo-Goldstone bosons is
similar to the search for Higgs particles.  Indeed, many models contain a
color-singlet charged boson $P^+$ which decays preferentially into 
the heaviest fermions available.  This experimental signature is 
identical to that of the charged Higgs boson, and is easily detected.
 Technicolor models often contain
CP-odd bosons that decay to $\gamma\gamma$ and can therefore be produced
singly in $\gamma\gamma$ collisions.  The rate for this production process
is similar to that for a standard Higgs boson \cite{Tandean}.
We have explained in Section 5.4 how to discover such a particle in 
$\gamma\gamma$ collisions and how to measure its coupling and CP 
properties.

More exotic scenarios are not only possible but preferred by technicolor
enthusiasts.  Lane and Ramana have proposed that walking technicolor
leads to  ``multiscale technicolor,'' in which the technifermion
flavor symmetry is strongly broken \cite{LandR}.
Then the vector mesons of the technicolor theory are not degenerate, and
some of them can be quite light.  The original Lane-Ramana model proposed
technirho resonances at 400 and 550 GeV, though somewhat higher values may
now be required.
  At these points, one finds resonances in $\ee\to W^+W^-$ of the
type discussed in Section 7.3, and also resonant enhancement of 
pseudo-Goldstone boson pair production.  

Randall \cite{RandallTC}
 and Georgi \cite{GeorgiTC} have proposed solving the flavor-changing
neutral current problem of technicolor by incorporating a GIM mechanism.
The resulting models have a proliferation of gauge groups at 1 TeV, 
leading to huge multiplets of pseudo-Goldstone bosons.  The phenomenology of 
these particles is quite complex \cite{Skiba}.

Finally, many of the proposals for reconciling the idea of technicolor 
with the precision electroweak measurements depend on contributions to 
electroweak radiative corrections from light pseudo-Goldstone bosons 
\cite{DuganR,SundandH,LutyS} or light uncolored technifermions
\cite{AT93}.  In either case, in order to give large electroweak
corrections, the masses of these particles must be of the order 
of 100 GeV.  The required pseudo-Goldstone bosons decay mainly to 
$\tau\nu_\tau$.  They can be studied using the techniques described in 
Section 6.3 for the scalar $\tau$.
 
\subsection{\it The Top Quark and Higgs-Sector Strong Interactions}
 
\intro{
In which we review the effect of Higgs sector strong interactions on the
gauge couplings of the top quark, and the visibility of this effect at
$\ee$ colliders.  (2 pages)}

The dynamics of fermion mass generation has the biggest effect on the
property of the heaviest fermion, namely the top quark.  Therefore we
expect that a detailed study of the top-quark properties will give us hints
about this dynamics.  This is illustrated in technicolor models, for which
the ETC particles that mediate the interaction between the top quark and the
technifermions can be light enough to have significant effects.

  In the simplest schemes for top-quark mass generation, the ETC particles  
are light enough to be observed in bound states with 
technifermions \cite{ArnoldW,AppelE}.
These states may have masses in the range 0.5--1 TeV, and
can be produced singly in association with $t$, $b$, or $\tau$.
In models with topcolor \cite{topcolor,topassist},
 both the composite states and the elementary top 
quark may exhibit couplings to the new gauge bosons of these models.

\begin{figure}
\centerline{\psfig{file=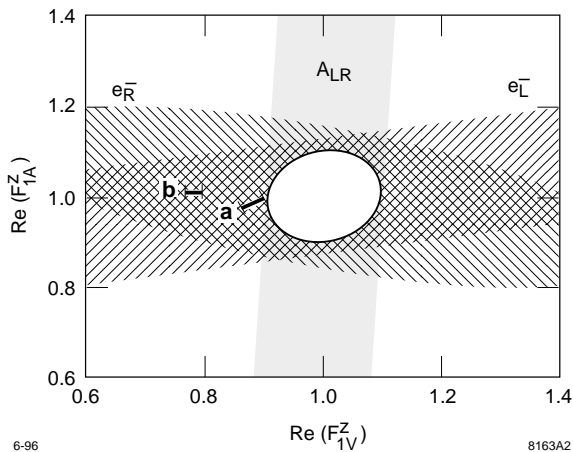,width=0.7\textwidth}}
\caption[topforms]{Expected 95\% confidence limits
 on top-quark anomalous couplings to the $Z^0$,
  from \protect\cite{SB}, for 100 fb$^{-1}$ of data at 400 GeV in the center
of mass.  These bounds are 
  compared to the expectations from the technicolor models of 
(a) \protect\cite{CST94} and (b) \protect\cite{HK95}.}
\label{topforms}
\end{figure}

  The light ETC bosons can also affect the top-quark couplings to 
standard-model gauge bosons.  Typically, the effect of new strong interactions
on the top-quark form factors of Eq. \leqn{topfs} is proportional to 
$(m_t/4\pi v)^2$, leading to effects of order 1\%.  However, the ETC
contribution to the vector and axial-vector form factors of $t$ and $b$
turns out to contain only one power of $m_t$ and thus can be a 10\% 
correction.  This effect was invoked in \cite{CSS92} for the $b$ couplings
to account for the 
observed anomaly in the branching ratio for $Z\to b\bar b$ \cite{Hagiwara95};
however, in the simplest ETC model, it gives an effect of the wrong sign.
More complicated ETC models can repair the sign problem and naturally
give an effect 
on the $b$ couplings of the correct magnitude \cite{CST94,HK95}.  These 
models predict similar anomalies in the
top-quark couplings and thus give an idea of the size of interesting effects
on these couplings from new strong dynamics.  In Figure~\ref{topforms}, we
display the predictions of the models \cite{CST94,HK95} for the vector and
axial-vector form factors in  the top-quark coupling to the $Z$.  These 
predictions are compared to the expected
95\% confidence limits on these form 
factors, according  to the simulation study of \cite{SB}. 

 Thus, both in the study of new particles that couple to the third 
generation, and in the 
precision study of the top quark properties, $\ee$ linear colliders can 
make significant tests of the couplings of new strong interactions to the 
heaviest quarks and leptons.

\section{OTHER NEW PARTICLES AND INTERACTIONS}
 
\intro{
In which we give lightening review of other exotic particle searches
proposed for $\ee$ linear colliders, including $Z^{0\prime}$ and
exotic fermion searches.  (1 page)}
 
In many extensions of the standard model, there exist new particles at
the TeV scale
which may not necessarily be related to physics of electroweak
symmetry breaking.  There are numerous examples: leptoquarks,
dileptons, diquarks, fourth generation, excited electrons, excited
quarks, excited $W$ and  $Z$ bosons, and the gauge bosons
of extended gauge symmetry groups.  Thanks to the
democracy of the linear-collider environment, all these particles can
be produced at rates comparable to the standard model backgrounds.  In
general, it is rather easy to discover such new particles if they are
present within the kinematic reach, unless they have 
vanishing electroweak couplings or decay completely invisibly.
Once the new particles are discovered, their standard-model quantum
numbers can be worked out from their production cross sections and 
asymmetries, and their couplings to lighter states from their branching
ratios.  A recent survey of  exotic particles can be found in 
\cite{DjouR}.   Specific examples that have been discussed in detail
include heavy neutral leptons \cite{heavynu}, excited leptons and quarks
\cite{excitede}, and leptoquarks \cite{leptoquarks}.  We should also note
that electron colliders can  incisively probe into quark and lepton 
compositeness; for example, the study of $e^-e^-\to e^-e^-$ at 1 TeV
with 50 fb$^{-1}$
of data can place a 95\% confidence
limit on the compositeness scale $\Lambda$ of 140 TeV \cite{BarklowEE}

\begin{figure}
\centerline{\psfig{file=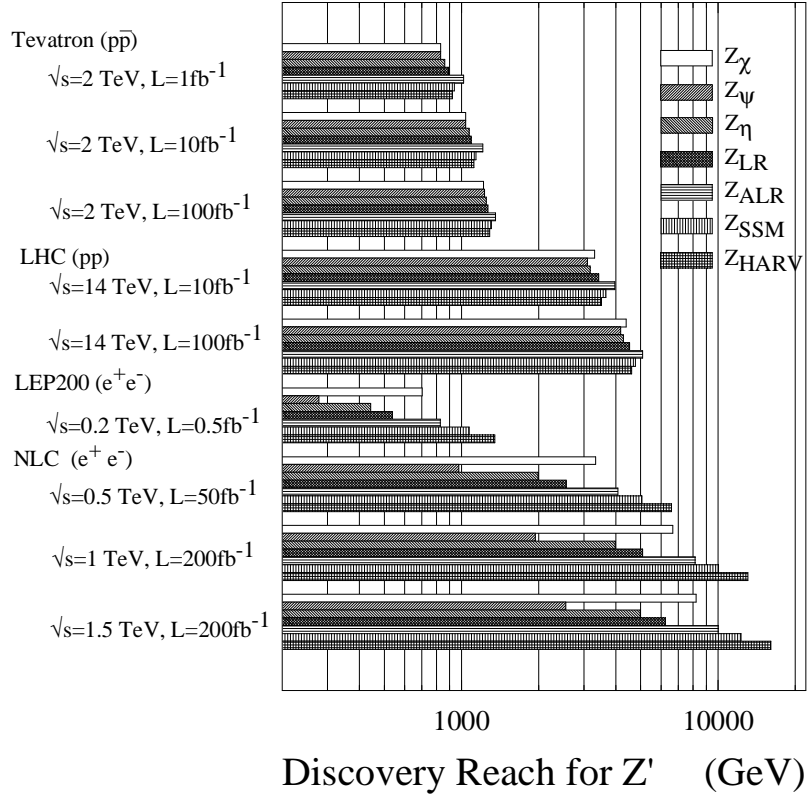,width=\textwidth}}
\caption[newZ]{Comparison of the sensitivity of various colliders to 
        $Z'$ bosons, in seven different theoretical models, 
from \protect\cite{ZpDPF}. The Tevatron and LHC bounds are based on 
       10 events in the $\ee$ and $\mu^+\mu^-$ channels.  The $\ee$ collider
       bounds are 99\% confidence limits obtained from 
          cross sections and polarization
         asymmetries.}
\label{newZ}
\end{figure}
  
An example that  deserves particular attention 
is the case of a new gauge boson $Z'$ which couples to 
a $U(1)$  symmetry which extends the standard model gauge group.
Recent surveys of the phenomenology of such bosons are given in 
\cite{Hewett-Hawaii,CveticDel,ZpDPF}.  Such a boson can be discovered
at the LHC, as a peak in the invariant mass distribution of lepton pairs,
up to a mass of several TeV.  On the other hand, the few diagnostic tools
available at the LHC to determine the couplings of a $Z'$ are effective
only up to about 1 TeV.  A linear collider at $\ECM =1$ TeV would not 
be able to observe the resonance peak for such a heavy boson. However, it 
could
measure the couplings of this boson to each fermion species,
 given the known mass
value  supplied by the LHC, by measuring the interference effect of the 
boson on forward-backward and polarization asymmetries in the fermion
pair production, just as experiments at PEP, PETRA, and TRISTAN measured 
the couplings of fermions to the $Z^0$.  
The expected 
sensitivities in these two different types of measurements are shown
in Figure~\ref{newZ}.
 This gives a  particularly clear example of the 
potential synergism of $\ee$ and $pp$ experiments
at the TeV scale.
 
\section{CONCLUSION}
 
In this review, we have surveyed the expected experimental program of an 
$\ee$ linear collider operating in the energy region up to 1.5 TeV in the
center of mass.  We have described how this collider will be able to 
perform precision studies on the heaviest  particles of the standard model,
the $W$ boson and the top quark, and we have used these  examples to 
demonstrate the power of $\ee$ experimentation to give a  concrete picture
of a new physical system.

We then discussed the  potential of this collider
to explore new and undiscovered sectors of physics.  In our arguments,
 we have concentrated
our attention on the new physics that must be present at the TeV scale, the
physics that explains the spontaneous breaking of the electroweak symmetry.
We  surveyed proposed models of electroweak symmetry breaking and showed
that, for each case, the linear collider makes possible unique experiments
that are essential for understanding the new particles and interactions
that appear.  We showed how the analytical tools that are available  for
particles of the standard model also work  to illuminate states that 
lie outside the standard model.   We  considered the interplay
expected in these models between the results of $\ee$ and $pp$ experiments,
and  showed that the $\ee$ experiments typically supply  crucial  ingredients
needed  to interpret signals seen in the hadronic environment.

We do not know  what physics is waiting for us  at the next step in energy.
That is the puzzle that we must solve.  We have argued here that $\ee$
linear colliders are well matched to this task and will play a central 
role in this solution.

\bigskip
\noindent ACKNOWLEDGEMENTS
 
\medskip

We thank many friends at KEK,  LBL, and SLAC, and
in the broader community, who have educated us on the 
issues we have discussed here.  We are particularly grateful to 
Tim Barklow, David Burke, Keisuke Fujii, Howard Haber, Kaoru Hagiwara, 
Ken-ichi Hikasa, Gordon Kane, Akiya Miyamoto, and 
Peter Zerwas, who have influenced our perspective
through many years of arguments and conversations.
This work was supported by the National Science Foundation
under grant PHY-90-21139 and by the
Department of Energy under
contracts DE-AC03-76SF00098 and DE--AC03--76SF00515.

\end{document}